\documentclass[aps,notitlepage,superscriptaddress,pre]{revtex4-1}
\usepackage{ucs}
\usepackage[utf8x]{inputenc}
\usepackage{amsmath}
\usepackage{amsfonts}
\usepackage{amssymb}
\usepackage{amsthm}
\usepackage{fontenc}
\usepackage[dvips,pdftex]{graphicx}
\usepackage{setspace}
\usepackage{color}
\usepackage[dvipsnames]{xcolor}
\usepackage{graphicx}

\usepackage{hyperref}

\usepackage{multirow}
\usepackage{array}
\usepackage{caption}
\usepackage{subcaption}
\usepackage{tikz}
\usepackage{multimedia}
\usepackage{float}
\usepackage{rotating}
\usepackage{epstopdf}

\makeatletter
\@addtoreset{chapter}{part}
\makeatother

\newcommand{\dt}{\,\text{d}}

\newcommand{\pd}{\partial}
\newcommand{\Tr}{\text{Tr}}
\newcommand{\avg}[1]{\langle #1 \rangle}

\DeclareGraphicsExtensions{.pdf,.jpg}

\hypersetup{linktocpage}

\begin{document}
	\begin{abstract}
		We study the shape and shape fluctuations of incompatible, positively curved ribbons, with a flat reference metric and a sphere-like reference curvature. Such incompatible geometry is likely to occur in many self assembled materials and other experimental systems. Ribbons of this geometry exhibit a sharp transition between rigid ring and an  anomalously soft spring as a function of their width. As a result, the temperature dependence of these ribbons' shape is unique, exhibiting a non-monotonic dependence of the persistence and Kuhn lengths on the temperature and width. We map the possible configuration phase space and show the existence of three phases- at high temperatures it is the Ideal Chain phase, where the ribbon is well described by classical models (e.g- worm like chain model);  The second phase, for cold and narrow ribbons, is the Plain Ergodic phase - a ribbon in this phase might be thought of as made out of segments that   gyrate within an oblate spheroid with extreme aspect ratio; The third phase, for cold, wide ribbons, is a direct result of the residual stress caused by the incompatibility, called Random Structured phase. A ribbon  in this phase behaves on large scales as an Ideal Chain, however the segments of this chain are not straight, rather they may have different shapes, mainly helices (both left and right handed) of various pitches.
	\end{abstract} 
	\title{ Shape and Fluctuations of Positively Curved Ribbons}
	\author{Doron \surname{Grossman}}
	\email[]{doron.grossman@mail.huji.ac.il}
	\affiliation{Racah Institute of Physics, Hebrew University, Jeruslaem 91904, Israel}
	\author{Eran \surname{Sharon} }
	\email[]{erans@mail.huji.ac.il }
	\affiliation {Racah Institute of Physics, Hebrew University, Jeruslaem 91904, Israel}
	\author{Eytan \surname{Katzav}}
	\email[]{eytan.katzav@mail.huji.ac.il }
	\affiliation{Racah Institute of Physics, Hebrew University, Jeruslaem 91904, Israel}	
	\date{\today}
	\maketitle

	\section{Introduction}
	The shape and the shape transitions of elastic nano-ribbons/polymers in a thermal environment is of interest in many disciplines, from the study of DNA and protein folding \cite{Bouchiat1998}, via the evolution of bio-molecular structures such as amyloids \cite{Adamcik2011} and cholesterol aggregates \cite{Smith2001}, to the way synthetic polymers and self assemblies may serve in drug deliveries, food additives or templates for the production of nano structures \cite{Zhan2005,Oda2008,Ziserman2011}. In a thermal environment, such slender structures fluctuate (gyrate) and their shape may vary widely depending on the temperature, size, chemical potential and other control parameters. Nano-ribbons and polymers are usually studied using simplified models such as the worm-like chain model, freely jointed chain and their variants {\cite{Sadowsky1930,Panyukov2000a,Giomi2010, Rubinstein2003}}, all assume no residual elastic stresses i.e., the ribbon/rod on its equilibrium configuration is stress free. However, many  (in fact most) nano-ribbons and polymers are made of complex  elements, which  likely do not fit perfectly to each-other when forming extended aggregates such as sheets and ribbons. They are incompatible, and as such the resulting structure would be residually stressed. Recent theoretical and experimental works \cite{Armon2011,Armon2014,Guest2011,Levin2016} show that incompatible slender structure undergo various non trivial shape transformation and that their mechanics could differ a lot from seemingly similar, compatible, structures. These characteristics are likely to affect the resulting shape fluctuations of nano-metric self-assemblies \cite{Aggeli1997,Ziserman2011,Oda1999}.
	
	In this work we apply the theory of incompatible elastic sheets to study the statistical mechanics of thermal ribbons that have spontaneous isotropic positive curvature. We use our recently developed effective 1D elastic model  \cite{Grossman2016} to derive expressions for the persistence length $\ell_p$, the Kuhn length $\ell_k$ and the gyration radius $R_g$ and obtain the phase diagram, which characterizes different types of ribbon conformation statistics unique to such ribbons.

\begin{figure}[h!]
		\centering
		\includegraphics[width=0.9\textwidth]{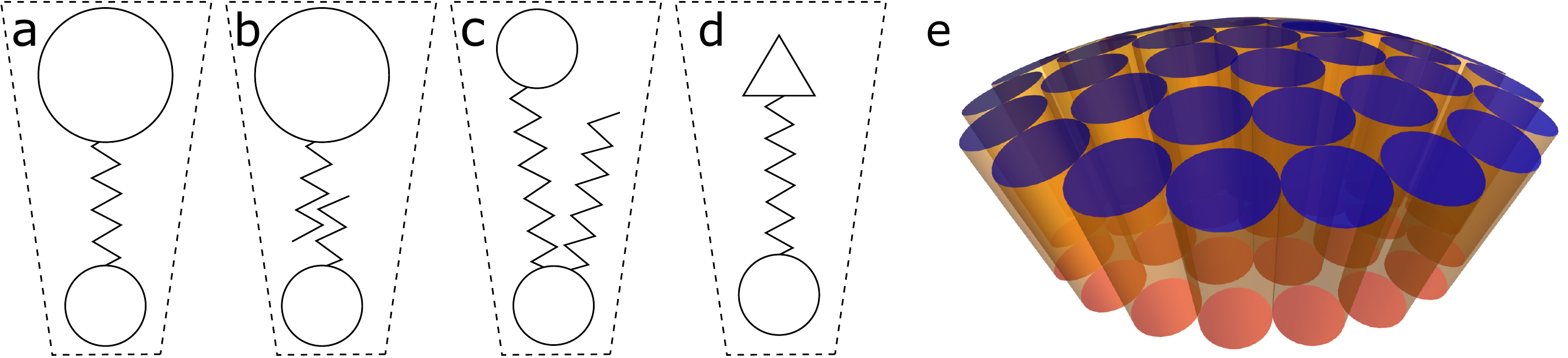}
		\caption{Illustration of systems with positive spontaneous curvature (as indicated by the "form" wedge - dashed lines). Zigzag lines correspond to carbon chains, different head groups are marked by triangles and circles. a- Asymmetric bola-amphiphile mono-layer; b- Asymmetric bi-layer; c- Other shape asymmetries of the constituent molecules (in this case another carbon "tail"; d- Also chemical differences (in this case different surface tension); e- 3D visualization  of the way such systems self assemble into positively curved surfaces.  }\label{fig:spherical_chem}
	\end{figure}

	Ribbons with spontaneous positive curvature may be produced in the lab (see for e.g \cite{Guest2011,Hu2016,Zhang2017}), or arise naturally in self assembled systems  with broken symmetry. Such systems may be  uneven  semi-solid bi-layers \cite{Yesylevskyy2014}, or  asymmetric bola-amphiphile mono-layer with two distinct sides to it (say different size head groups with strong same-type affinity), e.g- systems such as in
	\cite{Masuda2004,Zhan2005,Marson2014}. Microscopically, such a geometry arises in asymmetric semi-solid mono/bi-layers as a result of two competing geometries: the (in-plane) bond-induced geometry, i.e- the spatial arrangement that best satisfies the intermolecular bonds; and the molecular shape  geometry. The former depends solely on the number and direction of bonds between different molecules, the latter only on the shape of a single molecule. In this paper, we consider only systems whose bond-geometry is a flat, Euclidean geometry (as is ,in fact, very common- e.g,  all defect-free crystalline structures). This is to say that the preferred distances between neighbouring elements are satisfied (everywhere) by setting them in a planar geometry. In contrast, the molecules' shape  may prescribe a very different geometry-   molecular asymmetry (corresponding to the bi/mono-layer asymmetry ), which may arise due to different sized headgroups, or difference in surface tension, results with a curved geometry (Figure \ref{fig:spherical_chem}), as the intermolecular distance on one sige are larger than those on the other.  Unless anisotropy is introduced, the system has the same preferred  curvautre, $k_0$ is all directions.

	We begin this paper by studying the (mechanical) equilibrium shape of positively curved ribbons (section \ref{s_ch:Euilbrium}), deriving their unique conformational behaviour which includes shape transition at a critical width, abnormal floppiness of wide ribbons and dominance of boundary layer in setting ribbon's rigidity. We then (section \ref{s_ch:StatMech}) show that these unique results manifest themselves as unusual statistical behaviour. We divide the problem into bending dominated and stretching dominated regimes (section \ref{sub:NarrRibb} and \ref{sub:WideRibb}), and study common measures of conformation in elastic ribbons at these regimes. These include the persistence length $\ell_p$, describing the  decay tangent-tangent correlations, the Kuhn length $\ell_k$ and the gyration radius $R_g$.  We show that these measures on their own fail to wholly capture the phase diagram, and derive new measures to assist in this  task. We conclude the paper in \ref{ch:Conc}, where we calculate the phase diagram of such ribbons and discuss the meaning and implications of the results.  We suggest that in general, incompatible ribbons will have a phase diagram richer than compatible ones, since in addition to high/low temperature regimes, for such ribbons there are the wide/narrow width regimes. The transitions between them and the resultant abnormal mechanics determine a wider {configurational}  phase space.

	\section{Elastic Modeling of Positively Curved ribbons}\label{ch:PosRibbons}
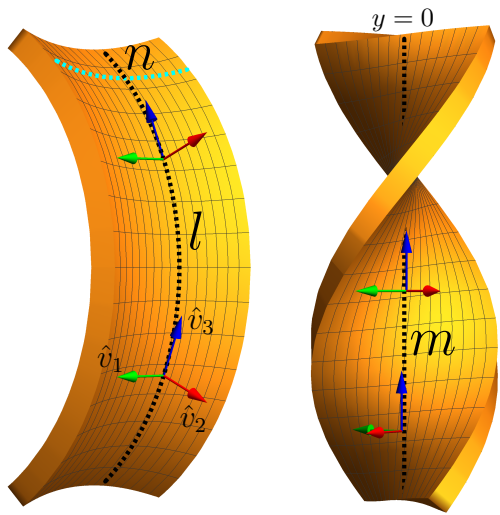
\begin{figure}[!h]
		\centering
		\begin{tikzpicture}[scale=1]
		\node[above, right] (ln) at (0,0) {\includegraphics[clip=true, trim={0.5cm 0.1cm 0.4cm .1cm}, width=.2\textwidth]{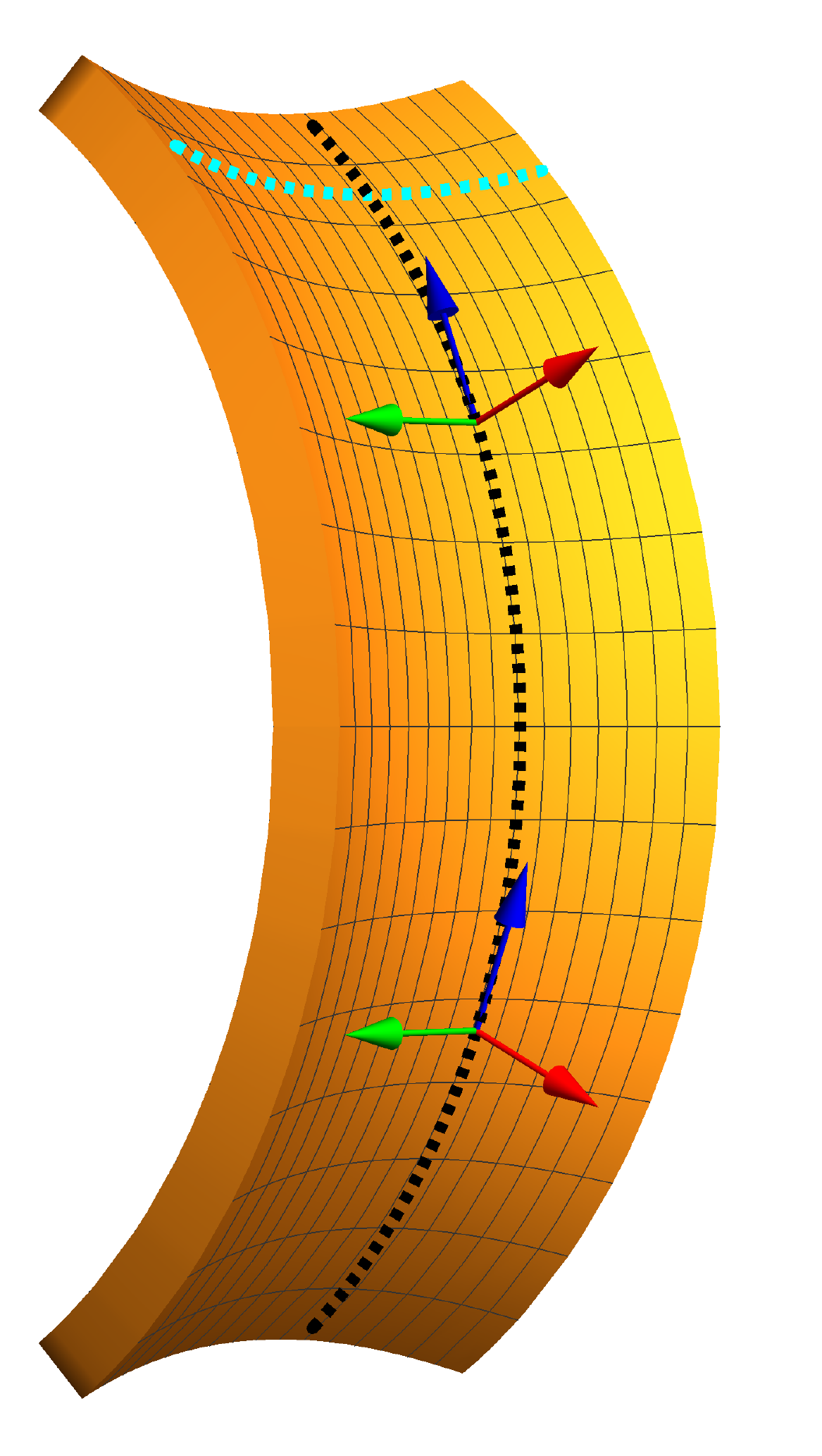}};
		\node[above, right] (m) at (3.5,0) {\includegraphics[clip=true, trim={0.1cm 0.1cm 0.1cm 0.1cm}, width=.2\textwidth]{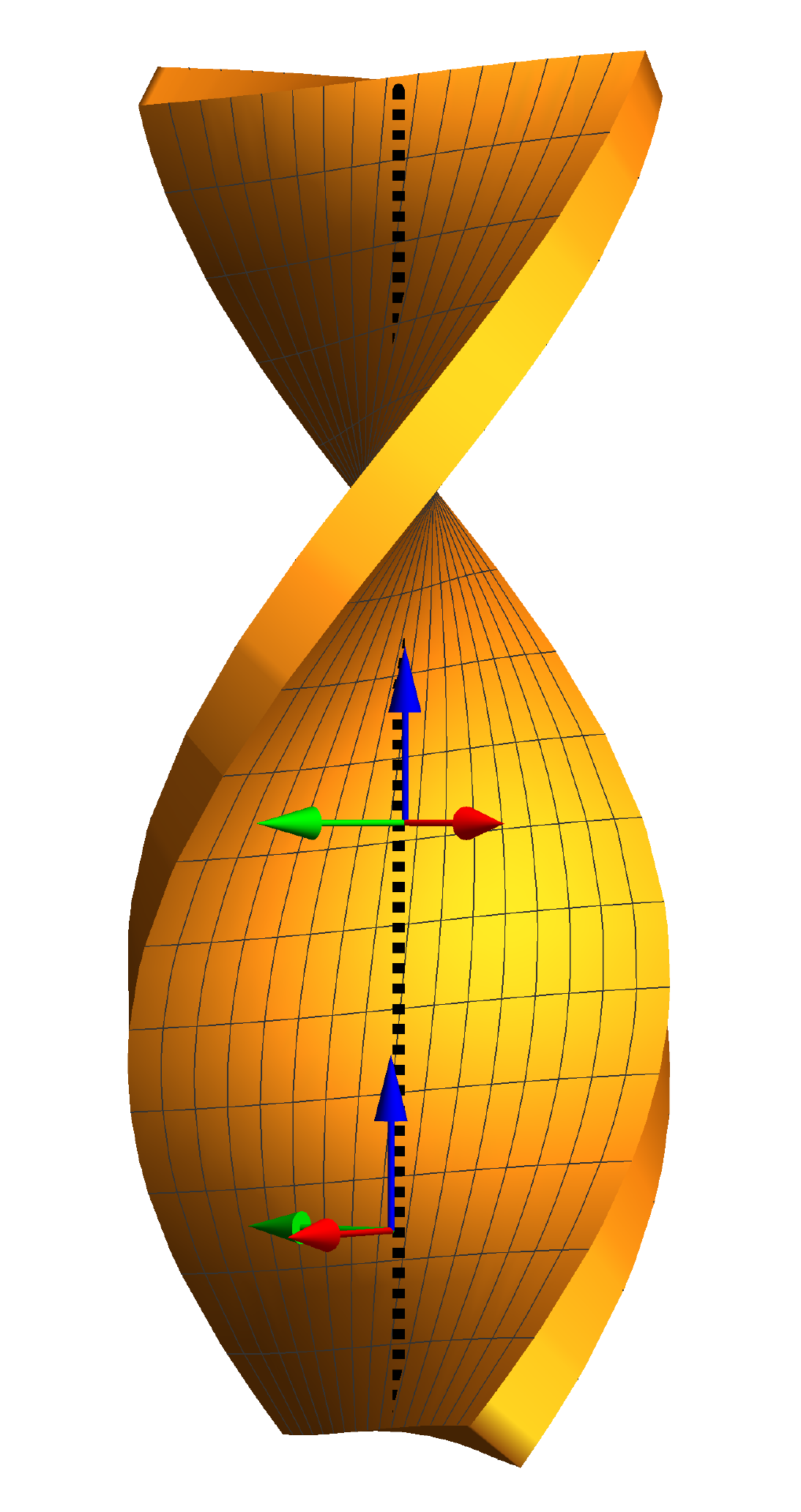}};
		\node[above, right] (binorm) at (1.2,-1.2) {\large$\hat{v}_1$};
		\node[above, right] (norm) at (2.3,-2) {\large$\hat{v}_2$};
		\node[above, right] (tang) at (2.4,-.7) {\large$\hat{v}_3$};
		\node[above, right] (y0) at (4.85,3.3) {$y=0$};
		\node[above, right] (L) at (2.4,0.5) {\huge $l$};
		\node[above, right] (N) at (1.55,2.8) {\huge $n$};
		\node[above, right] (M) at (5.35,-1) {\huge $m$};
		\end{tikzpicture}
		\caption{Visualization of the curvatures. Black line is the mid-line ($y=0$). Left- a ribbon with $l$ (bent mid-line) and $n$ (the transverse curvature along the cyan, $x=const$, line) with opposite sign ($m=0$). Arrows correspond to the local frame at two different positions along the ribbon mid-line (Blue for tangent vector- $\hat{v}_3$, Green for bi-normal - $\hat{v}_1$ and Red for the normal vector $\hat{v}_2$). $l$ corresponds to curvature along the tangent, and  $n$ along the bi-normal. Right- A ribbon with pure twist, $m$, around the mid-line ($l=n=0$).} 
		\label{fig:lmn}
	\end{figure}

	Within a continuum mechanics description, a ribbon is a thin, narrow, and long sheet with thickness, width and length which satisfy $t\ll W \ll L$ respectively. We choose coordinates on the ribbon $(x,y)\in \left[0,L\right]\times\left[-\frac{W}{2},\frac{W}{2}\right]$, such that the line $y=0$ parametrizes the ribbon's mid-line. Following \cite{Panyukov2000a} and \cite{Grossman2016} we assume that  the ribbon's shape $\vec{r}(x,y)$ is  well described by the curvatures at the mid-line. These are the  normal curvature ($l(x)$), the twist ($m(x)$) describing the mid-line's shape  and transverse curvature ($n(x)$) describing the ribbon's profile (see Fig \ref{fig:lmn}).

	Formally, we assign a Darboux frame at the ribbon's mid-line ($\left\{\hat{v}_1,\hat{v}_2,\hat{v}_3\right\} $) such that $\hat{v}_3(s)$ is tangent to the ribbon's mid-line, $\hat{v}_2(s)$ is  normal to the ribbon (pointing outside the ribbon) and the bi-normal $\hat{v}_1 (s)$ pointing along the width (narrow dimension) of the ribbon. This Darboux frame satisfies the generalized Frenet-Serret equations (see e.g \cite{Panyukov2000})
	
	\begin{subequations} \label{eq:Gen_FrenetSerret}
		\begin{align} 
		\hat{v}'_3&= l \, \hat{v}_2 \\ \nonumber 
		\hat{v}'_2&= -l \, \hat{v}_3  - m \, \hat{v}_1 \\ \nonumber 
		\hat{v}'_1&= m\, \hat{v}_2. \\ \nonumber 
		\end{align}
		
		where $X' \equiv \pd_x X $ is the derivative of $X$ along the ribbon. In principle, there is also contribution from the geodesic curvature $k_g$ of the mid-line, however, we assume it is zero. The general solution of these equations, given the frame at some position $x'<x$ along the ribbon is
		
		\begin{align}\label{eq: Solution}
		\hat{v}_i(x) & =O_{ij}(x,x') \, \hat{v}_j(x'), 
		\end{align}
		where $O_{ij}$ is an element of the  a rotation matrix 
		\begin{align}
		\mathbf{O}(x,x') &= T_x\left[e^{-\int_{x'}^x \mathbf{\Omega} \dt x''}\right] = \lim\limits_{N \rightarrow \infty} e^{-\mathbf{\Omega}(x_{N}) \Delta x}e^{-\mathbf{\Omega}(x_{N-1}) \Delta x}...e^{-\mathbf{\Omega}(x_1) \Delta x},
		\end{align}
		where $T_x$ is the position ordering operator, $\Delta x = \frac{x-x'}{N}$ and $x\geq x_N > x_{N-1} > \dots >x_1 \geq x'$.
		$\mathbf{\Omega}$ is the generator of rotations, given by
		\begin{align}
		\mathbf{\Omega}(x) &= \left(\begin{array}{ccc}
		0 & -m(x) & 0 \\
		m(x) & 0 & l(x) \\
		0 & -l(x) & 0
		\end{array}\right).
		\end{align} 
		The mid-line's configuration is then given by
		\begin{align}
		\vec{r}(x,0) &= \int_0^x \hat{v}_3(x') \dt x' ,
		\end{align}
		and the configuration of ribbon in 3D Euclidean space is 
		\begin{align}\label{eq:Rib_Shape}
		\vec{r}(x,y) &\simeq \vec{r}(x,0) + y \, \hat{v}_1(x) + \frac{1}{2} y^2 n(x) \hat{v}_2   	
		\end{align}  	
	\end{subequations}
	where  $-\frac{W}{2}\leq y \leq \frac{W}{2}$ along the ribbon's width. 
	
	\colorlet{deepg}{green!20!black}
	\colorlet{deepb}{blue!50!black}
	The shape of a an elastic ribbon is described  by  its metric ($a$) describing distances between neighboring material elements and curvature tensor ($b$) describing  relative angles between neighboring elements. The actual shape is then determined by minimizing the the elastic energy \cite{Efrati2009a} which is schematically given by 
	\begin{align}
	H &= \int  {\color{deepb} t \left|a- \bar{a}\right|^2} +{\color{deepg} t^3 \left|b-\bar{b}\right|^2 }\dt S 
	\end{align}
	Where $\dt S$ is an area element of the ribbon. This energy is composed of two terms-  the first is the stretching energy (dark blue), which is linear in the thickness and penalizes metric ($a$) deviations from the {reference metric ($\bar{a}$)} (describing preferred in-plane distances between material elements), while the second is the bending energy (dark green), cubic in $t$, penalizes deviations from the reference (or spontaneous) curvature \cite{Efrati2009a}. In the reduced 1D model, the energy functional takes the form: (\cite{Grossman2016})
	\begin{align}\label{eq:hamiltonian1D}
	\nonumber
	H =& \frac{Y}{8 \left(1-\nu ^2\right)} \int
	\left\{ {\color{deepb} \frac{1}{80} t W^5 \left(\bar{K}- n \,l + m^2\right)^2 } 
	\right. \\  &
	+{\color{deepg} \frac{1}{3} t^3 W \left[2 (1-\nu ) \biggl( \frac{W^2}{12} \left(n'\right)^2 \biggr. \right.}  \\ \nonumber  
	& {\color{deepg} \left. \biggl.  
		-  \left( \bar{l} - l\right) \left(\bar{n} - n \right)+\left(\bar{m} - m \right)^2  \biggr) 
		\right. }\\ \nonumber 
	& \left. {\color{deepg} \left.
		+ \left( \bar{l} + \bar{n}-n-l \right)^2 
		+\frac{W^2}{12} \left(m'\right)^2
		\right] } \right\}  \dt x,
	\end{align}
	where $Y$ is Young's modulus, $\nu$ is Poisson's ratio, $\left(\bar{l}(x), \bar{m}(x),\bar{n}(x)\right)$ are the \emph{reference} curvature components; 
	$\bar{K}(x)$ is the \emph{reference} Gaussian curvature which depends only on $\bar{a}$ and is related to preferred distances between neighbouring material elements. When $\bar{K} = \bar{l}\bar{n}-\bar{m}^2$  (i.e. the reference metric and curvature fulfill Gauss's \emph{Theorema Egregium}), the ribbon is said to be compatible since in this case setting the trivial solution $l=\bar{l},m=\bar{m}, n=\bar{m}$ is the only local and global solution for every width $W$ and thickness $t$. Generally, however, the ribbon is incompatible, $\bar{K} \neq \bar{l}\bar{n}-\bar{m}^2$ and there is no solution that can satisfy simultaneously both the stretching and bending terms, giving rise to residual stresses. It's important to note that the above energy functional accounts for various reference geometries, as well as mechanical limits. For example, in the case of un-stretchable, flat, compatible ribbon,  it reduces to the  well known Sadowsky functional (\cite{Sadowsky1930,Giomi2010}).
	
	We now turn to the specific case of positive spontaneous curvature and zero reference Gaussian curvature. Specifically $\bar{K}=\bar{m}=0$ and constant $\bar{l}=\bar{n}= k_0$. Such a geometry is evidently incompatible as $\bar{l}\bar{n}-\bar{m}^2= k_0^2 \neq\bar{K}$, and it describes a material which "wants" to have the shape of a sphere on one hand (prescribed by $\bar{l}$ and $\bar{n}$ in the bending term), while keeping it's in-plane distances as in a flat sheet on the other (as prescribed by the stretching term).  As mentioned before, this geometry is likely to arise naturally in self-assembled, uneven bi-layers and bola-amphiphile mono-layers in the $L_\beta$ (gel-like) phase. The preferred planar geometry is encompassed by  $\bar{K}=0$, while the isotropic difference between the two side is expressed by  $\bar{l}=\bar{n}=k_0$, $\bar{m}=0$.
	
	We begin in solving the mechanical equilibrium equations in subsection \ref{s_ch:Euilbrium}, and then move to the statistical behaviour of these ribbons in subsection \ref{s_ch:StatMech}.
	\subsection{Mechanical Equilbrium}\label{s_ch:Euilbrium}
	We begin by simplifying the Hamiltonian (\ref{eq:hamiltonian1D}) via changing of variables
	\begin{subequations}
		\begin{align}
		l &\equiv h+z\cos(\theta),\\
		n &\equiv h-z\cos(\theta),\\
		m &\equiv z\sin(\theta).
		\end{align}
		Thus we defined
		\begin{align}
		h &\equiv \frac{1}{2}\left(l+n \right),\\
		z &\equiv\sqrt{\frac{1}{4}\left(l-n \right)^2+m^2},\\
		\theta&\equiv \arctan\left( 2 \frac{m}{l-n}\right).
		\end{align}
		$h(x)$ is the mean curvature of the mid-line, $z(x)=\sqrt{h^2- K}$ (where $K= n l -m^2$ is the Gaussian curvature at the mid-line) measures the a-sphericity of the ribbon's geometry (in  the case of a spherical geometry $z=0$), and $\theta/2$ measures (for $z\neq 0 $ ) the  relative angle between the $x$ coordinate (ribbon's long axis) and the principal curvatures axes.
	\end{subequations}
	
	Using these variables (and neglecting derivatives) the Hamiltonian of a positively curved ribbon assumes the form 
	
	\begin{align}\label{eq:hamiltonian_spherical}
	\nonumber
	H =& \frac{Y}{8 \left(1-\nu ^2\right)} \int
	\left\{ \frac{1}{80} t W^5 \left(h^2 -z^2\right)^2  \right.
	\\  & \left.
	+ \frac{2}{3} t^3 W \left[(h-k_0)^2(1+\nu)+z^2(1-\nu)
	\right] \right\} \dt x.
	\end{align}

	Note, that the Hamiltonian, describing the bulk energy of a ribbon, is independent of $\theta$. {This implies that continuous deformations of the ribbon via change of theta (stretching the ribbon spring) cost no bulk energy and the ribbon is expected to be infinitely floppy \cite{Guest2011,Pezzulla2016}}. Such is the case for every "locally spherical" reference geometry (i.e- $\bar{m}=0$, $\bar{l}=\bar{n}=k_0(x)$). The degeneracy is partially lifted by the derivative terms, and by possible boundary layer (see \cite{Efrati2009} and Eq. \ref{eq:boundEner})-  at this point both are neglected, we will later see how they affect the result.
	
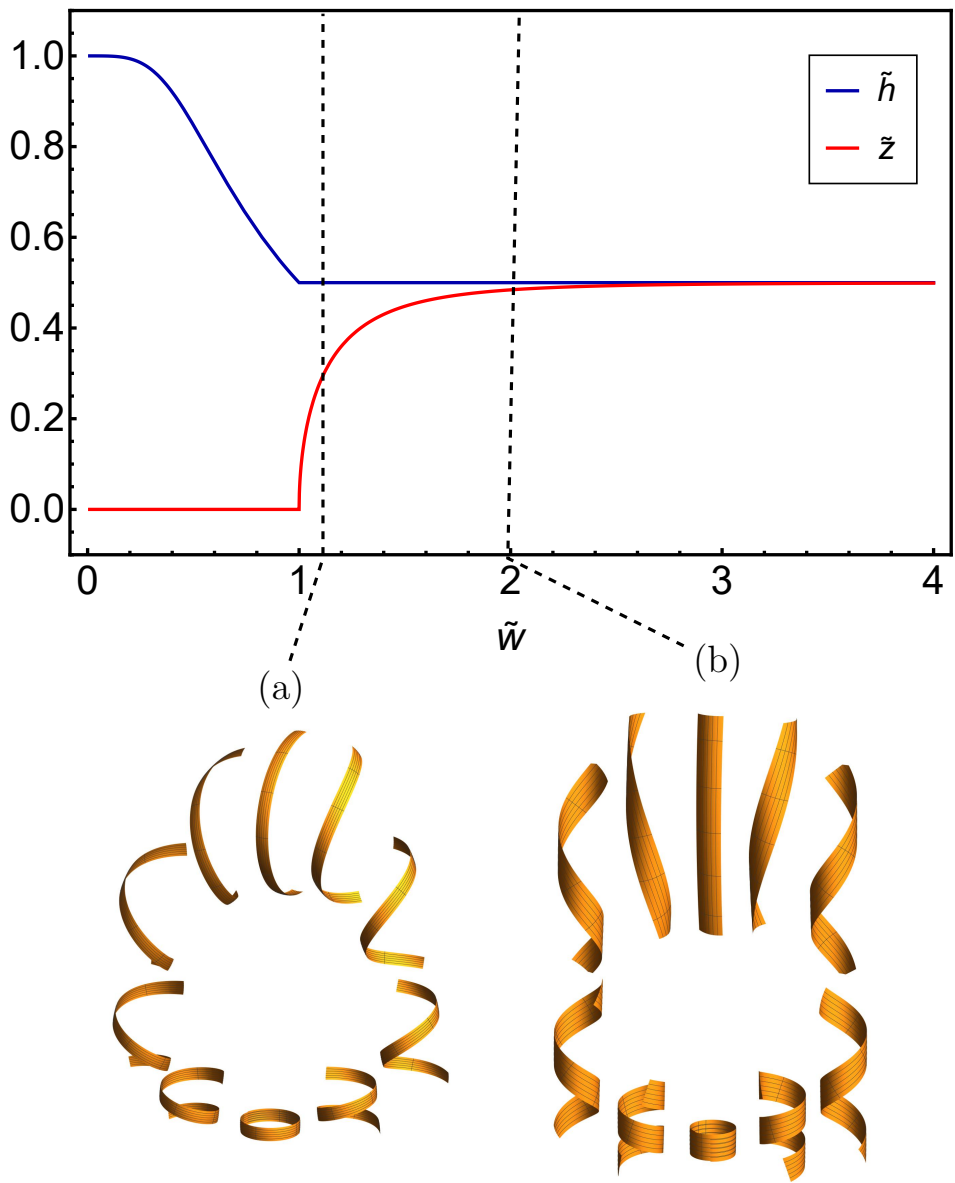
\begin{figure}
		\centering
		\begin{tikzpicture}{scale=1}
		\node[above] at (0,0){\includegraphics[width=.7\textwidth]{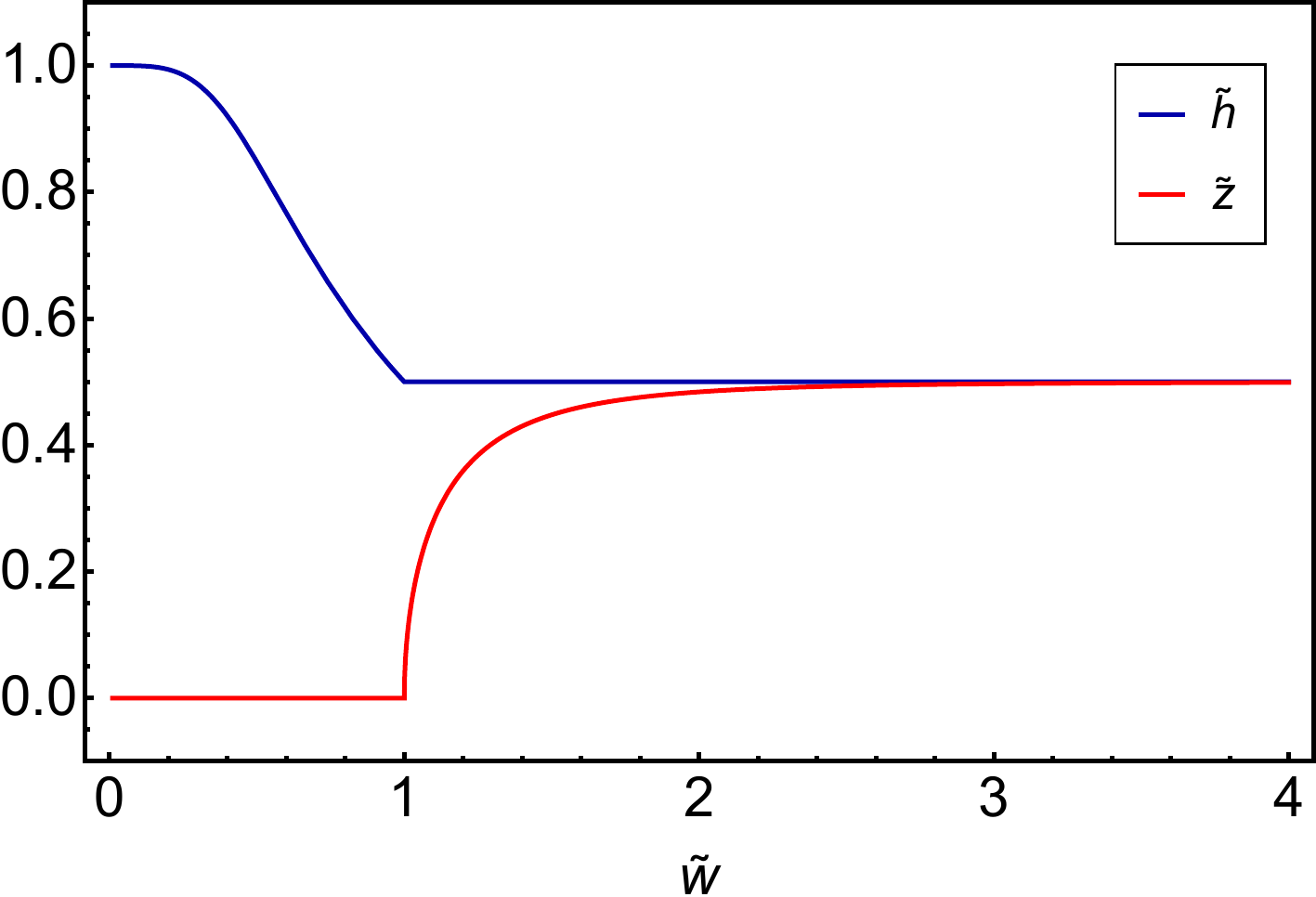}};
		\node[below](a) at (-3.8*0.7,0){\mbox{\Large (a)}};
		\node[below](b) at (4.5*0.7,.5*0.7){\mbox{\Large (b)}};
		\node[below] at (-4*0.7,-1*0.7){\includegraphics[clip, trim={0cm 0cm  0 5cm },width=0.35\textwidth]{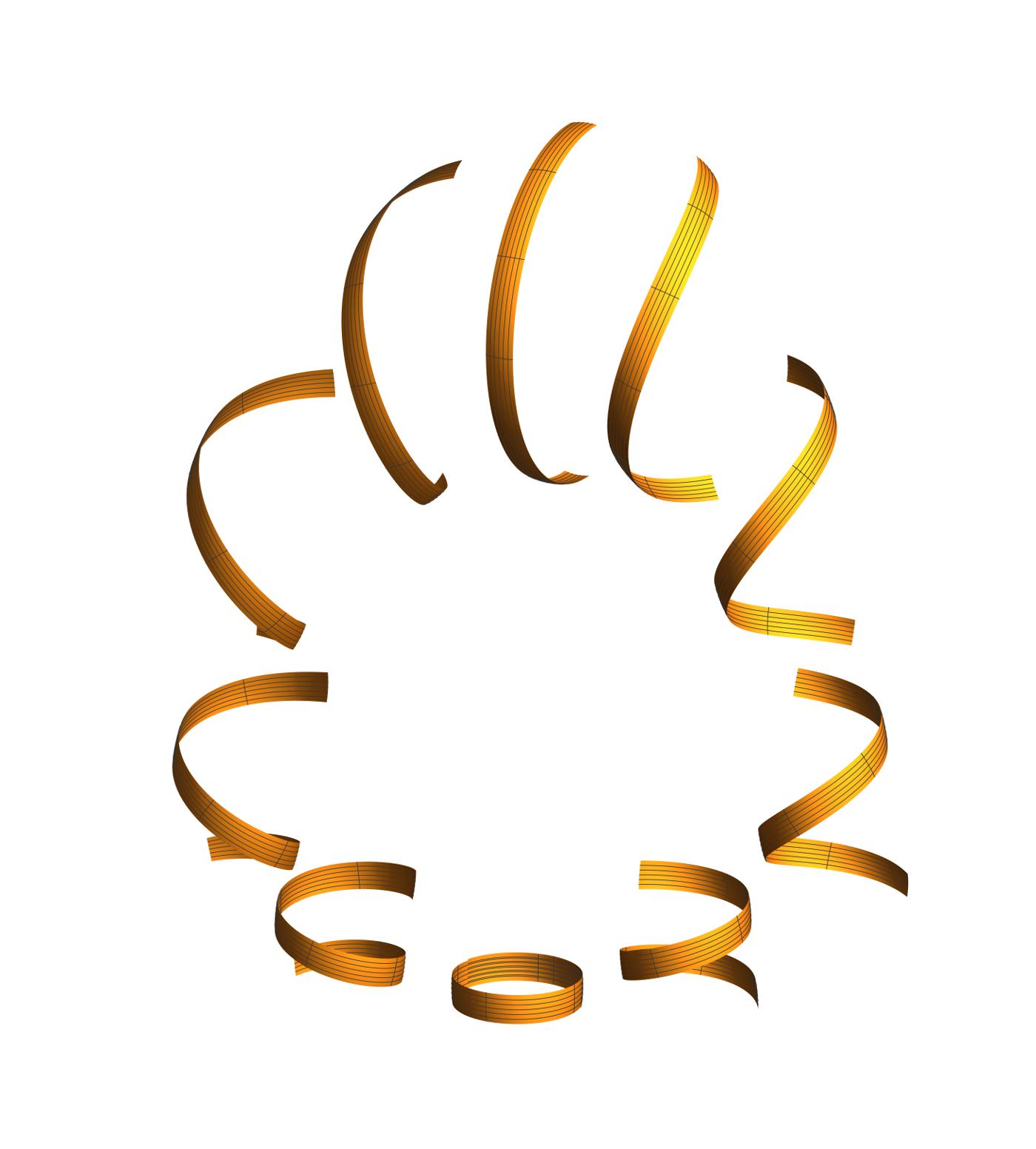}};
		\node[below] at (4.5*0.7,-0.3*0.7){\includegraphics[width=0.35\textwidth]{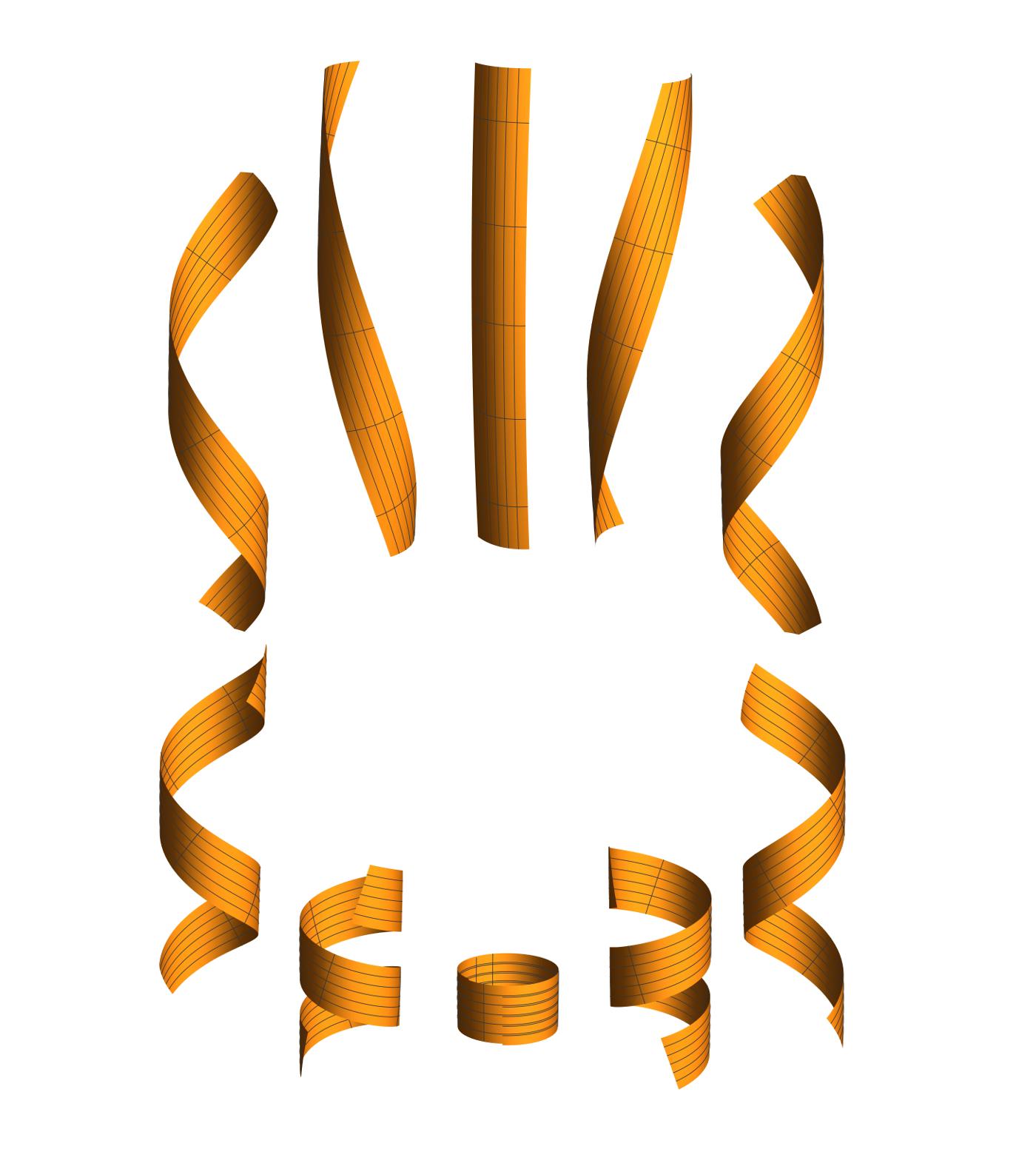}};
		\draw[dashed, very thick] (a)--(-3*0.7,1.95*0.7) -- (-3*0.7,12.3*0.7);
		\draw[dashed,very thick] (b)--(.51*0.7,1.95*0.7) -- (.51,12.3*0.7);
		\end{tikzpicture}
		
		\caption{Plots of the solution given by Eq. \ref{eq:equi} for the $\nu=0$ case. $\tilde{h}$ in Blue and $\tilde{z}$ in Red. Larger/smaller values of $\nu$ result with higher/lower asymptotic values.  A narrow ribbon ($\tilde{w}<1$ ) has a ring like shape with radius which depends on the width. A wide ribbon, at any $\tilde{w}>1$ may have a degenerate family of shapes related by a single parameter $\theta$. These shape are visualized at two given widths $\tilde{w}=1.075$ and $\tilde{w}=2$ (markers (a) and (b) respectively). Below  the main figure.}\label{fig:curvs}
	\end{figure}

	By defining the length scale (critical width), and energy scale.
	\begin{subequations} 
		\begin{align}
		W^*&= 2^{3/2} \left(\frac{5\left(1-\nu\right)}{3}\right)^{1/4} \sqrt{\frac{t}{(1+\nu)k_0}}\\
		E&=\frac{5^{1/4} ~Y t^{7/2} k_0^{1/2}}{ 3^{5/4}\sqrt{2}\left(1-\nu\right)^{3/4}(1+\nu)^{7/2}},
		\end{align}
		whose significance will be shown immediately, we may transform into dimensionless variables-
		\begin{align}
		\tilde{h} &= h/k_0\\
		\tilde{z} &= z/k_0\\
		\tilde{w}&=W/W^*\\
		\tilde{x}&= k_0 x,\\
		\end{align}
		where $\tilde{h}$ is the dimensionless mean curvature of the mid-line, $\tilde{z}$ is the dimensionless a-sphericity, $\tilde{w}$ is dimensionless width and $\tilde{x}$ is dimensionless arc-length along the mid-line.
	\end{subequations}
	We may then rewrite the Hamiltonian 
	\begin{align}\label{eq:hamiltonian_spherical_dimless}
	\nonumber
	H =&  E \int
	\left\{2(1-\nu) \tilde{w}^5 \left(\tilde{h}^2 -\tilde{z}^2\right)^2  \right.
	\\  & \left.
	+ \left(1+\nu\right)^2\tilde{w} \left[(\tilde{h}-1)^2(1+\nu)+\tilde{z}^2(1-\nu)
	\right] \right\} \dt \tilde{x} \\ \nonumber
	\end{align} 
	The equilibrium equations obtained by variation of $H$ with respect to the mean curvature $\tilde{h}$, and a-sphericity $\tilde{z}$:
	
	\begin{subequations}
		\begin{align}
		4\left(1-\nu\right)\tilde{w}^4\tilde{h}\left(\tilde{h}^2-\tilde{z}^2\right)+ \left(1+\nu\right)^3 \left(\tilde{h}-1\right)=0
		\end{align}
		\begin{align}
		2\left(1-\nu\right)\tilde{z}\left[4 \tilde{w}^4\left(\tilde{h}^2-\tilde{z}^2\right)-\left(1+\nu\right)^2\right]=0
		\end{align}
	\end{subequations}
	The solution is given by-
	\begin{subequations} \label{eq:equi}
		\begin{align}
		\tilde{h}_0&= \left\{ \begin{array}{cc}
		\frac{1}{2}(1+\nu) \frac{\left(\tilde{\Xi}^2(\tilde{w})-(1-\nu^2) 3^{1/3}\right)}{3^{2/3}(1-\nu)\tilde{\Xi}(\tilde{w})\tilde{w}^2} & \tilde{w}\leq 1 \\
		\frac{1}{2} \left(1+\nu\right) & \tilde{w}>1
		\end{array}\right.	
		\\
		\tilde{z}_0&= \left\{\begin{array}{cc}
		0 & \tilde{w} \leq 1 \\
		\frac{1}{2} \left(1+\nu\right) \frac{\sqrt{\tilde{w}^4-1}}{\tilde{w}^2} & \tilde{w} >1
		\end{array}\right.,
		\end{align}
		where
		\begin{align*}
		\tilde{\Xi}(\tilde{w})&=\left[9\left(1-\nu\right)^2 \tilde{w}^2+\sqrt{81\left(1-\nu\right)^4 \tilde{w}^4+3\left(1-\nu^2\right)^3}\right]^{1/3}.
		\end{align*}
	\end{subequations}
	As can be seen $\tilde{w}=1$ is the critical dimensionless width where a sharp, yet continuous transition between two regimes occurs:  bending dominated, narrow regime ($0<\tilde{w}<1$), and stretching dominated, wide regime  ($1<\tilde{w}$, Fig. \ref{fig:curvs}). Also note, that these equations are identical to those written in \cite{Grossman2016} for ribbons with negative (saddle) reference curvature,
	under the map: $\tilde{h} \leftrightarrow \tilde{m}$, $\tilde{z} \leftrightarrow \tilde{l}$, $\nu \leftrightarrow -\nu$.

	When narrow ($\tilde{w}<1$, $\tilde{z}=0$), the ribbon is shaped as a ring taken around the equatorial of a sphere, who's radius grows monotonically with  the width (around $\tilde{w}=0$, $R(\delta\tilde{w})-R(0) \propto  \delta\tilde{w}^4$ respectively), up until the critical width. Above it (wide regime), the system is locally an ellipsoid (hence non-zero a-sphericity), where the mean curvature remains constant, but the principal curvatures change such that one of them reaches zero at the limit $\tilde{w} \gg 1$ (geometry of a cylinder, see  Fig. \ref{fig:curvs}).
	
	The angle $\theta$ has no physical meaning in the bending dominated (narrow) regime, as the system is locally a sphere (as indicated by the zero a-sphericity, $z=0$ ). However, in the stretching dominated (wide) regime it relates to the local shape, the degeneracy in $\theta$ means there is a degenerate continuous family of configurations all related by a local change in $\theta$. Physically, however, this huge degeneracy is partially lifted, as variation along the ribbon's length cost energy ($\Delta E _{der} \propto Y t^3 W^3 (m')^2$ and a similar expression for $n$), thus the ground states are uniform along the ribbon. They correspond to shapes varying continually between a ring and different helices  (both left and right handed). For a very wide ribbon, a straight mid-line (infinite pitch) is also a possible solution. In  Fig.\ref{fig:curvs} we see some of the possible configurations of a wide ($\tilde{w}>1$) ribbon, close to the transition (a), and far from it (b). For a given width, all the configurations have the same energy according to Eq. \ref{eq:hamiltonian_spherical_dimless}, and are related by different values of $\theta$. The limit $\tilde{w} \gg 1$  recovers the pre-stressed metal ribbons studied by Guest et al.(\cite{Guest2011})
	
	The remaining degeneracy (as depicted in Fig. \ref{fig:curvs}) is lifted by a boundary layer \cite{Efrati2009}, where the local geometry is bending dominated (rather than the bulk which is stretching dominated at $\tilde{w}>1$). In other other words, the local geometry is set by the spontaneous curvatures ($\bar{l},\bar{m}, \bar{n}$) rather than the reference {metric} ($\bar{K}$). This results with a boundary layer along the ribbon's width, such that exactly at the ribbon edge $\tilde{n}=\bar{n}$. The boundary layer's energy  for $\tilde{w}\gg1$ and $\nu=0$  is given by (for a general expression for any $\tilde{w}>1$ and $\nu$ see appendix \ref{app:Bound_Layer})
	\begin{align}\label{eq:boundEner}
	H_{bound}&\propto -Y t^{7/2}k_0^{3/2}  \int |\cos(\theta/2)|^3  \dt x. 
	\end{align}
	Therefore, the boundary layer defines the actual shape of a wide ribbon at mechanical equilibrium (a "ring" configuration ($\theta=0$) is preferential). The resulting shape in such cases is therefore sensitive to the shape and orientation of the edges of the elastic sheet (\cite{Pezzulla2016}). Thus by controlling the shape of the edge one is able to produce many different shape from the same material.  
	
	However, as can be seen from  Eq.(\ref{eq:boundEner}), the boundary layer's energy (and also size- see  \cite{Efrati2009,Levin2016})  is independent  of the ribbon's width (in contrast with the bulk energy), and vanishes at the infinitely thin limit. Therefore, in the context of statistical mechanics, we may consider the degeneracy of a global  angle $\theta$ only weakly broken. It can be shown (see section \ref{sub:curv_fluct} ) that for wide enough ribbons we may always work at temperatures where the boundary layer's contribution to the statistical behaviour of the ribbon is negligible.

	\subsection{Shape Fluctuations}\label{s_ch:StatMech}
	
	In a thermal environment, our ribbons fluctuate, causing their configuration to deviate from their (mechanical) equilibrium one. Characterizing these deviations is of great importance.  In principle, both boundary layer and derivatives contribute to the statistical properties of these ribbons. In practice, however, boundary layer contributes only to wide ribbons (no boundary layer for thin ribbons) and is negligible at the thin ribbon limit ($t\rightarrow0$, see the end of section \ref{sub:curv_fluct}). In what follows we assume a thin enough ribbon such that the boundary layer may be neglected. In other words, we assume that the temperatures are high enough so that the system is practically indifferent to the boundary layer.
	The derivatives contribution, is somewhat more subtle, as it affects at every temperatures range (there is always small enough scale such that fluctuations on that scale will be suppressed). Nevertheless, their effect (apart of a finite correlation length) is usually quantitative rather than qualitative. Such is the case in every ribbon who's equilibrium shape is extended (in contrast to compact, which is our case), such as in \cite{Panyukov2000a,Giomi2010,Ghafouri2005,Grossman2016}.  As will be seen, in positively curved ribbons, the derivatives affect some  aspects of  the thermal behaviour of the ribbon qualitatively. For simplicity, in what follows we neglect both boundary layer and derivatives from the analysis, unless otherwise mentioned.

	\subsubsection{Curvature Fluctuations}\label{sub:curv_fluct}
	One way to describe the statistical nature of elastic ribbons is to calculate the fluctuations in their curvatures which are the variables in the Hamiltonian. As such they are a natural choice to describe the statistical behaviour of a ribbon. Calculation is done in a similar manner to the one described in \cite{Grossman2016}, we limit ourselves to  the Gaussian approximation, i.e around the solutions in Eqs. \ref{eq:equi}. We start by expanding $H$ to second order about the equilibrium values
	\begin{align}
	H \simeq H_{0} + H_{(2)}.
	\end{align}
	In principle, $H_{(2)} =H_{(2)}\left[z(x),h(x),\partial_x z,\partial_x h,\partial_x \theta\right]$, however, as mentioned earlier, we start by neglecting derivatives. The average of a quantity Q is then given by the functional integral
	\begin{align}
	\langle Q \rangle = \frac{1}{\mathcal{Z}}\int Q e^{-\beta H_{(2)}\left[z(x),h(x)\right]}  \prod_x  z(x)\dt z(x) \dt h(x) \dt \theta(x). 
	\end{align}
	Where we defined the partition function  $\mathcal{Z}= \int e^{-\beta H_{2}\left[z(x),h(x)\right]}\prod_x  z(x)\dt z(x) \dt h(x) \dt \theta$. 
	At the given approximation, integration over the angle is trivial. Neglecting the energetic contribution of derivatives results with no correlation at different positions. i.e-
	$ \langle \Delta Q(\tilde{x}) \Delta Q(\tilde{x}')\rangle = \delta(\tilde{x}-\tilde{x}')\langle  \Delta Q^2 \rangle$. Hence, the averages and fluctuations of the curvatures are given by
	\begin{subequations} \label{eq:avgs}
		\begin{align}
		\langle \tilde{l} \rangle& = \langle \tilde{h} \rangle+ \langle \tilde{z}\rangle \langle \cos \theta \rangle = \langle \tilde{h} \rangle  = \langle \tilde{n} \rangle \equiv \tilde{h}_{eq}\\
		\langle \tilde{m} \rangle &=0 \\
		\langle  \Delta\tilde{l}(\tilde{x})  \Delta \tilde{l}(\tilde{x}')\rangle &=\langle \tilde{l}(\tilde{x}) \tilde{l}(\tilde{x}') \rangle - \langle \tilde{l} \rangle^2 =  \delta (\tilde{x}-\tilde{x}') \langle \Delta \tilde{l}^2 \rangle\\ \nonumber
		\langle \Delta \tilde{l}^2 \rangle &= \langle\Delta \tilde{h}^2 \rangle + \frac{1}{2} \left(  \langle\Delta \tilde{z}^2 \rangle \right) = \langle \Delta \tilde{n}^2 \rangle\\
		\langle \Delta \tilde{m}^2 \rangle &=  \frac{1}{2}   \langle\Delta \tilde{z}^2 \rangle \\	\nonumber
		\langle \Delta \tilde{h}^2\rangle&= \langle \left(\tilde{h}- \tilde{h}_{eq} \right)^2 \rangle = \langle \left(\tilde{h}- \tilde{h}_{0} \right)^2 \rangle- \left(\tilde{h}_{eq}-\tilde{h}_0\right)^2  \\ \nonumber
		\langle \Delta \tilde{z}^2\rangle&= \langle \left(\tilde{z}- \tilde{z}_{eq} \right)^2 \rangle = \langle \left(\tilde{z}- \tilde{z}_{0} \right)^2 \rangle- \left(\tilde{z}_{eq}-\tilde{z}_0\right)^2,
		\end{align}	
	\end{subequations}
	Where $\tilde{h}_0, \tilde{z}_0$ are given in Eq \ref{eq:equi}.

	Due to the non trivial measure (a result from $z$ being a non- negative variable), the resulting averages are cumbersome. They are given fully in appendix \ref{app:Specific_Calcs} and depicted in Fig. \ref{fig:moments}. It is clearly seen in Fig. \ref{fig:moments} (a)  that  the thermal equilibrium values diverge near $\tilde{w}=1$ indicating that the calculation breaks near the critical width. Indeed we perform in appendix \ref{app:Saddle_Point} a more accurate calculation using a saddle point approximation (taking into account the non-trivial measure in the integral) in which these values are finite. At low enough temperatures the treatment shown in Fig. \ref{fig:moments} coincides with the saddle point approximation. In any case, at low enough temperatures (even near $\tilde{w}=1$), the more accurate treatment affect only quantitatively and not qualitatively on any of the following results. We therefore settle for clarity over accuracy in what we show hereafter. For practical use, low enough temperatures and  $\tilde{w} \neq 0,1$,  we may approximate
	\begin{subequations}
		\begin{align} 
		\tilde{h}_{eq} \simeq &\tilde{h}_0 \\ 
		\tilde{z}_{eq} \simeq &\tilde{z}_0 + \left\{ \begin{array}{cc}
		\frac{\sqrt{\pi}}{2\sqrt{(1-\nu)\Psi \tilde{w}\left((1+\nu)^2-4 \tilde{w}^4\tilde{h}_{eq}^2\right)}} & \tilde{w}<1 \\
		0 & \tilde{w}>1
		\end{array} \right. \\
		\langle \Delta \tilde{h}^2 \rangle \simeq & \left\{ \begin{array}{c c}
		\frac{1}{2 \Psi \tilde{w} \left((1+\nu)^3  + 12 (1-\nu) \tilde{w}^4 \tilde{h}_{eq}^2\right)} & \tilde{w}<1 \\
		\frac{1}{4 \Psi \tilde{w}  (1+\nu)^2} & \tilde{w} >1
		\end{array} \right.\\ 
		\langle \Delta \tilde{z}^2 \rangle \simeq & \left\{ \begin{array}{c c}
		\frac{1}{ \Psi (1-\nu) \tilde{w}  \left((1+\nu)^2 \tilde{w} - 4 (1-\nu) \tilde{w}^5 \tilde{h}_{eq}^2\right)} & \tilde{w}<1 \\
		\frac{1+{(1-\nu)\tilde{w}^4}}{4 \Psi \tilde{w} (1-\nu) (1+\nu)^2 \left(\tilde{w}^4-1\right)} & \tilde{w} >1
		\end{array} \right.
		\end{align}\label{eq:flucs_simp}
	\end{subequations}
	Where $\Psi= \frac{E}{k_B T}=\frac{5^{1/4}}{ 3^{5/4}\sqrt{2}\left(1-\nu\right)^{3/4}(1+\nu)^{7/2}}	 \frac{Y t^{7/2} k_0^{1/2} }{k_B T}$, and we kept deviations of $\tilde{z}_{eq}$ from $\tilde{z}_0$ for the narrow ribbon as $\tilde{z}_0(\tilde{w}<1)=0$ and  they are of order $\sqrt{T}$ ( not $T$) and therefore important. It is worth to note that  even at high temperatures $\langle \Delta \tilde{h}^2 \rangle\propto 1/\Psi $, $\langle \Delta \tilde{z}^2 \rangle \propto 1/\Psi$, yet with different dependence on $\tilde{w}$.
	\begin{figure}
		\centering
		\begin{tabular}{cc}
			\begin{subfigure}{0.45\textwidth}
				\centering
				\includegraphics[width=\textwidth]{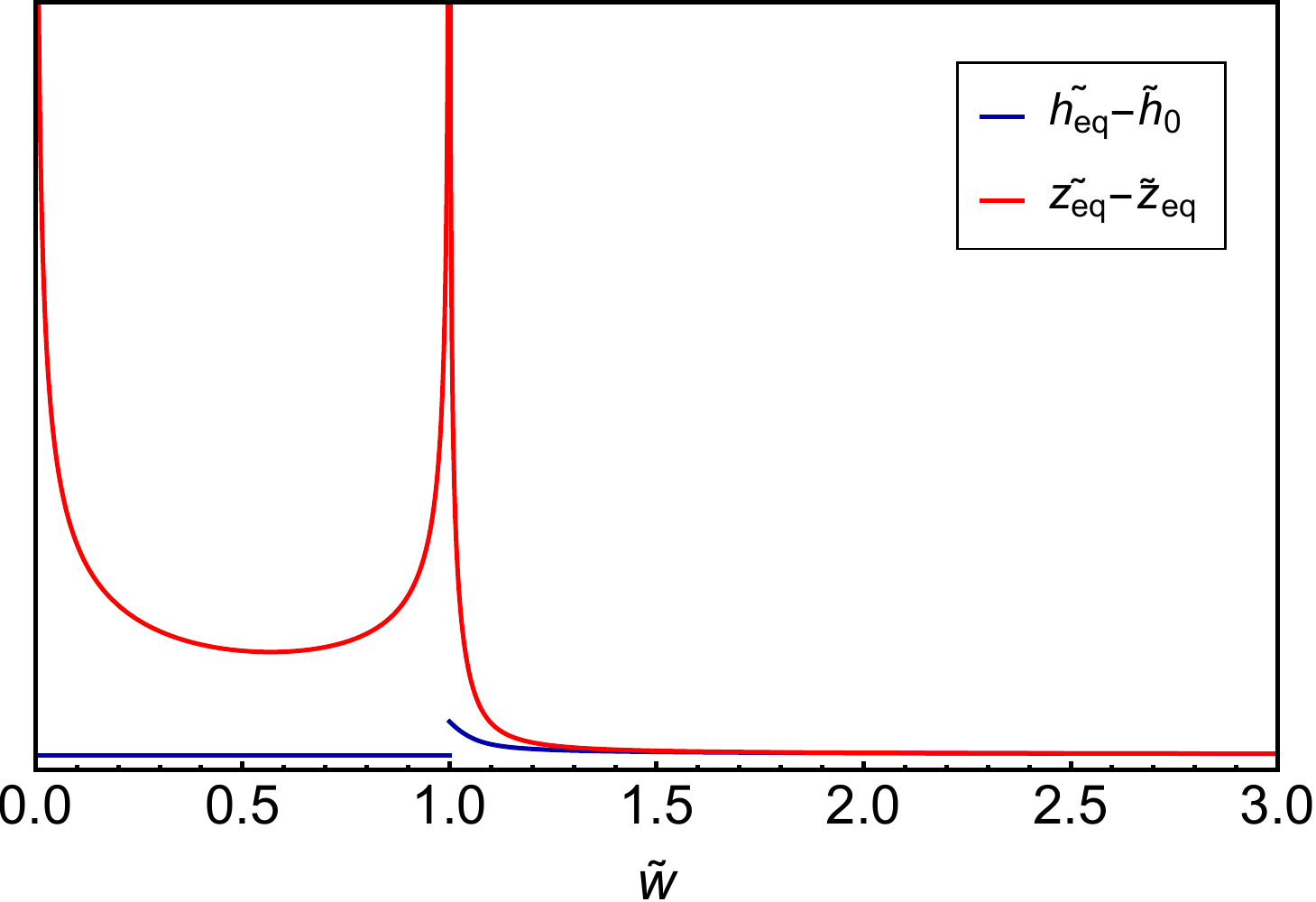}
				\caption{}\label{}
			\end{subfigure} &
			\begin{subfigure}{0.45\textwidth}
				\centering
				\includegraphics[width=\textwidth]{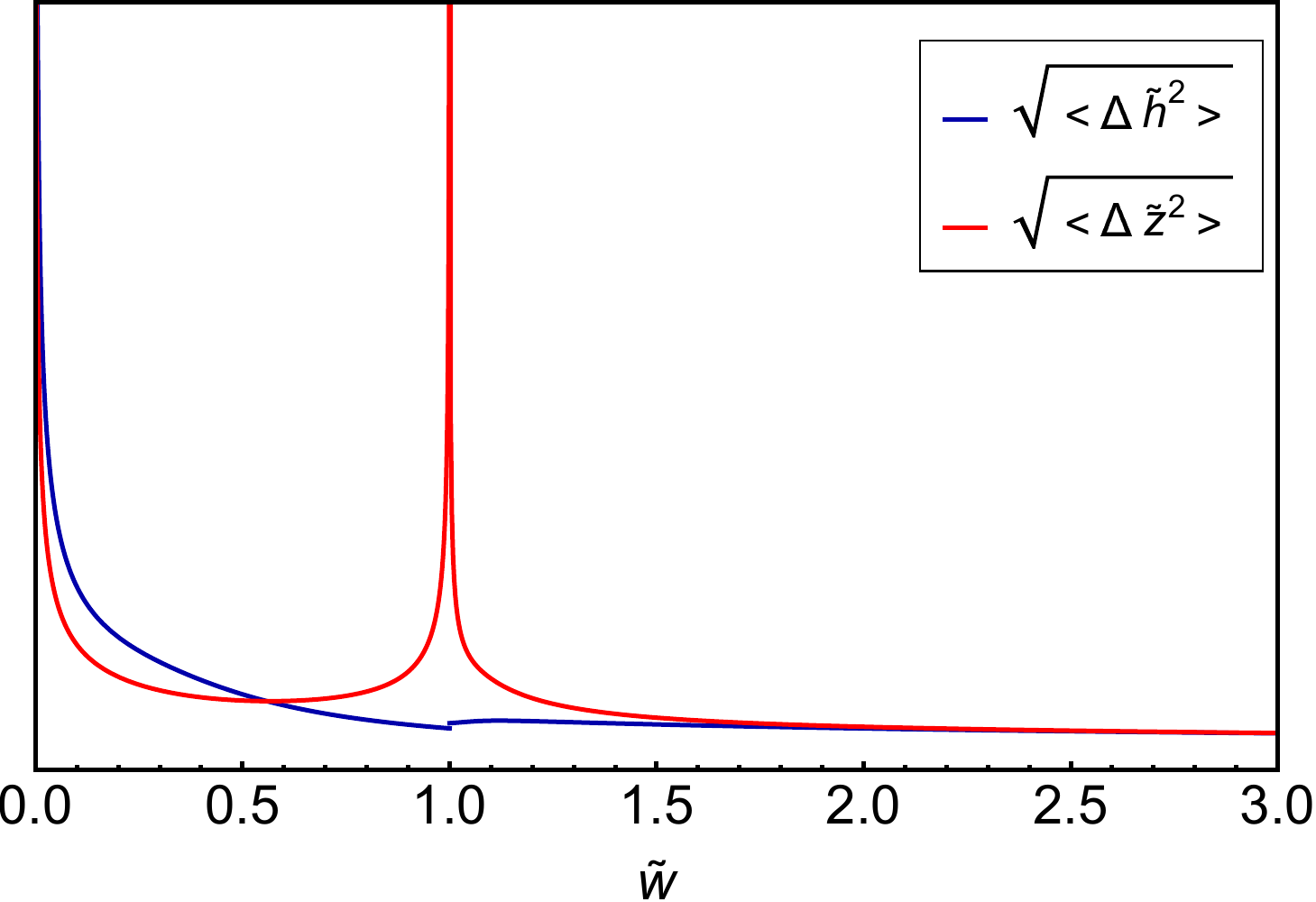}
				\caption{}\label{}
			\end{subfigure} \\
		\end{tabular}
		\caption{Moments of $\Delta \tilde{h}$ and $\Delta \tilde{z}$, for the case of $\nu=0, \Psi=\frac{E}{k_B T}=10$.  In Red are the moments of the mean curvature $\tilde{h}$, in Blue those of the a-sphericity $\tilde{z}$. Different values of $\Psi$ or $\nu$ results is similar graphs (differences are numerical only).} \label{fig:moments}
	\end{figure}
	
	While the statistics of a narrow ($\tilde{w}<1$) ribbon  are well captured even without including derivatives, some of the wide ($\tilde{w}>1$) ribbon's statistics  are governed by those terms. Therefore, we include here results of the correlations including those terms as they will come in handy soon enough. Expanding the Hamiltonian to 2$^{nd}$ order, and changing into Fourier space, we end up with an angle ($
	\theta$) dependent expression. However, to our needs, as we eventually integrate the angle out, we may approximate it without introducing any significant error as (see appendix \ref{app:Theta_avg} for a more detailed analysis)
	\begin{align}\label{eq:second_order_fourier}
	\Psi H_{(2)}&= \Psi  \int \dt q ~ \frac{1}{2} \left\{ \left[\left(8 \tilde{w}^5\left(1-\nu\right)\left(3 \tilde{h}^2_{eq} -\tilde{z}^2_{eq}\right)+2\tilde{w} \left(1+\nu\right)^3 \right)|\Delta \tilde{h}(q)|^2  \right.\right. \\ \nonumber 
	&\left. +\left(8 \tilde{w}^5\left(1-\nu\right)\left(3 \tilde{z}^2_{eq} -\tilde{h}^2_{eq}\right)+2\tilde{w} \left(1-\nu\right)\left(1+\nu\right)^2 \right)|\Delta \tilde{z}(q)|^2 - 32 \tilde{w}^5 \left(1-\nu\right)\tilde{h}_{eq} \tilde{z}_{eq}\Re(\Delta \tilde{z}(q) \Delta \tilde{h}^\dagger(q))\right] \\ \nonumber
	&  \left.+ 16 \sqrt{ \frac{5}{3} \left(1-\nu\right) } \frac{k_0 t}{1+\nu} \left[\frac{(1+\nu)^2\tilde{w}^3 q^2}{48} \left(4\left(1-\nu\right)  |\Delta{\tilde{h}}(q)|^2+ \left(3-2\nu\right)\left(\tilde{z}_{eq}^2 |\Delta \theta(q)|^2 + |\Delta{\tilde{z}(q)}|^2\right)\right)\right] \right\}.
	\end{align}
	
	The correlations and fluctuations of the mean curvature $\tilde{h}$ and a-sphericity $\tilde{z}$ result with the usual finite correlation lengths $\xi_h(\tilde{w}),~\xi_z(\tilde{w})$,  with the slight exception (as in \cite{Grossman2016}, stemming from the fact that the shape transition is function of $\tilde{w}$ and not $T$)
	that they are temperature independent. Such that 
	\begin{subequations}
		\begin{align}\label{eq:FiniteCorrsLength}
		\langle \Delta \tilde{h}(\tilde{x}) \Delta \tilde{h}(\tilde{x}')  \rangle &= \frac{\langle \Delta \tilde{h}^2 \rangle}{2 \xi_h}e^{-\frac{\left|\tilde{x}-\tilde{x}'\right|}{\xi_h}} \\
		\langle \Delta \tilde{z}(\tilde{x}) \Delta \tilde{z}(\tilde{x}')  \rangle &= \frac{\langle \Delta \tilde{z}^2 \rangle}{2 \xi_z}e^{-\frac{\left|\tilde{x}-\tilde{x}'\right|}{\xi_z}}, 
		\end{align}
		where $\langle \Delta \tilde{h}^2 \rangle,~\langle \Delta \tilde{z}^2 \rangle$ are as given above in Eq. \ref{eq:flucs_simp}. This result is true for any finite width, except at $\tilde{w}=1$, where technically $\xi_z\rightarrow \infty$ (and $\langle \Delta \tilde{z}^2 \rangle$). 
	\end{subequations} 
	The expressions in Eqs. \ref{eq:avgs} now change 
	\begin{subequations} 
		\begin{align}\label{eq:newAvgs}
		\langle  \Delta\tilde{l}(\tilde{x})  \Delta \tilde{l}(\tilde{x}')\rangle &= \frac{\langle\Delta \tilde{h}^2 \rangle}{2 \xi_h}e^{-\frac{\left|\tilde{x}-\tilde{x}'\right|}{\xi_h}} + \frac{1}{2} e^{-\frac{\left|\tilde{x}-\tilde{x}'\right|}{\xi_\theta}} \tilde{z}_{eq}^2+ \frac{\langle\Delta \tilde{z}^2 \rangle}{4\xi_z} e^{-\frac{\left|\tilde{x}-\tilde{x}'\right|}{\xi_z} }= \langle \Delta{n}(\tilde{x}) \Delta{n}(\tilde{x'})\rangle \\ 
		\langle \Delta{m}(\tilde{x}) \Delta{m}(\tilde{x'})\rangle &=\frac{1}{2} e^{-\frac{\left|\tilde{x}-\tilde{x}'\right|}{\xi_\theta}} \tilde{z}_{eq}^2+ \frac{\langle\Delta \tilde{z}^2 \rangle}{4\xi_z} e^{-\frac{\left|\tilde{x}-\tilde{x}'\right|}{\xi_z} }
		\end{align}
	\end{subequations}
	where $\xi_\theta=16 \Psi \sqrt{ \frac{5}{3} \left(1-\nu\right) } \frac{k_0 t}{1+\nu}\frac{(3-2\nu) (1+\nu)^2\tilde{w}^3 \tilde{z}_{eq}^2}{48}$. 
	By taking the limit $\xi_z,\xi_h,\xi_\theta \rightarrow0$ we retrieve the previous results. Thus far, inclusion of derivatives has changed the nature of the correlation by adding finite correlation lengths. In the following sections we will see how (and when) these correlation length affect the shape of the ribbon {(see section \ref{sub:WideRibb})}.
	
	Finally, as a wide ribbon also has energetic contribution from a boundary layer, it is worth to see when is the boundary layer important. From Eq. \ref{eq:boundEner}, the boundary layer's contribution scales as $\Psi$, therefore for $\Psi \ll 1$  we may neglect it completely. From Eq. \ref{eq:avgs} it is clear that the low temperature limit is achieved for $\Psi \gg \frac{1}{\tilde{w}}$ , hence in the wide regime we may choose to work in regimes so that $\frac{1}{\tilde{w}} \ll \Psi \ll 1$. From this point on we neglect the boundary  layer.

	\subsubsection{Shape Fluctuations}	
	Experimentally the curvatures are hard to measure, and they are not an intuitive tool to understand the ribbon's shape. In polymer science, other measures are used to describe the ribbon's stiffness and shape -  most common  \cite{Rubinstein2003} are the persistence length $\ell_p$, the gyration radius $R_g$ and the Kuhn length $\ell_k$. The persistence length, $\ell_p$  is the scale over which tangent - tangent correlation decay,
	\begin{subequations}
		\begin{align}
		\ell_p: & \langle \hat{v}_3 (x) \hat{v}_3 (x') \rangle \propto e^{-\frac{\left| x-x' \right|}{\ell_p}}.
		\end{align}
		The gyration radius $R_g$ approximates the shape enclosing volume of gyrating ribbon in a thermal environment as a sphere and is given by
		\begin{align}
		R_g^2&= \langle \int_0^L \frac{\dt x}{L} \left[r(x)^2 - 2\vec{r}(x) \vec{r}_0+ r_0^2 \right] \rangle=- \langle r_0^2 \rangle + \frac{1}{L}\int_0^L \langle r(x)^2\rangle \dt x
		\end{align}
		where $\vec{r}(L)$ is the end-to-end distance of the ribbon, and $\vec{r}_0$ is it's center of mass. The Khun length $\ell_k$ is  defined as the segment length of a random freely-jointed chain that reproduces  same end-to-end length. 
		\begin{align}
		\ell_k = \lim\limits_{L\rightarrow \infty }\frac{1}{L} \langle  r^2(L) \rangle.
		\end{align}
		Another, less common, measure is the torsional correlation length $\ell_\tau$ which is the scale over which bi-normal - bi-normal correlation decay \cite{Giomi2010}.
		\begin{align}
		\ell_\tau: & \langle \hat{v}_1 (x) \hat{v}_1 (x') \rangle \propto e^{-\frac{\left| x-x' \right|}{\ell_\tau}}.
		\end{align}	
		Nevertheless (as will be seen in the next sections), these quantities fail to encompass the shape of a gyrating ribbon. To this end we should calculate the gyration tensor which approximates the volume in which a ribbon gyrates as an ellipsoid (rather than a sphere).
		\begin{align} \label{eq:gyration_tensor}
		R_{ij}(L) = \frac{1}{L} \int_{0}^L  \left[\langle r_i(x) r_j(x) \rangle - \langle (r_0)_i(r_0)_j \rangle \right]\dt x.
		\end{align}
		where $r_i(x)= \vec{r}(x) \cdot \hat{v}_i(0)$. Note that $R_g^2= R_{11}+R_{22}+R_{33}$. As it turns out, $R_{ij}$ is hard to calculate (see appendix \ref{app:Stat_Meas} we therefore add to the list another measure to probe the shape of a gyrating ribbon, the Frame-Origin Correlations 	
		\begin{align}\label{eq:FOCM}
		\rho_{ij}(x) &=\langle v^i_3(x)v^j_3(x) \rangle
		\end{align}
		where $v^i_3(x)=\hat{v}_i(x) \cdot \hat{v}_3(0)$.
	\end{subequations}
	
	Using the 2$^{nd}$ order expansion of the Hamiltonian	\ref{eq:hamiltonian_spherical_dimless},  we find that for $x>x'$ (see appendix \ref{app:Stat_Meas})
	\begin{align} \label{eq:corrmat}
	\langle \hat{v}_i(x) \hat{v}_j(x')\rangle &= \langle e^{-\int_{{x'}}^{x} \mathbf{\Omega} \dt \xi''} \rangle_{ij} =\left[e^{-\mathbf{ \Lambda} (x-x')}\right]_{ij}
	\end{align}
	where (neglecting derivatives)
	\begin{align} \label{eq:Lambda}	
	\Lambda &=  \left( \begin{array}{ccc}
	0 & 0 & 0  \\
	0 & 0 & \langle l \rangle \\
	0 & -\langle l \rangle & 0
	\end{array}\right) + \frac{1}{2}\left( \begin{array}{ccc}
	\langle \Delta m^2 \rangle & 0 & 0\\
	0 & \langle\Delta m^2\rangle + \langle\Delta l^2\rangle & 0 \\
	0& 0 &   \langle\Delta l^2 \rangle 
	\end{array}\right).		
	\end{align}
	The averages are given above in Eqs. \ref{eq:avgs}, \ref{eq:flucs_simp}. Retrieving dimensionality is straightforward (note that $\langle\Delta m^2\rangle = k_0 \langle \Delta \tilde{m}^2 \rangle$, and similarly for $\Delta l^2$). The $\Lambda$ matrix is central for calculations of the entities presented above.
	
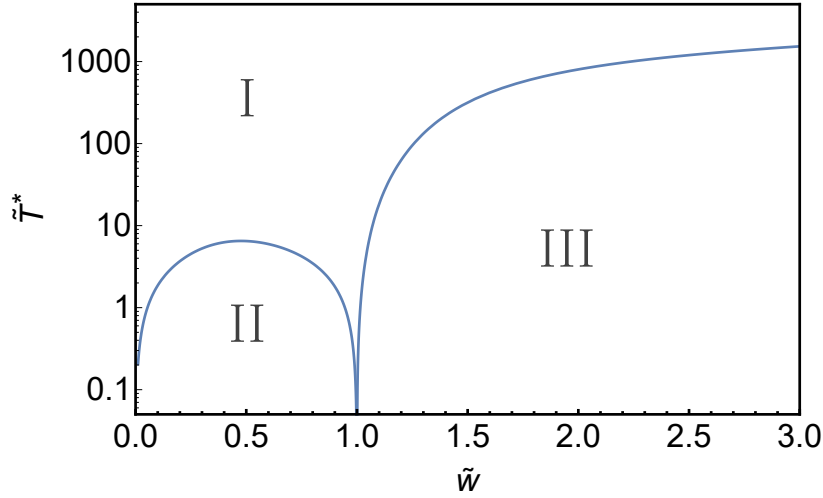
\begin{figure}
		\centering
		\begin{tikzpicture}
		\node[above] at (0,0){\includegraphics[width=0.6\textwidth]{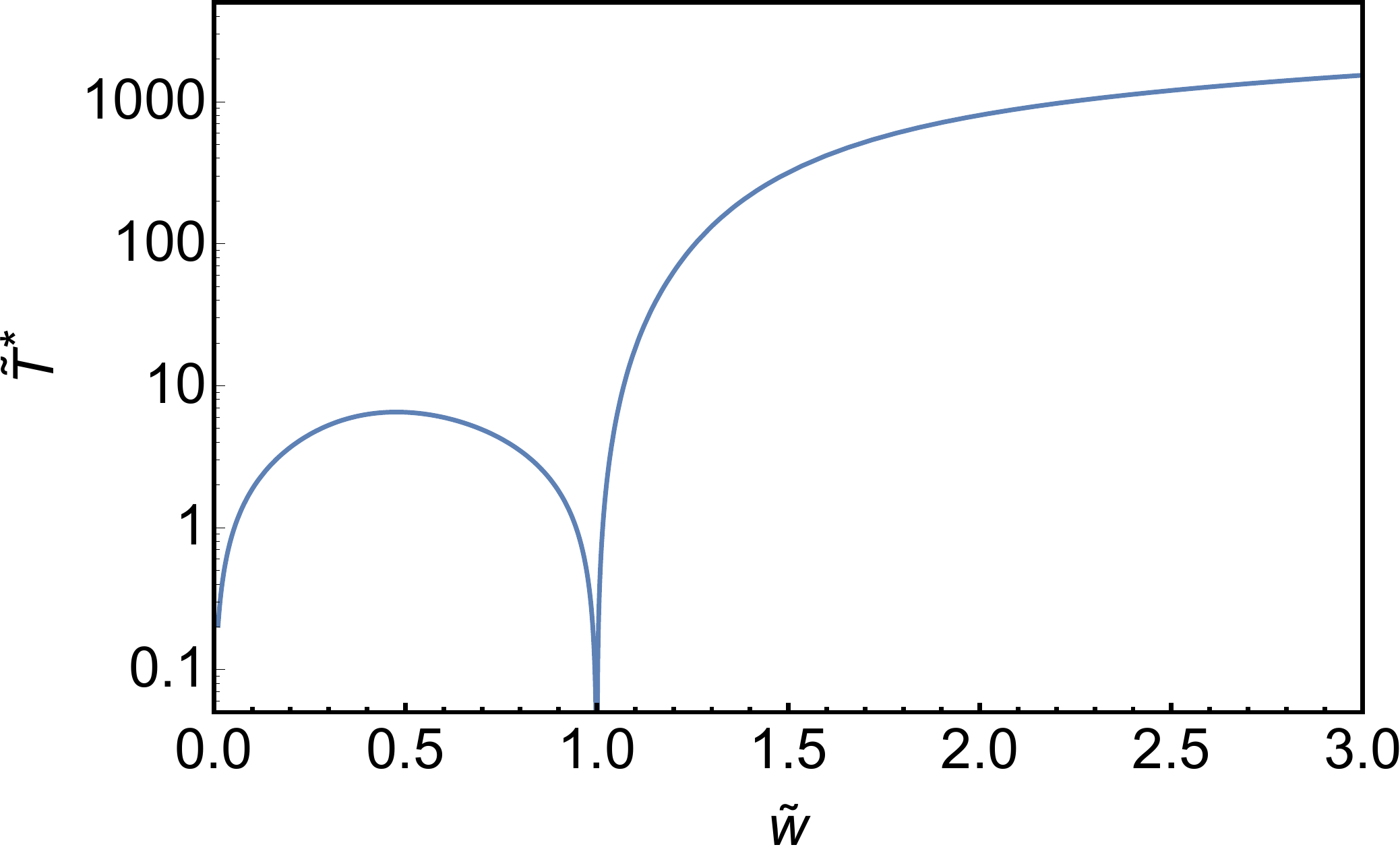}};
		\node[above] at (-2.25,5){\fontfamily{arev}\color{darkgray} \huge I};
		\node[above] at (-2.25,2){\fontfamily{arev}\color{darkgray}\huge II};
		\node[above] at (2,3){\fontfamily{arev}\color{darkgray}\huge III};
		\end{tikzpicture}
		
		\caption{$\tilde{T}^*=\frac{1}{\Psi^*}$ , the temperature above which a ribbon behaves like a random coil as a function of ribbon width $\tilde{w~}$ ($\nu=0$). $T^* =\frac{5^{1/4} }{3^{5/4} \sqrt{2} (1-\nu)^{3/4}(1+\nu)^{7/2} } \frac{Y t^{7/2} k_0^{1/2}}{k_B} \tilde{T}^*$. At $\tilde{w}=0,1$, $\tilde{T}^*=0$. Area $I$ corresponds to the Ideal Chain phase, area II is the Plane Ergodic phase and area II is the Random Structured phase (see subsections \ref{sub:NarrRibb}, and \ref{sub:WideRibb} respectively)  \label{fig:Tstar}}
	\end{figure}

	As $\Lambda$ encompasses significant information (yet not all) regarding the ribbon's thermal behaviour and structure, it is useful to study some of it's properties. Specifically, it's eigenvalues are
	\begin{align} \label{eq:eigen}
	\lambda_0 &= \Lambda_{11} =\frac{1}{2}  \langle \Delta m^2 \rangle \\ \nonumber
	\lambda_{\pm} &= \frac{1}{4} \left(\langle \Delta m^2 \rangle +2 \langle \Delta l^2 \rangle \pm \sqrt{\varpi} \right). \\ \nonumber
	\text{\small $\varpi (T,\tilde{w})$} &= \text{\small $ \langle \Delta m^2 \rangle^2 - 16 \langle l \rangle^2$}
	\end{align}
	At any given width, $\langle \Delta m^2 \rangle$, and $\langle \Delta l^2 \rangle$ roughly scale as $T$ (\cite{Bigtheta}, see also Eq. \ref{eq:flucs_simp} for low temperature approximation).  This is in contrast to $\langle l \rangle$ which is a non-zero constant as $T\rightarrow 0$, and (depending on $\tilde{w}$) at hight $T$ scales as $T$. Therefore, at low $T$, $\varpi < 0$ and $\lambda_{\pm}$ are complex conjugates. At some $T^*$, $\varpi(T^*)=0$, and $\lambda_{\pm}$ are equal. Above $T^*$,  they are positive reals, signifying that the ribbon lost all structure, and behaves as a random coil.  Since for every $\tilde{w}$ there exists only such one (positive) temperature, $T^*(\tilde{w})$ is well defined. It is important to note that experimentally the ribbon will seem to lose structure at lower temperatures. This will happen roughly as the typical fluctuation will be of magnitude similar to the curvature of the ribbon. In other words, when either $\langle \Delta l^2 \rangle \sim \langle l \rangle $ or $\langle \Delta m^2 \rangle \sim \langle l \rangle $. As it turns out, in this case at least, the former happens earlier than the latter. Nevertheless, $T^*$ remains a better defined parameter. Solving for $T^*$ analytically is hard, we therefore plot it in Fig \ref{fig:Tstar} for the case $\nu=0$. We see the critical temperature divides the graph into three  regions. It is essentially the phase diagram of the ribbon, which we are yet to characterize. Region I is the high temperature limit- or the Ideal Chain phase, where the ribbons behaves as a random coil. We named regions II and III  the Plane Ergodic and the Random Structured phases respectively. In the following subsections we characterize them.
	
	\subsubsection{Narrow Ribbon $\tilde{w}<1$}\label{sub:NarrRibb} 
	In this section we characterize the thermal dependence of a narrow, cold $T<T^*$ ribbon. In all calculations in this section we omit derivatives  as their contribution is negligible in the limit of $T\rightarrow 0$ (as they give rise to finite, temperature independent,  correlation lengths).
	
	As in \cite{Giomi2010}, $\ell_\tau$, and $\ell_p$ hold important information about the structure of the ribbon. Specifically they are the scales on which bi-normal - bi-normal and tangent - tangent correlation decay exponentially. From $\Lambda$ (Eq. \ref{eq:Lambda}) we can easily extract $\ell_\tau = 2/\langle\Delta m^2 \rangle$ and $\ell_p=2/\langle \Delta l^2 \rangle$.  Asymptotically
	\begin{subequations}\label{eq:ell_tau_ell_p_ass_narrow}
		\begin{align}
		\ell_\tau/\ell_p & \xrightarrow{\tilde{w}\rightarrow 0} \frac{8-\pi(1+\nu)}{(4-\pi)(1+\nu)}.
		\end{align}
	\end{subequations}
	
\begin{figure}[h]
		\centering
		\includegraphics[width=0.6\textwidth]{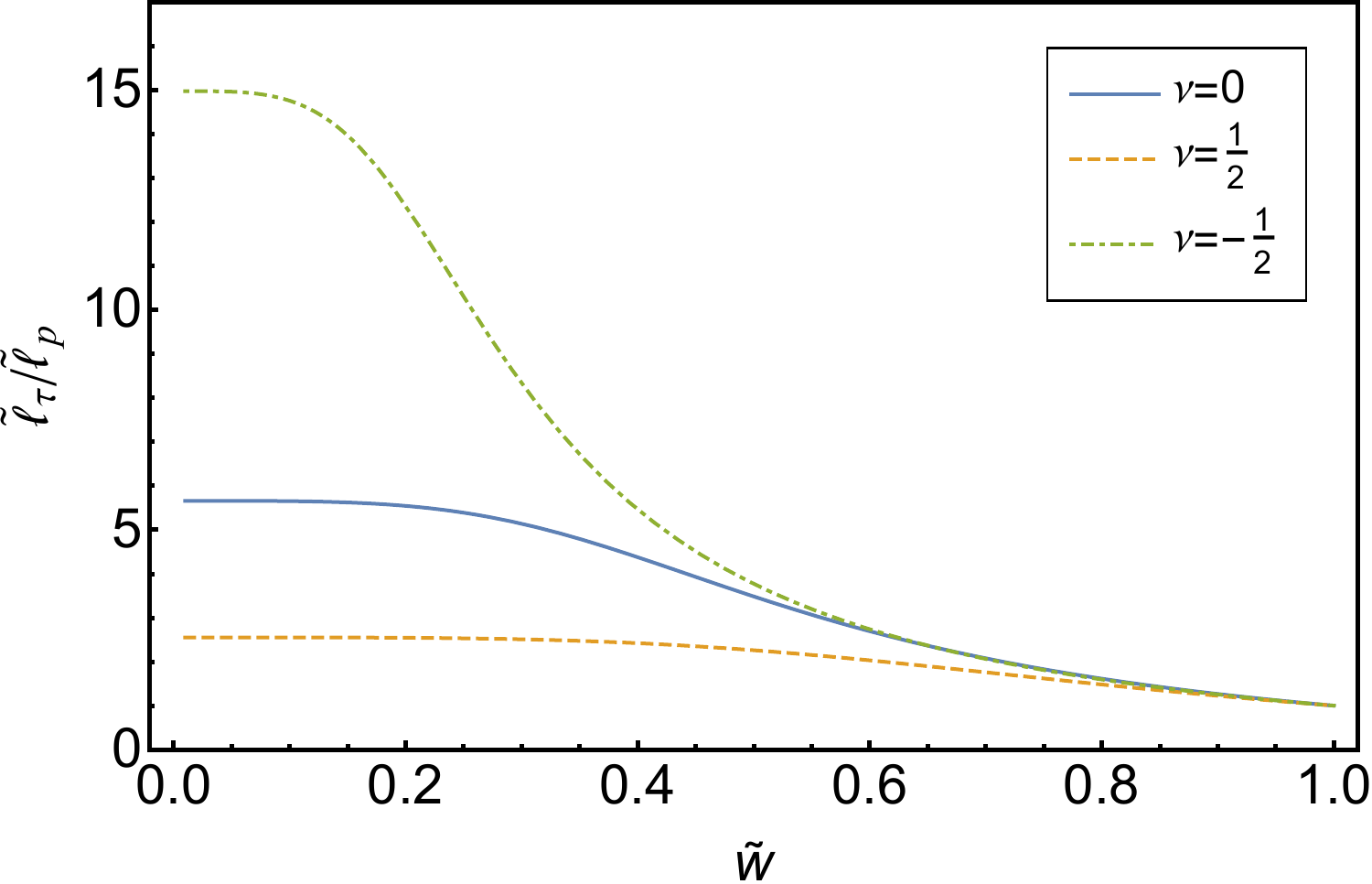}
		\caption{The ratio, $\ell_\tau/\ell_p $ for a narrow ribbon is independent of temperature, but varies with $\tilde{w}$, plotted here for the cases $\nu=0,\pm \frac{1}{2}$. At different values of $\nu$, the narrow limit  follows ${\ell_\tau}/{\ell_p}|_{\tilde{w}=0}=\frac{8-\pi(1+\nu)}{(4-\pi)(1+\nu)}$. At $\tilde{w}=1$, ${\ell_\tau}/{\ell_p}|_{\tilde{w}=1}=1$.
			\label{fig:ltauOnlp}}
	\end{figure}

\begin{figure}[h!]
		\hspace*{-1em}\begin{tabular}{c c c}
			\begin{subfigure}{0.3\textwidth}
				\centering
				\includegraphics[width=\textwidth]{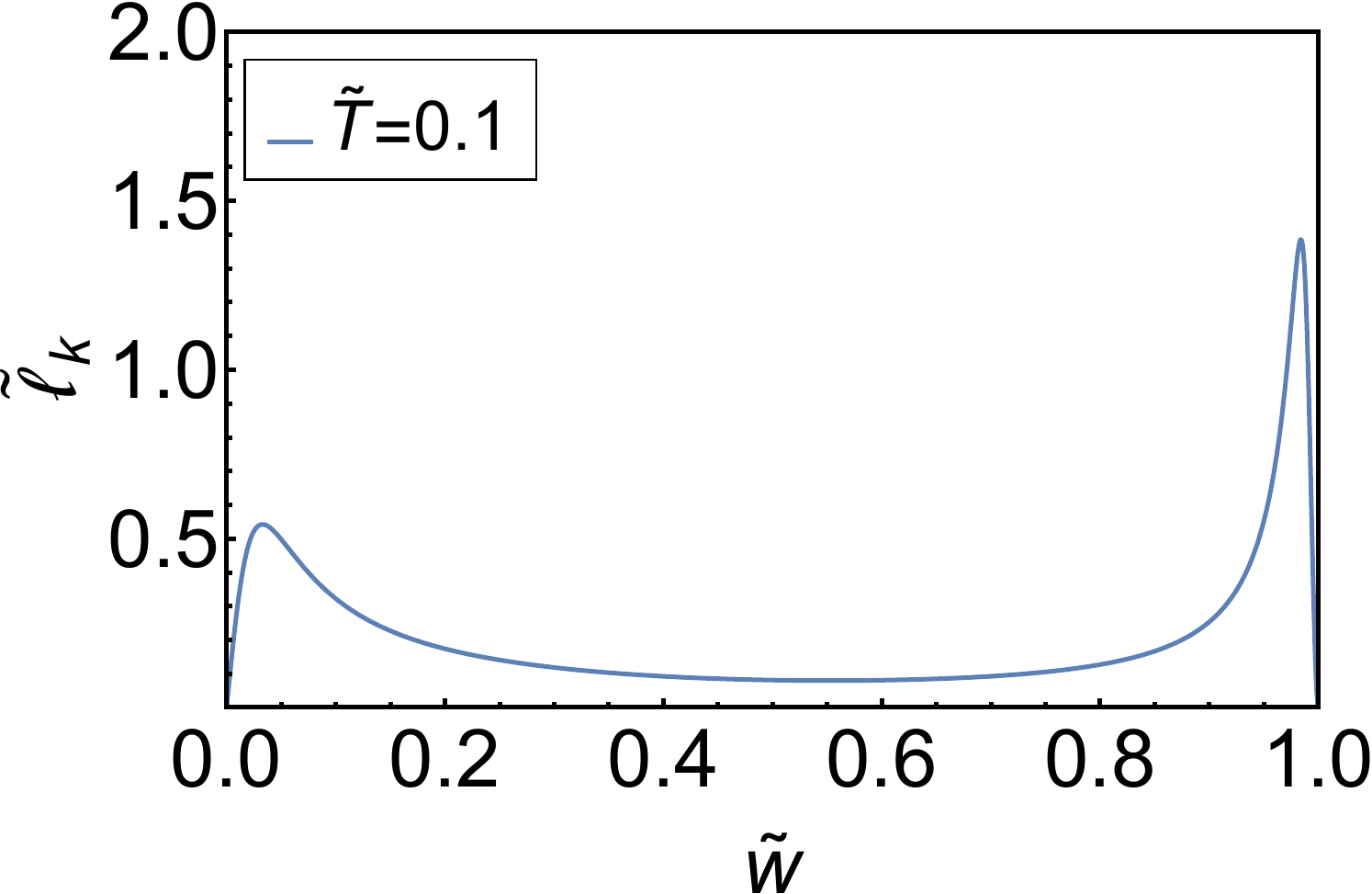}
				\caption{$\tilde{T}=.1$}\label{fig:KunhLengths_sub10}
			\end{subfigure}
			& 
			\begin{subfigure}{0.3\textwidth}
				\centering
				\includegraphics[width=\textwidth]{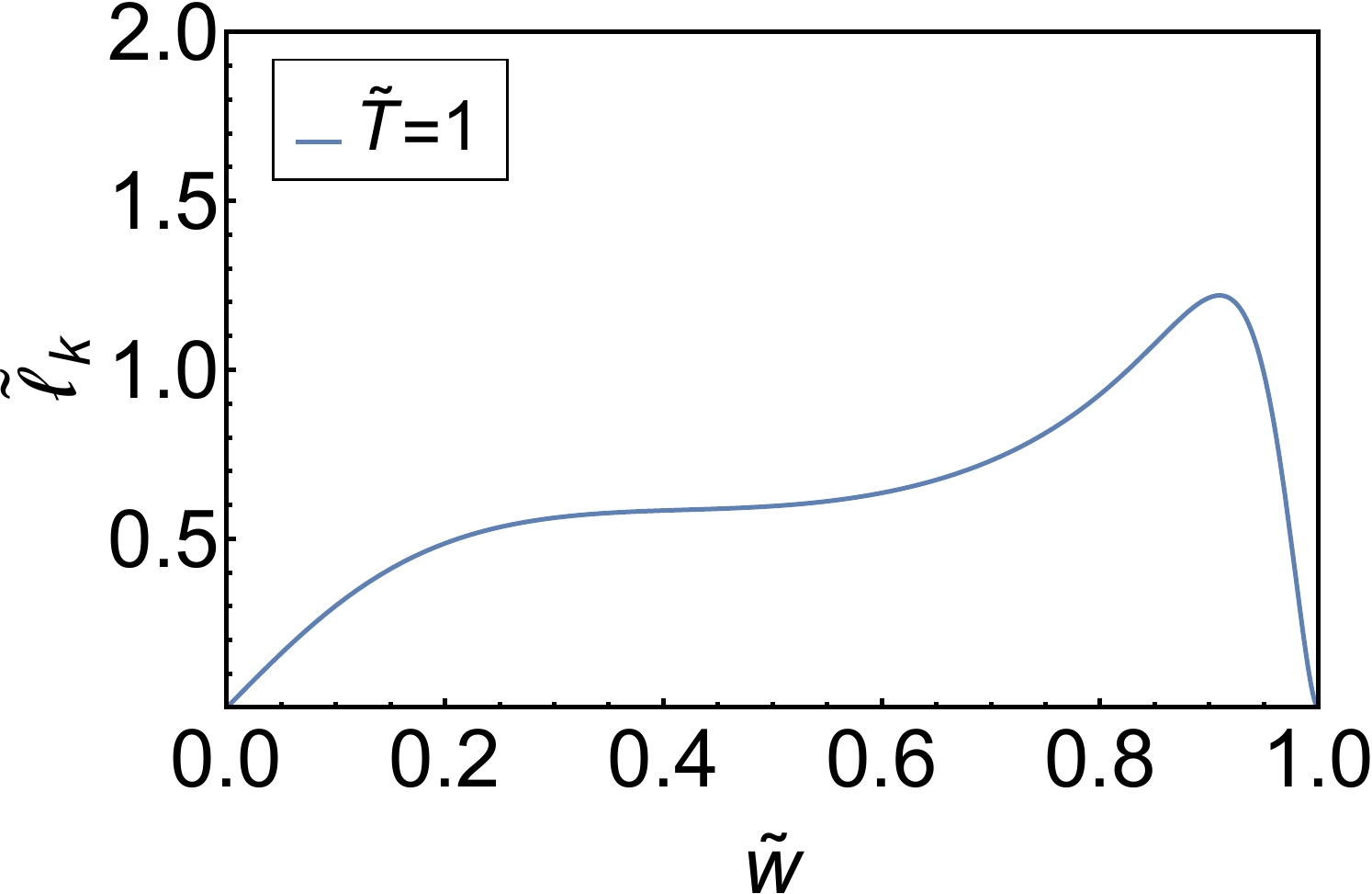}
				\caption{$\tilde{T}=1$}\label{fig:KunhLengths_sub01}
			\end{subfigure}&
			\begin{subfigure}{0.3\textwidth}
				\centering
				\includegraphics[width=\textwidth]{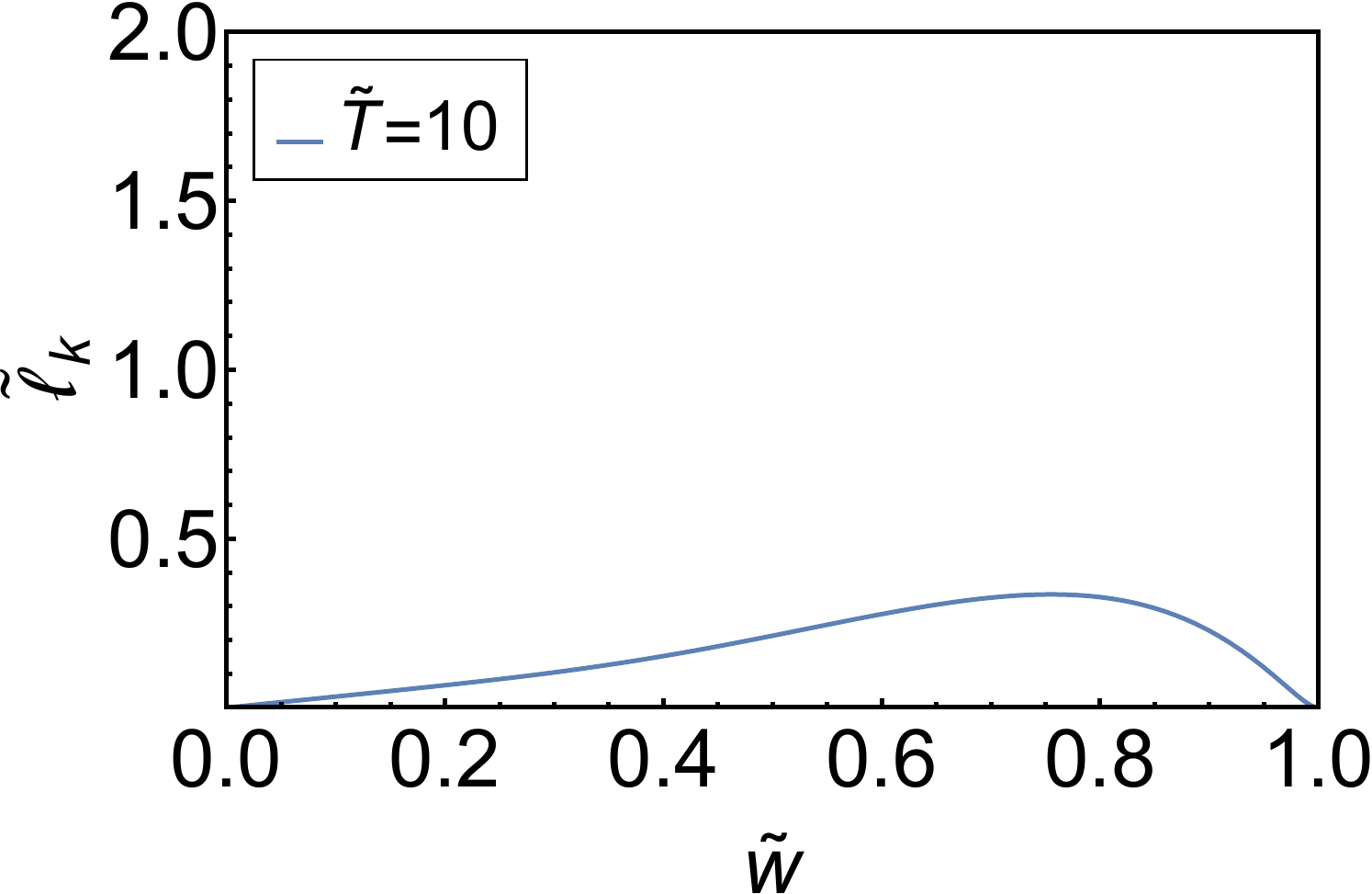}
				\caption{$\tilde{T}=10$}\label{fig:KunhLengths_sub1}
			\end{subfigure} \\
			\multicolumn{3}{ c}{\begin{subfigure}{0.3\textwidth}
					\raggedright
					\includegraphics[width=\textwidth]{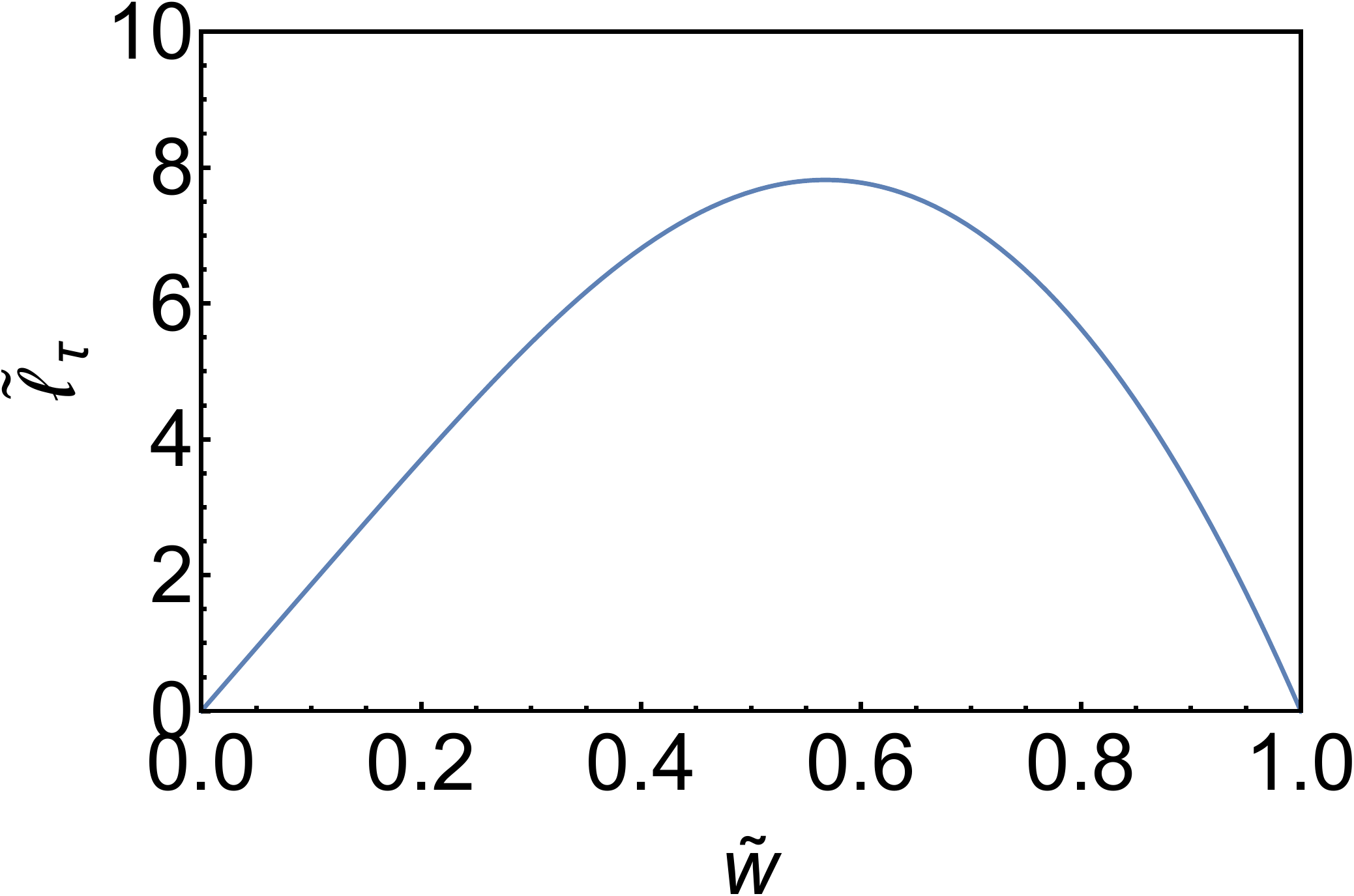}
					\caption{} \vspace*{2em}\label{fig:l_tau_sub10} 
				\end{subfigure}\hspace*{1em}}
		\end{tabular}
		\caption{Dimensionless Kuhn length, $\tilde{\ell}_k= \ell_k k_0$ (a-c),  and dimensionless torsional correlation length, $\tilde{\ell}_\tau = \ell_\tau k_0/\Psi$ (d), for $\nu=0$ (note the difference in definition).  While, $\ell_\tau$ is independent of temperature (up to scaling by $\Psi=\frac{1}{\tilde{T}}$), $\ell_k$ is very sensitive to temperature change.  At $\tilde{w}=1$ both $\ell_\tau, \ell_k=0$ as the ribbon is anomalously soft at the critical width.\label{fig:l_tau_vs_l_k} }
	\end{figure}

	While $\tilde{\ell}_\tau$ and $\tilde{\ell}_p$ have a slight dependence on temperature, their ratio $\ell_\tau/\ell_p$ is independent of it. $\tilde{\ell}_\tau$ and $\ell_\tau/\ell_p$ are depicted in Figs. (\ref{fig:l_tau_vs_l_k} ,\ref{fig:ltauOnlp}) respectively, for the cases $\nu=0,\pm \frac{1}{2}$.  The result remain qualitatively the same for different values of $\nu$.

	A direct calculation (\ref{app:Stat_Meas}) shows that the Kuhn length  is given by $(\Lambda^{-1})_{33}$. The inverse of $\Lambda$ is
	\begin{align}
	\Lambda^{-1}= \left(\begin{array}{ccc}
	\frac{2}{\langle \Delta m^2 \rangle} & 0 & 0 \\
	0 & \frac{ 2 \langle \Delta l^2 \rangle}{\varsigma} &\frac{- 4 \langle l \rangle}{\varsigma} \\
	0 & \frac{4 \langle l \rangle}{\varsigma} & 2\frac{ \langle \Delta l^2 \rangle+\langle \Delta m^2 \rangle}{\varsigma}
	\end{array}\right),&~~~
	\text{\small $\varsigma$}= \text{\small $\langle \Delta l^2 \rangle(\langle \Delta l^2 \rangle+\langle \Delta m^2 \rangle)+ 4 \langle l \rangle^2$}\\ \nonumber
	\Downarrow &\\ 
	\ell_k = & 2\frac{\langle \Delta l^2 \rangle+\langle \Delta m^2 \rangle}{\varsigma}
	\end{align} 
	$\ell_k$ is plotted at different  temperatures is in Fig. (\ref{fig:l_tau_vs_l_k}) (for the case of $\nu=0$, sub figures (a)-(c)), the result stands in stark contrast to $\ell_\tau$ (Fig. \ref{fig:l_tau_sub10}) and $\ell_p$. Not only that $\ell_k$ has a non-monotonic dependence on the ribbon's width, its temperature dependence is also very unique (in contrast with \cite{Grossman2016}) as it converges to $0$  also at low temperatures ($T \rightarrow 0$), not only high ($T \rightarrow \infty$), while it reaches a maximum at some intermediate temperature. This unique dependence of the Kuhn length $\ell_k$ on the temperature, in contrast to that of the persistence and torsional correlation lengths ($\ell_p$ and $\ell_\tau$) directly relates to the fact that $\ell_p$ and  $\ell_\tau$ measure arc-distance (\emph{on the ribbon}), while $\ell_k$ measures distance in the embedding space.

\begin{figure}
		\centering
		\includegraphics[width=0.7\textwidth,clip,trim={0 0 0 6.7cm}]{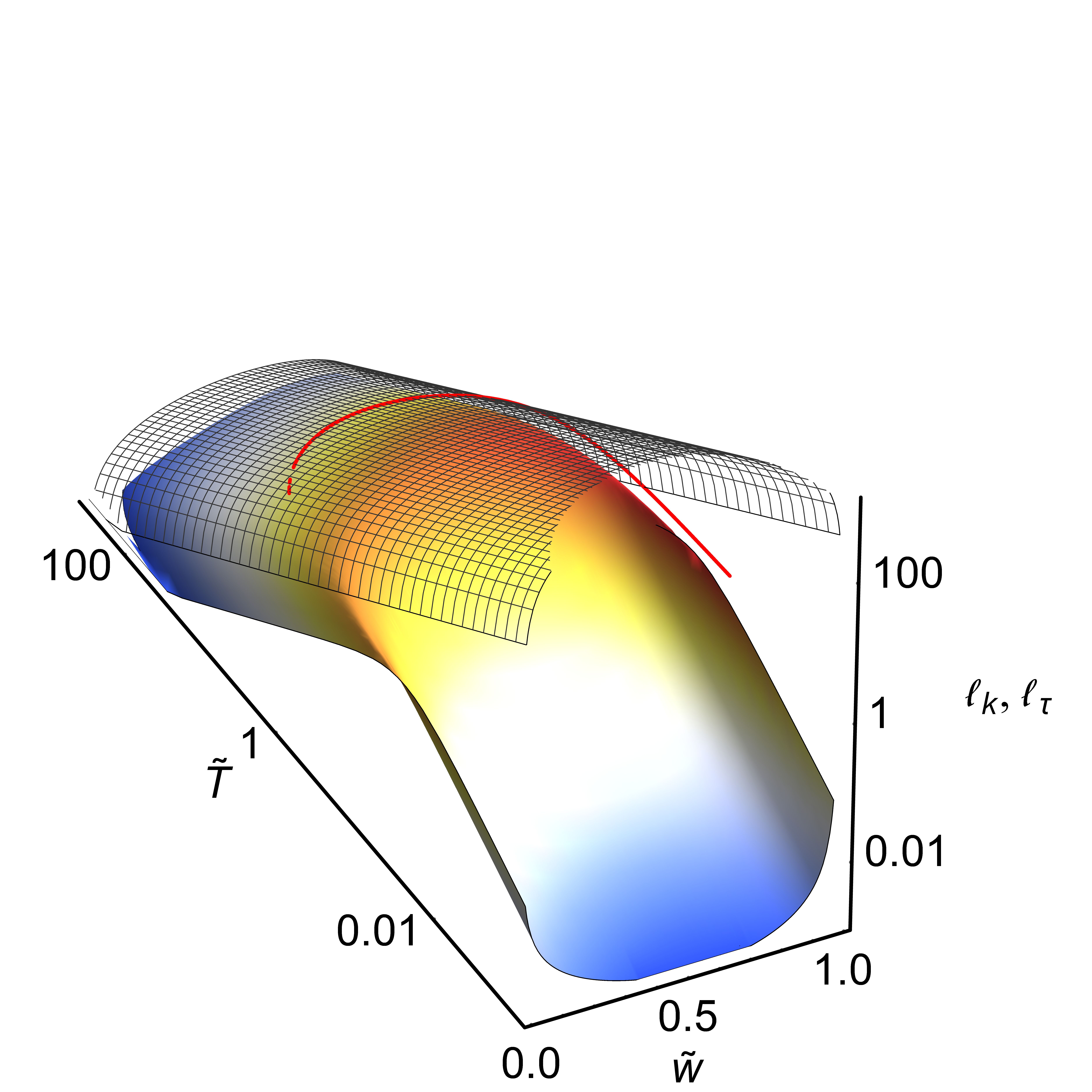}
		\caption{$\ell_k$ (coloured surface) and $\ell_\tau$ (black mesh)  as a function of $\tilde{w}$ and $\tilde{T}$. The red curve marks $\ell_\tau(\tilde{T}^*)$. The orange peaks in $\ell_k$ occur at lower temperatures, marking an ``apparent melting" of the ribbon.}\label{fig:lp_ltau}
	\end{figure}

	At high temperatures ($T\gg T^*$), this result is easily understandable as the ribbon loses all structure, and reverts to the classical behaviour of a freely jointed chain where $\ell_k \sim \ell_p \propto \frac{1}{T}$. This result is merely the "ideal chain" phase that every ribbon of any kind exhibits at high enough temperatures. At low temperatures ($T\ll T^*$),  the ribbon mostly retains it's structure (large $\ell_p$, $\ell_\tau$). Since at $T=0$ a narrow ($\tilde{w}<1$) ribbon is ring shaped, the end-to-end distance has a maximal value, $\langle r^2 \rangle \leq 4 R^2$ (where $R$ is the ring's radius). Therefore at the limit we get 
	\begin{align*}
	\lim\limits_{L \rightarrow \infty} \frac{1}{L} \langle r^2(L)\rangle_{T=0} = 0.
	\end{align*}
	At finite low temperatures, the ribbon retains its shape for a long, but finite, arc-length. This results with $\ell_k (T\ll T^*) \propto T$. The local maxima in Figs. (\ref{fig:KunhLengths_sub10}, \ref{fig:KunhLengths_sub1}), occur roughly around the widths such that $T\sim T^*(\tilde{w})$. Indeed, at $\tilde{T}=4$, a narrow ribbon may be considered "completely melted" (i.e- lost all structure) as at any $0<\tilde{w}<1$, $T^*(\tilde{w})\leq T$.  At $\tilde{w}=1$ the ribbon is infinitely soft, indicated by the fact that $\ell_\tau,~ \ell_p,~\ell_k$, and $T^*(\tilde{w}=1)$ are all equal to zero. Counter intuitively, the ribbon gets softer as it widens and approaches the transition. Both $\ell_\tau$, and $\ell_k$ are depicted as a function of $\tilde{T}$, and $\tilde{w}$ in Fig. (\ref{fig:lp_ltau}). In it, also $\ell_\tau(T^*)$ is plotted to visualize $T*$ (red line), it is immediately seen that the apparent "melting" of the ribbon (peaks in $\ell_k$) occur when $T$ is significantly smaller than $T^*$. Nevertheless, $T^*$ remains a better defined parameter.
	
	Following these results, the ribbon's shape comes into question- what should we expect to see when looking into a thermal soup of such ribbons? The fact that $\ell_p \leq \ell_\tau$ at most widths, suggests that the ribbon gyrates non-spherically, at least for some range of lengths. However, $\ell_\tau$ and $\ell_p$ are "internal" measures, in the sense that they tell us how long we should measure the arc-length of a ribbon before we see some change. $\ell_k$ supplies an "external" measure, yet it turns out to be insufficient as there is no randomly jointed chain of finite length rods that can reproduce the statistical behaviour of the end-to-end vector in the cold limit (as indicated by the fact that $\ell_k=0$ at this limit). 
	
	As discussed earlier, one way to quantify the shape of a gyrating ribbon is the gyration tensor $R_{ij}$ \ref{eq:gyration_tensor}. In appendix (\ref{app:Stat_Meas}) we develop a formal expression of this tensor, yet even a numerical approximation of it is difficult to calculate as it requires an infinite sum without any immediate cut-off parameter. We therefore decide to use other measures to shed light on the overall shape of the ribbon. First we calculate the gyration radius $R_g^2= \Tr(R_{ij})$, that is essentially a spherical approximation of the ribbon's shape. We also calculate the Frame-Origin Correlation Matrix,  $\rho_{ij}$, (Eq. \ref{eq:FOCM}) analytically. Together these two measures enable us  explore the overall shape ($R_g$) and any anisotropies ($\rho_{ij}$) of the ribbon, allowing us to estimate the ribbon's (anisotropic) shape.
	
\begin{figure}[h!]
		\centering
		\begin{tikzpicture}[scale=1]
		\node (pic) at (0,0) {\includegraphics[width=0.7\textwidth]{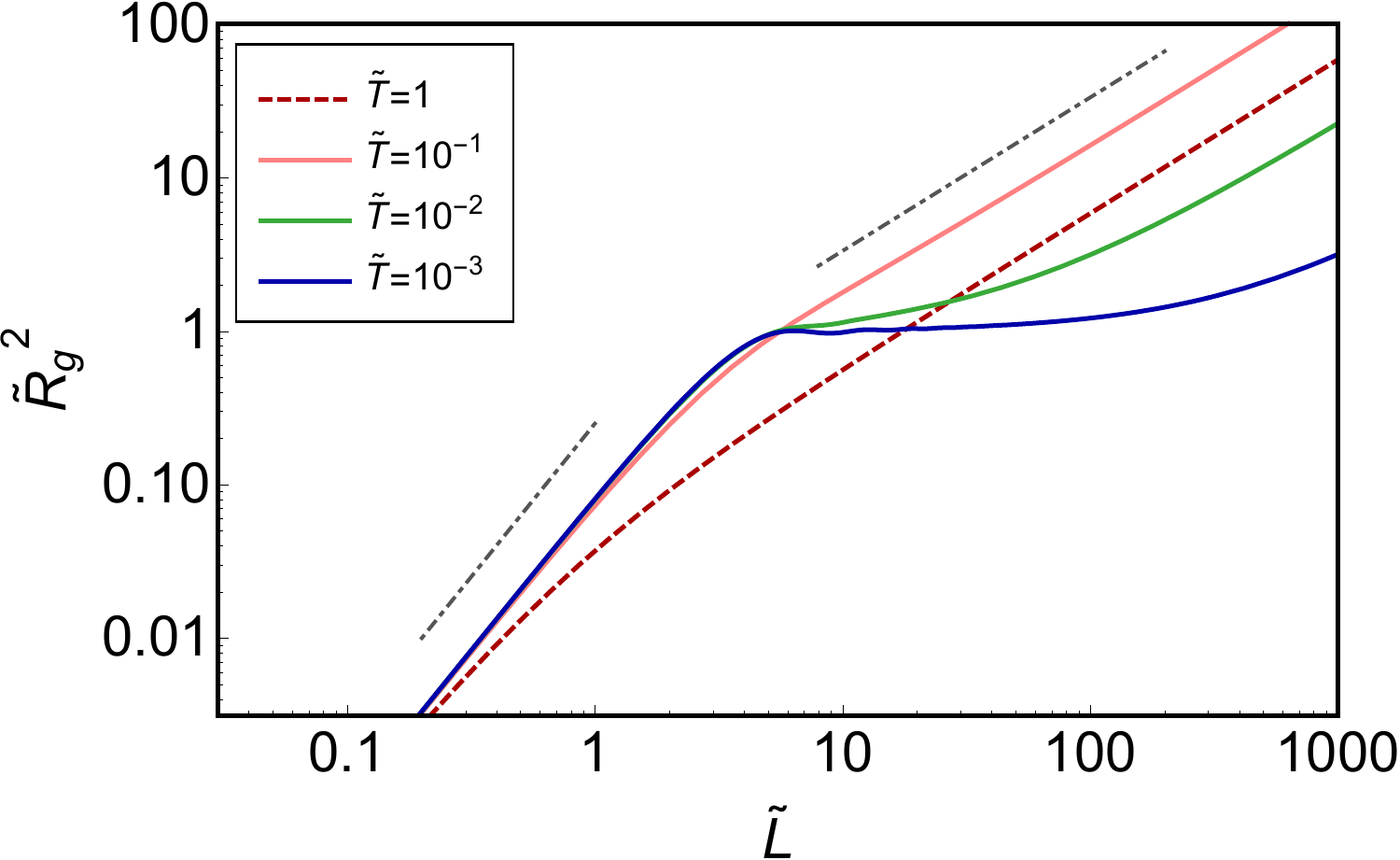}};
		\node (L2) at (-2.1,-.7) {$\sim \tilde{L}^2$};
		\node (L1) at (2.2,2.6) {$\sim \tilde{L}$};
		\end{tikzpicture}
		\caption{The squared gyration radius $\tilde{R}_g^2= k_0^2 R_g^2$   as a function of $\tilde{L}=k_0 L$. At different  temperatures  (as indicated in the figure), for $\tilde{w}=0.055$, $\nu=0$. Gray dash-dotted lines mark asymptotics  of $\tilde{R}_g^2$, as indicated in figure. At this width $\tilde{T}^*=\frac{1}{\Psi^*}=1$, note that the (narrow) ribbon seems to lose structure already at $T=1/10$ (i.e. lose the $R_g^2 \sim const$ region between the $\sim L^2$ and $\sim L $ regions). Also note that at $T^*$ the gyration radius (and $\ell_k$) already start to decrease. The graphs remain qualitatively the same for different values of $\nu$ and $\tilde{w}$. At $T\rightarrow0$, $R_g^2 \xrightarrow{\tilde{L}\geq \tilde{L}_{min}} const.$ }\label{fig:gyr_rad}
	\end{figure}

	A direct calculation (see appendix \ref{app:Stat_Meas}) yields 
	\begin{align}\label{eq:gyration_radius}
	R_g^2 = \left[\frac{1}{3}\Lambda^{-1} L -  \Lambda^{-2} + \frac{2}{L}\Lambda^{-3}-\frac{2}{L^2}\Lambda^{-4}\left(1-e^{-\Lambda L}\right)\right]_{33}
	\end{align}
	$\tilde{R}_g^2= k_0^2 R_g^2$ is plotted in Fig, \ref{fig:gyr_rad}, as a function ribbon's length $\tilde{L}=k_0 L$ for different temperatures. At low temperatures, one can describe
	\begin{align}\label{eq:Rg_spherical}
	R_g^2 \propto \left\{ \begin{array}{cc}
	L^2 & 0\leq L < L_{min} \\
	const & L_{min}<L <L^*\\
	\ell_k L & L^*< L
	\end{array}\right.
	\end{align}
	$L^*(T)\propto \ell_p$ is the scale at which ribbons start to behave as ideal chain, shorter ribbons  should be considered as stiff. This scale depends on the temperature and width of the ribbon, but also on it's structure ($k_0,~t$), it exists for ribbons of any type- not just the ones studies here. $L_{min}$ is temperature independent and scales as the ribbon's curvature radius, it marks the scale at which the ribbon's ring-shape manifests itself. Ribbons shorter than $L_{min}$ may be considered, practically at least, as straight and featureless. Since $L^*$ decreases as $T$ increases, there exists some temperature such $L^* \sim L_{min}$ beyond which the ribbon will appear "melted" and structureless.  At the chosen width in Fig. \ref{fig:gyr_rad} $T^*=1$, however we already see that the ribbon has an apparent "melting" point around $T\sim T^*/10$ (at which point the typical magnitude of fluctuations is of similar size to the curvature scale, $\langle \Delta m^2 \rangle  \sim \langle l^2 \rangle$). It is worth noting that { a similar dependence of $R_g$ should be expected also for non compact ribbons  (e.g- helical ribbons).  The main difference is the behaviour at the mid range $L_{min} < L< L^*$, where $R_g^2 \propto \sin(\phi) L^2 $, $\phi$ being ribbons' pitch angle}

\begin{figure}[h!]
		\begin{tabular}{ccc}
			\begin{subfigure}{0.33\textwidth}
				\centering
				\includegraphics[width=.95\textwidth]{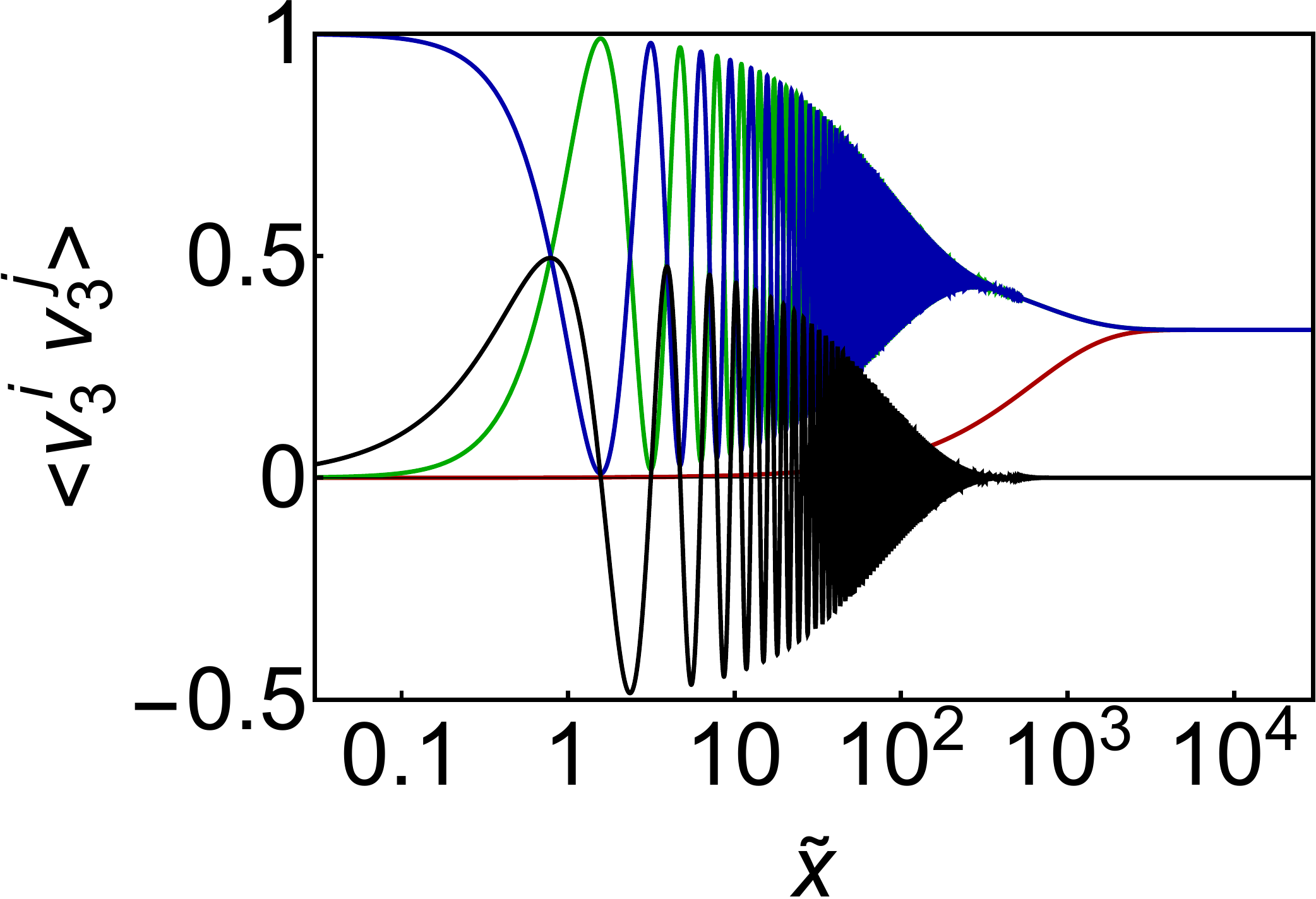}
				\caption{$\Psi=10^{3}$, $\tilde{w}=0.1$}\label{fig:corrs_psi_1000_w_1}
			\end{subfigure} &
			\begin{subfigure}{0.33\textwidth}
				\centering
				\includegraphics[width=.95\textwidth]{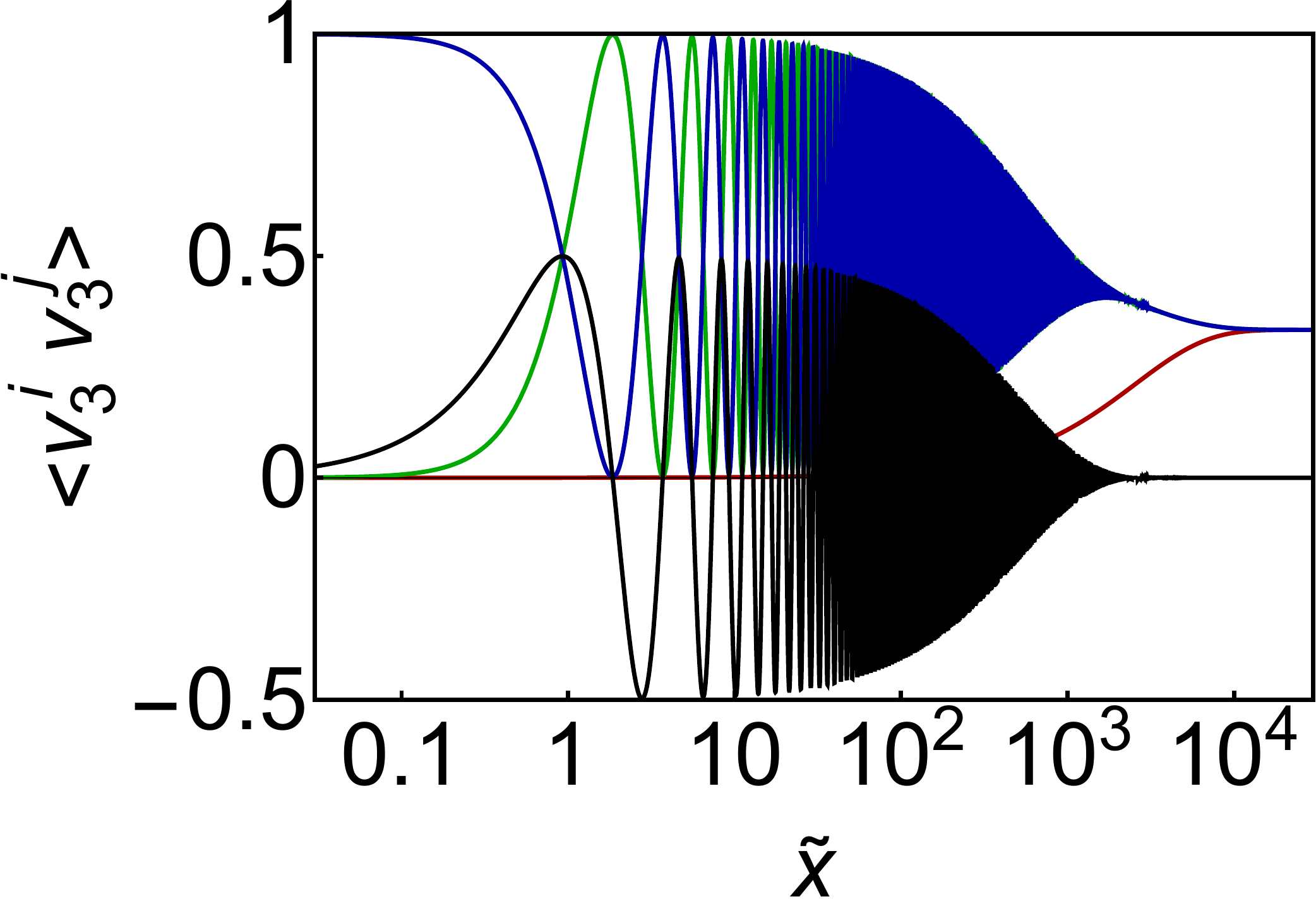}
				\caption{$\Psi=10^{3}$, $\tilde{w}=0.5$}\label{fig:corrs_psi_1000_w_5}
			\end{subfigure} &
			\begin{subfigure}{0.33\textwidth}
				\centering
				\includegraphics[width=.95\textwidth]{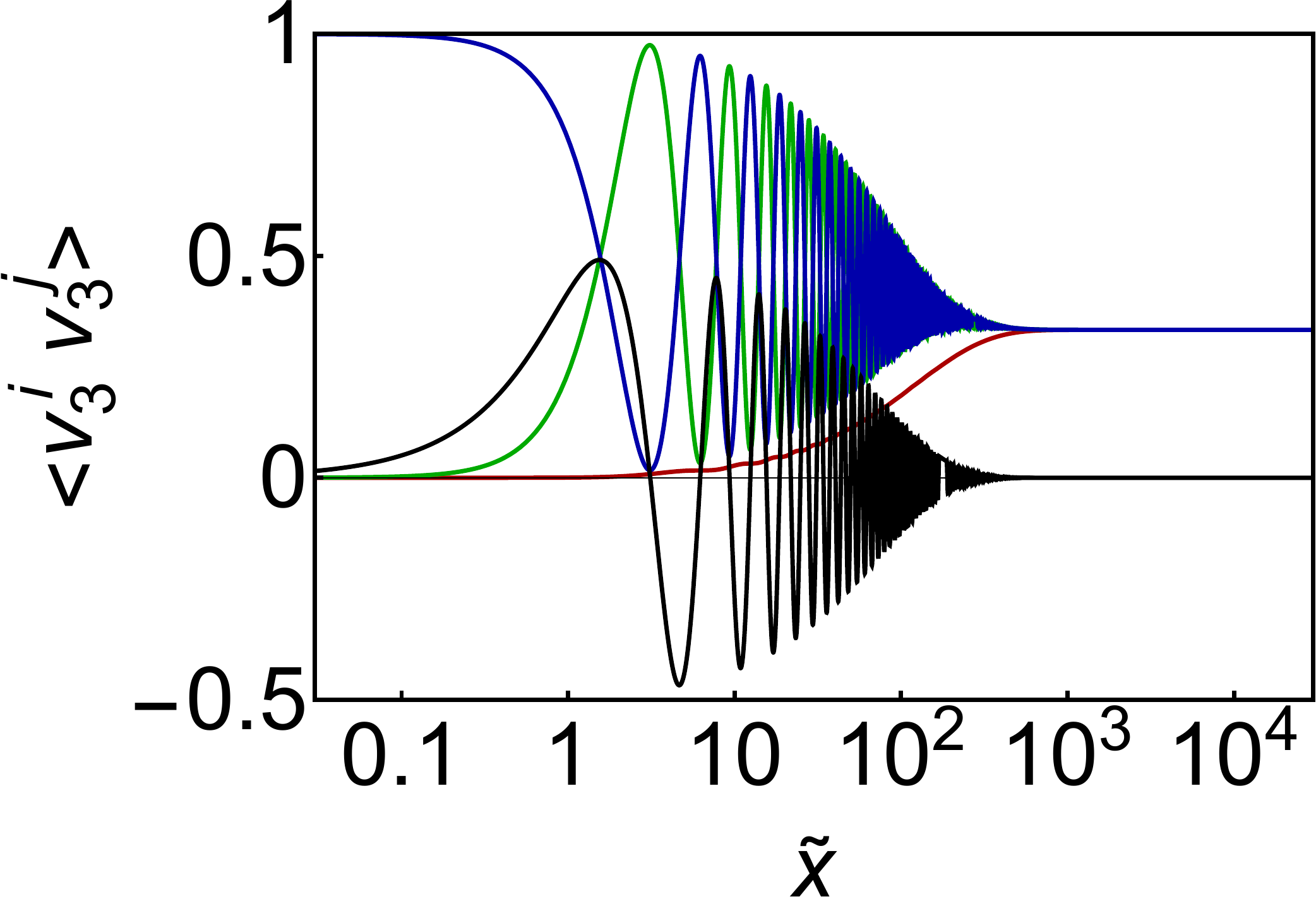}
				\caption{$\Psi=10^{3}$, $\tilde{w}=0.9$}\label{fig:corrs_psi_1000_w_9}
			\end{subfigure} \\
			\begin{subfigure}{0.33\textwidth}
				\centering
				\includegraphics[width=.95\textwidth]{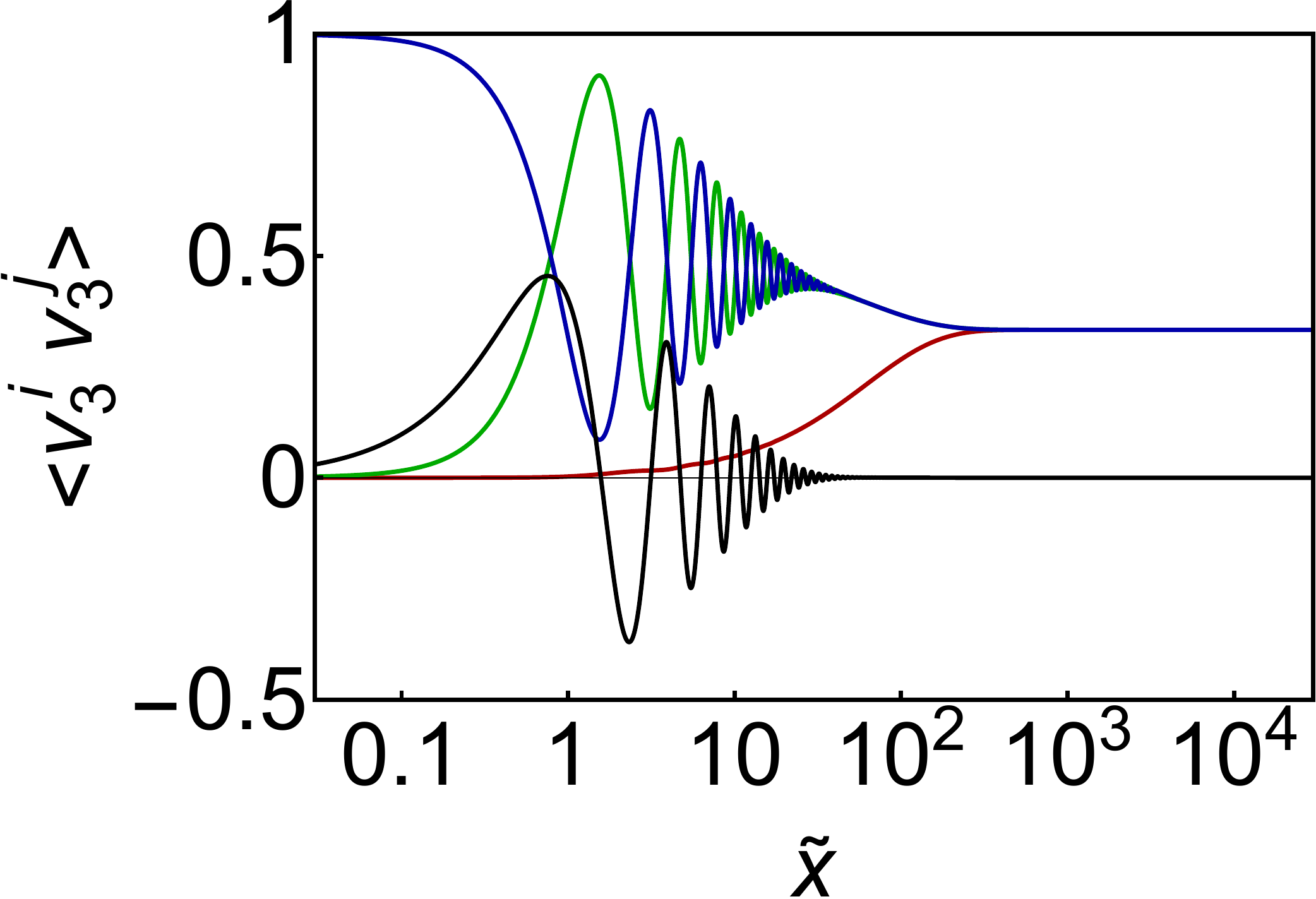}
				\caption{$\Psi=10^{2}$, $\tilde{w}=0.1$}\label{fig:corrs_psi_100_w_1}
			\end{subfigure} &
			\begin{subfigure}{0.33\textwidth}
				\centering
				\includegraphics[width=.95\textwidth]{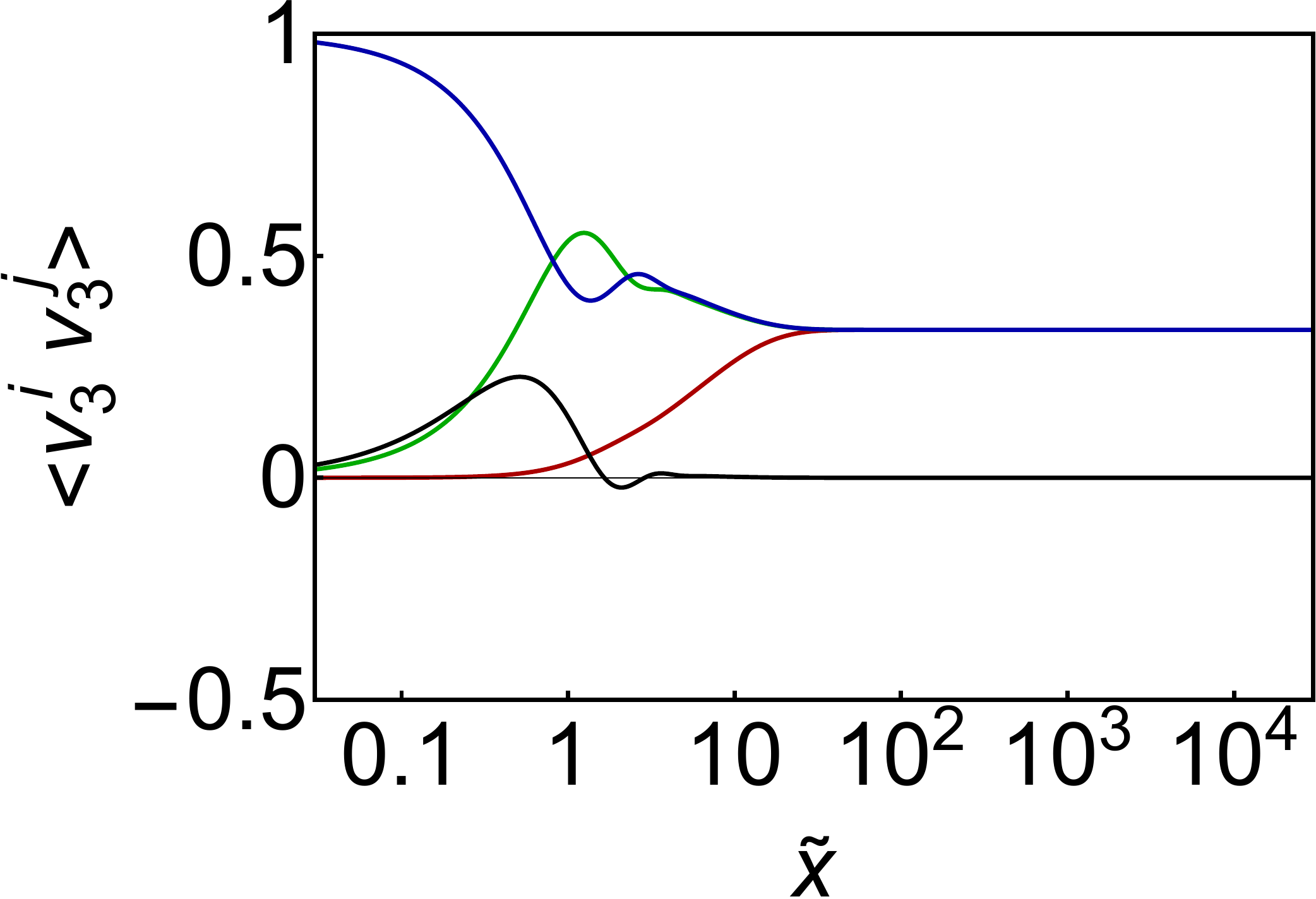}
				\caption{$\Psi=10$, $\tilde{w}=0.1$}\label{fig:corrs_psi_10_w_1}
			\end{subfigure}& \\
		\end{tabular}
		\caption{Correlation at different temperatures and width. Subfigures (a)-(e): The components of $\rho_{ i  j}(x,x)$, semi-logarithmic scale. $\rho_{11}$-red,  $\rho_{22}$-blue, $\rho_{33}$-green, $\rho_{23}$-black	 }\label{fig:tangent_corrs}
	\end{figure}

	Being a scalar quantity, $R_g^2$ cannot provide information about statistical structural properties of the the ribbon. A direct example to this is the the fact that at $L_{min} < L <L^*$ (especially at low temperatures), the ribbon is shaped like a ring. Such configurations are described by a bounding  oblate spheroid, while a spherical description using $R_g $is misleading. To see this we study the frame origin correlation matrix (FOCM) $\rho_{i j}(x)$ (as the gyration tensor $R_{ij}$ is difficult to calculate even in simplet cases). It measures correlations between components of the $i^{th}$ and $j^{th}$  frame vectors at $x$ which are  parallel to the ribbon's tangent at the origin $x=0$.
	\begin{align}
	\rho_{ij}(x) &= \langle {v}_3^i (x) {v}_3^j (x)\rangle=  \langle \left(\hat{v}_i(x)\cdot \hat{v}_3(0)\right)\left(\hat{v}_j(x)\cdot \hat{v}_3(0)\right)\rangle
	\end{align}
	This matrix is plotted in Figure \ref{fig:tangent_corrs}, Using our knowledge of $\ell_\tau$, $\ell_p$ the curvatures and the fact that on long distances $\rho_{ij} \rightarrow \frac{1}{3} \delta_{ij}$ ($\delta_{ij}$ being Kronecker's delta), we may heuristically evaluate $\rho_{ i  j}$ (which  turns out to be a good approximation)-
	\begin{align}
	\rho_{ij}(x) &= \left(\begin{array}{ccc}
	\frac{1}{3}- \frac{1}{3}e^{-\frac{x}{\ell_\tau/2}} & 0 & 0 \\
	0 & -\frac{1}{2}\cos(2 \langle l \rangle x)e^{-\frac{x}{\ell_p/4}} +\left(\frac{1}{2}-\frac{1}{3}\right)e^{-\frac{x}{\ell_\tau/2}}+\frac{1}{3} & \frac{1}{2}\sin(2 \langle l \rangle x)e^{-\frac{x}{\ell_p/4}}\\
	0 & \frac{1}{2}\sin(2 \langle l \rangle x)e^{-\frac{x}{\ell_p/4}} & \frac{1}{2}\cos(2 \langle l \rangle x)e^{-\frac{x}{\ell_p/4}} +\left(\frac{1}{2}-\frac{1}{3}\right)e^{-\frac{x}{\ell_\tau/2}}+\frac{1}{3}\
	\end{array} \right).
	\end{align}
	
	This evaluation, and the actual results (Figure \ref{fig:tangent_corrs}) may be explained as follows:
	At low temperatures, and at short distances ($x < \ell_p$), the ribbon behaves as a stiff ring, as indicated by the fact $\rho_{11}\simeq 0$, and by the strong oscillatory contributions in the remaining nonzero elements of $\rho$. At intermediate positions $\ell_p < x< \ell_\tau$, while $\rho_{11}\sim 0$ remains, we see the oscillations subside (depending on ${\ell_\tau}/{\ell_p}$) and we may roughly approximate (especially for large values of this ratio)
	\begin{align}
	\rho_{ij} &\sim \left(\begin{array}{ccc}
	0&0&0 \\
	0& \frac{1}{2} & 0\\
	0 & 0& \frac{1}{2}
	\end{array}\right),
	\end{align}
	suggesting the ribbon's tangent is randomly directed in 2D. That is, the ribbon occupies an oblate spherical volume with a large aspect ratio. At further positions $x >\ell_\tau $ we find that the gyrating ribbon occupies an evermore spherically shaped volume such that \begin{align}
	\rho_{ij}(x\rightarrow \infty) &= \frac{1}{3} \delta_{ij},
	\end{align}
	corresponding to  randomly directed tangent in 3D.  The fact that the tangents of the mid-line at distance $x$ from the origin are randomly distributed does not mean that the ribbon lacks any structure (as also indicated by the fact that some of the eigenvalues of $\mathbf{\Lambda}$ are complex). Rather, that there is no long range order (of the Darboux frame) in the system. Knowledge  at one point along the ribbon does not tell us anything about parts of the system far away from it.
	
	\begin{figure}
		\centering
		\includegraphics[width=0.6\textwidth,clip,trim={0cm 0cm 0cm  0cm}]{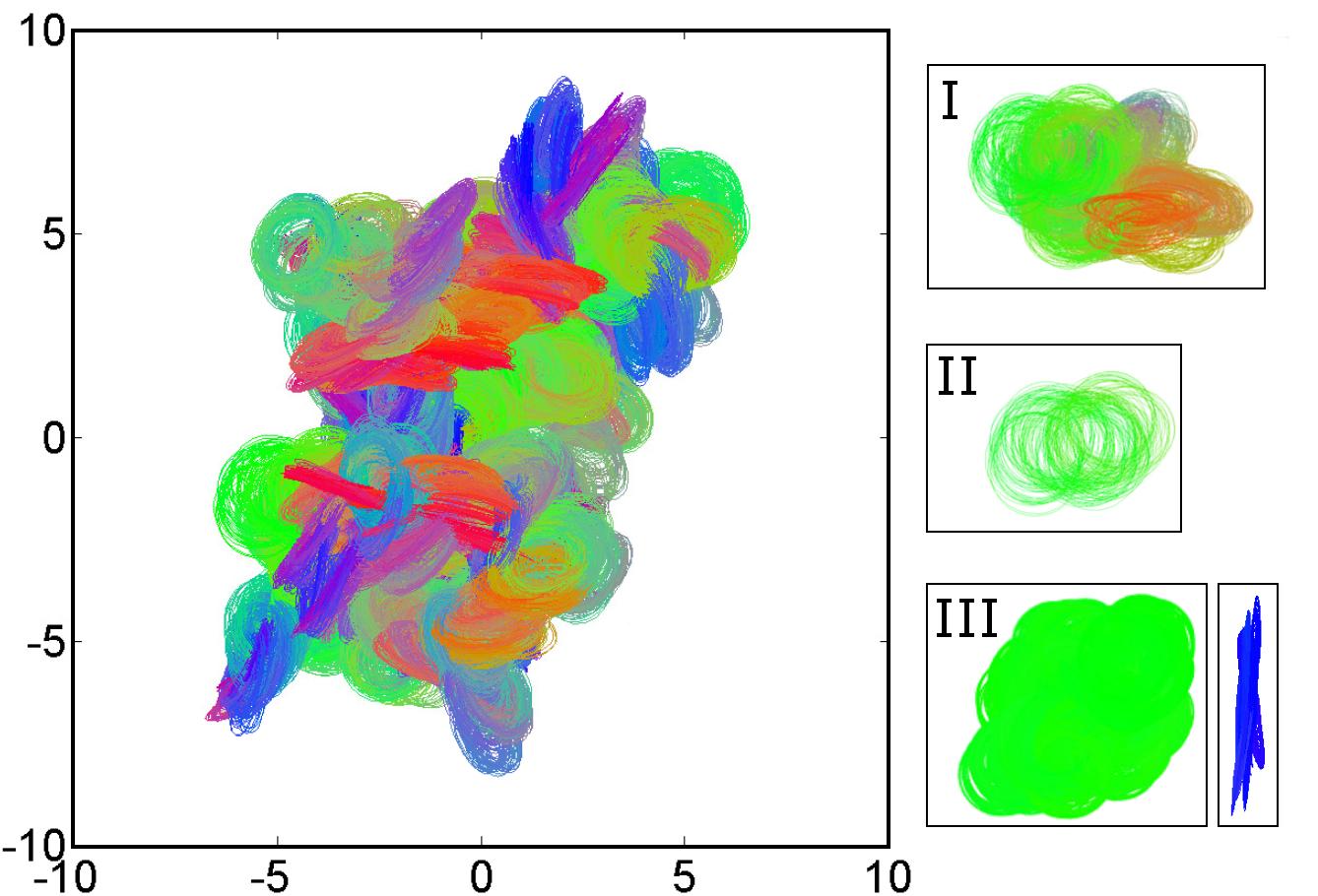}
		\caption{ Visualization mid-line configuration at $\tilde{T}=10^{-3}$ (except Inset III) and $\tilde{w}=0.1$ (all), obtained using MC simulation, all panels are to scale. Colours correspond to different direction of the bi-normal: red- up/down, blue- right/left, green-front/back. Main- full length of a ribbon $\tilde{L}=k_0 L= 10^5$; Inset (I)- a segment of $\tilde{L}=10^4$ of the same ribbon; Inset(II)- a segment of $\tilde{L}=10^3$; Inset(III)- Left- A segment of length $\tilde{L}=10^4$ at $\tilde{T}=10^{-4}$, Right- a similar segment facing sideways (hence coloured blue) to emphasize the high aspect ratio. Uniform colour indicates the  ribbon has not lost it's bi-normal correlation, yet the shape clearly shows it has lost any long range  tangent-tangent correlation.}\label{fig:MC_psi_103_w_1}
	\end{figure}
	
	We used Monte-Carlo (MC) simulations, to visualize a thermalized ribbon at different temperatures.  The results are plotted in Figure \ref{fig:MC_psi_103_w_1}, where different colours correspond to different directions of the bi-normal ($\hat{v}_1$). It is immediately seen that at lower temperatures, larger same-colour segments appear, indicating large correlation lengths of the bi-normal. We term this behavior (and hence the phase of the ribbon) "Plane Ergodic" (PE).  The existence of this state depends on the ratio of $\ell_\tau/ \ell_p$ and temperature. If $\ell_\tau/ \ell_p \gg 1$, at a given temperature,  the "planar" regions of the ribbon are made of longer segments and are more easily observed. If, however $\ell_\tau/ \ell_p = 1$, then this phase does not exist at all.  At higher temperatures, this phase is harder to observes as it spans shorter distances, while at temperatures larger than $T^*$ the ribbon has lost structure completely, therefore behaving as an ideal chain and no such phase may be observed.

	\subsubsection{Wide Ribbons $\tilde{w}>1$}\label{sub:WideRibb}
	
\begin{figure}[!h]
		\begin{tabular}{ccc}
			\centering
			\includegraphics[width=.45\textwidth,clip,trim={1cm .5cm 1cm  .5cm}]{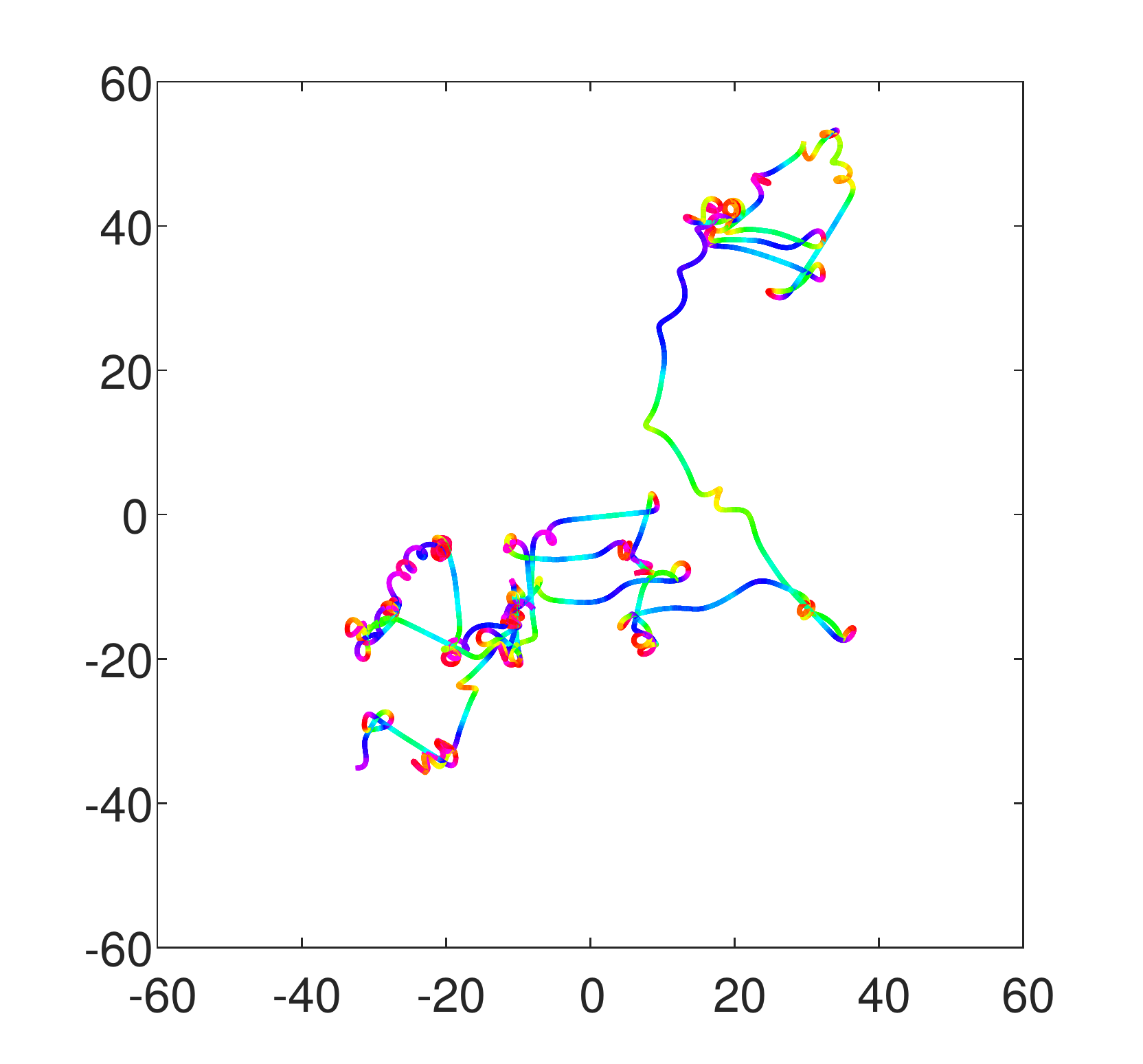}
			\includegraphics[width=.45\textwidth,clip,trim={1cm .5cm 1cm  .5cm}]{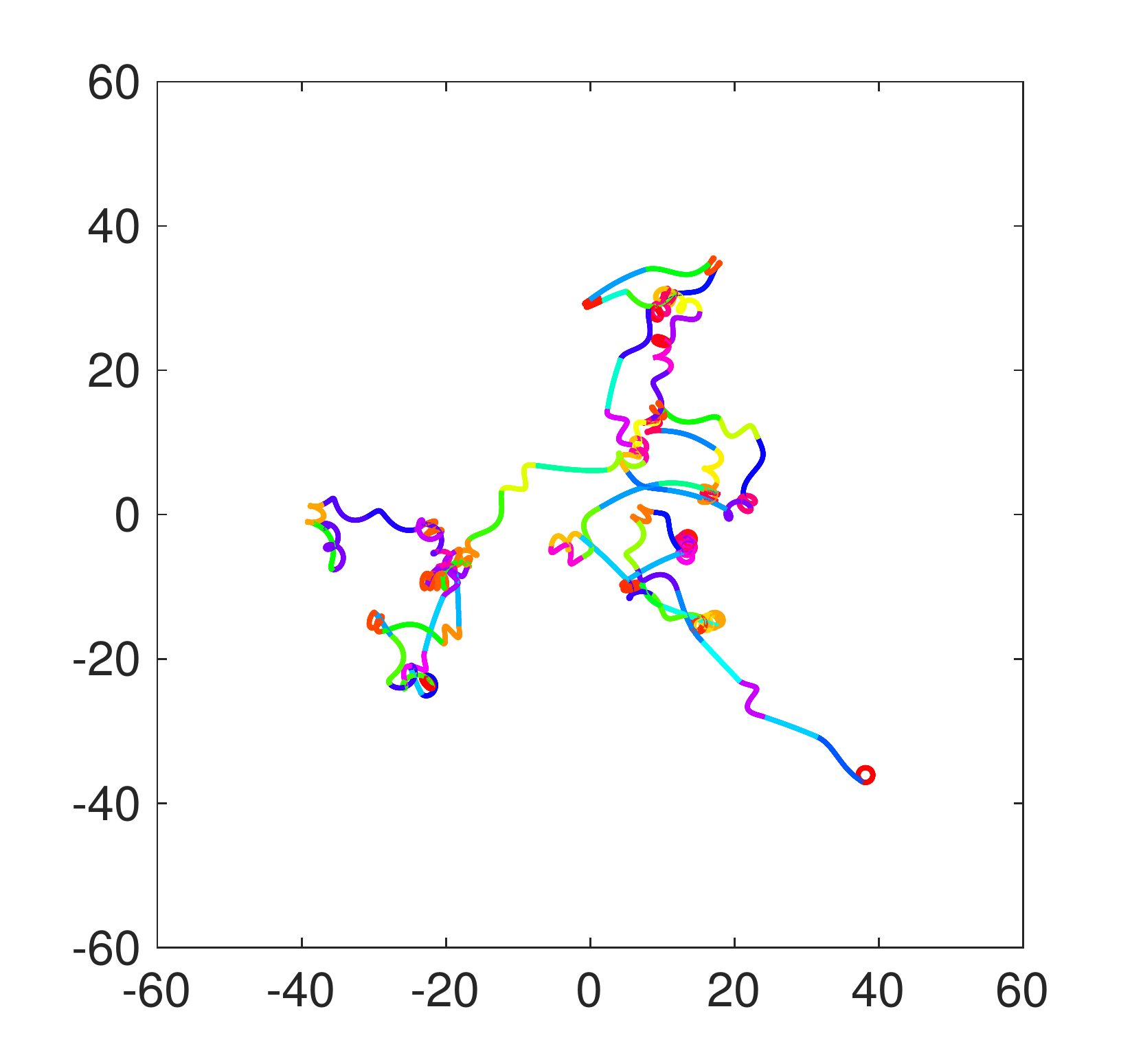}
		\end{tabular}
		\caption{ Left- a MC simulation of a possible mid-line configuration at the wide  regime with $\tilde{T}=10$, $\tilde{w}=10$ (random structured phase) of total length $\tilde{L} = 10^3$ ; Colours correspond to different values of $\theta$, $\xi_\theta \sim 100$  ; Right- a naive reproduction of a random structured ribbon. Made by assigning  a random angle to segment exactly  $\xi_\theta =100$ long  
		}\label{fig:MC_psi_w_3}
	\end{figure}

	We now turn to study the statistics of a wide ribbon.  As mentioned in \ref{s_ch:Euilbrium}, at the wide regime one cannot neglect derivatives even at mechanical equilibrium. As they are the only energy scale (regarding $\theta$ fluctuations), they affect the statistics significantly.  We therefore repeat the calculations of $\langle \hat{v}_i (x) \hat{v}_j (x') \rangle= \langle e^{-\int_{\tilde{x}'}^{\tilde{x}} \Omega \dt x} \rangle $ (Eq. \ref{eq:corrmat}), and find that  $\log\langle e^{-\int_{\tilde{x}'}^{\tilde{x}} \Omega \dt x} \rangle = \langle \Omega \rangle \Delta + O(\Delta^2)$, for small $\Delta=|\tilde{x}-\tilde{x}'|$, while  $\log\langle e^{-\int_{\tilde{x}'}^{\tilde{x}} \Omega \dt x} \rangle \simeq \Lambda_\xi \Delta$ for large $\Delta$. Where  to leading order
	\begin{align}\label{eq:Lambda_xi}
	\Lambda_\xi&= \langle \Omega \rangle +\frac{1}{2} \left(\begin{array}{ccc}
	\frac{1}{2}\left(\langle \Delta \tilde{z}^2 \rangle +\tilde{z}_{eq}^2 \xi_\theta\right) & 0 & 0 \\
	0 & \frac{1}{2} \langle \Delta \tilde{h}^2\rangle + \langle \Delta \tilde{z}^2 \rangle +\tilde{z}_{eq}^2 \xi_\theta  & 0 \\
	0 & 0 & \frac{1}{2}\left(\langle \Delta \tilde{h}^2\rangle + \langle \Delta \tilde{z}^2 \rangle +\tilde{z}_{eq}^2 \xi_\theta \right)
	\end{array}\right),
	\end{align}
	It is easily verified that $\lim\limits_{\xi_\theta \rightarrow 0} \Lambda_\xi = \Lambda$, thus retrieving  Eq. \ref{eq:Lambda}.
	At low temperatures ($\xi_\theta \rightarrow \infty$), $\Lambda_\xi$ reduces into 
	\begin{align}\label{eq:Lambda_xi_low}
	\Lambda_\xi=  \left(\begin{array}{ccc}
	\frac{ \tilde{z}_{eq}^2  \xi_\theta}{4} & 0 &0 \\
	0& \frac{\tilde{z}_{eq}^2  \xi_\theta}{2} & 0 \\
	0 & 0 & \frac{ \tilde{z}_{eq}^2  \xi_\theta}{4}
	\end{array}\right),
	\end{align}
	where we also omittede the $\langle \Omega \rangle $ term as it is negligible.
	This means that on scales much larger than $\xi_\theta$ the ribbon statistically behaves as an ideal chain $\ell_\tau=\ell_p=\ell_k$, while on smaller scales it behaves as rigid. However, unlike an ideal chain, the different Kuhn segments are not structureless. Rather, they have different shapes ranging from rings, through helices, to straight segment depending on the chosen angle and the width of the ribbon (as depicted in Fig \ref{fig:MC_psi_w_3}), we term this unusually soft phase "Random Structured". Due to the non trivial position dependence, even numeric calculation of $R_g$ and $\rho_{ i  j}$ is hard. Nevertheless we can compare the above interpretation with MC simulation. Indeed, in Figure \ref{fig:MC_psi_w_3}, we compare the result of a MC simulation (left) to "naive" model consisting of a random ,uniform, angle for segments exactly $\xi_\theta$ long. The similarity between plots is clear, thus our interpretation is supported by the simulation.
	
	As mentioned before, at very low temperatures $\tilde{T} \ll 1$ the  boundary layer dominates the ribbon's configuration giving rise to statistical behaviour identical to that of the Plane Ergodic phase. As  temperature get higher the ribbons smoothly transitions into the Random Structured phase. Direct calculation shows that the RS phase is describes the ribbon sufficiently at $\tilde{T}\gtrsim 0.1$.
	
	\subsubsection{Ribbon's Phase Diagram}
	
\begin{figure}[h!]
		\centering
		\includegraphics[width=1\textwidth]{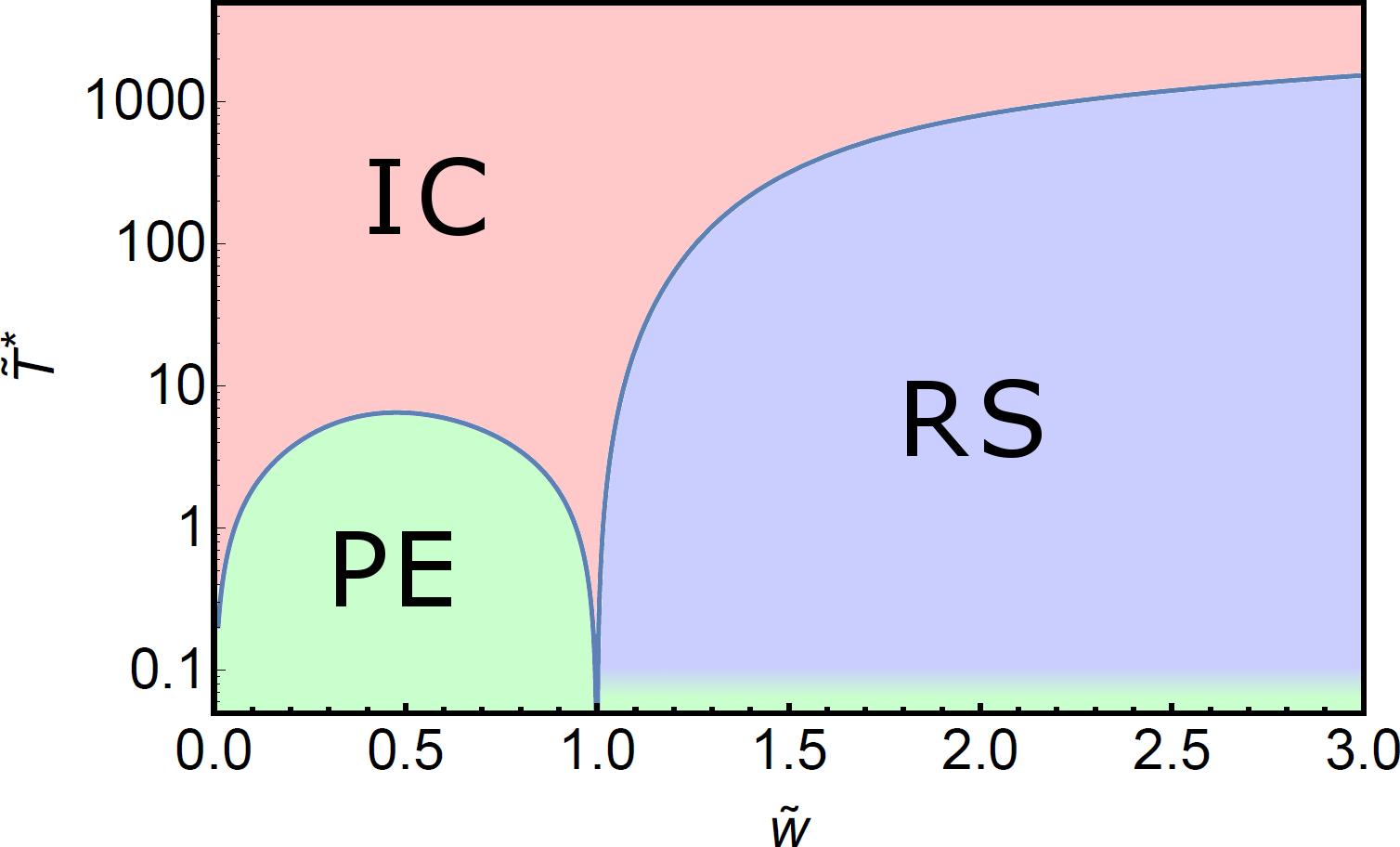}
		\caption{Ribbon phases. IC- Ideal Chain, RS- random structured, PE- plane ergodic. The RS phase does not exist in compatible ribbons, the thin green region at the bottom of the RS phase marks the temperature range at which the boundary layer also affects the statistic, exhibiting a similar behaviour to that of PE phase. PE exists even in compatible ribbons, but is different in the details.  IC- is the natural limit of every statistical theory of elastic ribbons.}\label{fig:Phases}
	\end{figure}

	We may now finally plot the phase diagram of our ribbons. In Figure \ref{fig:Phases}  we plotted all three phases we've described so far. The blue curve is $T^*(\tilde{w})$ separating the "structured" phases from the "melted" (ideal chain) phase (light red). This phase is termed Ideal Chain (IC) phase, since at high temperatures the ribbon is well described by the ideal chain model at all scales. Under $T^*$, in a continuous, yet sharp, manner, we find two other phases, separated at the critical width. A narrow ribbon will exhibit the Plain Ergodic (PE) phase, in which at intermediate scales we may find the ribbon coiled into rings which slowly drift away yet contained within a plane, while  at large scales the ribbon drifts out of plane. Thus the ribbon may be describe as made of "planar" segments that are randomly connected (both in orientation and position, as seen in Fig. \ref{fig:MC_psi_103_w_1}). Note that in this phase $\ell_p\propto \frac{1}{T}$ yet $\ell_k \propto T$, suggesting that the actual size of the planar segments varies only slightly with temperature. It is important to emphasize, that this phase is geometrical in nature and appears solely due to the ribbon's underlying structure (ring-like).
	Finally we have the Random Structured (RS) phase at $\tilde{w}>1,~ T<T^*$. This is a unique phase appearing due to the incompatibility of the ribbon. A ribbon in this phase is soft, and appears almost like an ideal chain on large scales. On intermediate scale one can see that the ribbon has a randomly selected structure- it is made of segments whose shapes are uniform along the segment and may either be right handed, or left handed helical of various pitches as visualized in Fig. (\ref{fig:MC_psi_w_3}). These segments' size scales as $\frac{1}{T}$. It is important to note that on intermediate temperatures ($T<T^*$ but not $T\ll T^*$) the segments shape is not completely uniform and may fluctuate. Nevertheless, they are still distinct, and well defined in their shape, losing it only when we get close to $T^*$. At extremely low temperatures ($\tilde{T} \lesssim 0.1$) this phase exhibits a PE-like behaviour, as the boundary layer becomes important also from a statistical point of view (and not only mechanical).
	
	\section{Conclusions}\label{ch:Conc}
	
	Using newly developed formalism to describe incompatible ribbons (\cite{Grossman2016}), we identified and quantified the phase space of incompatible ribbons with positive spontaneous curvature at different temperatures and widths. Such ribbons exhibit configuration transition at a critical width, even at zero temperature. At small widths, the ribbon is bending dominated- i.e its configuration is set by its spontaneous curvature, while at large widths, it is stretching dominated, and  its flat (Euclidean) in-plane geometry prescribes a range of developable configurations. We showed that the mechanics of such ribbons is nontrivial and strongly depends on ribbon width. In particular, wide ribbons are abnormally floppy, as their bulk energy is degenerate and their rigidity stems  solely from variations in boundary layer energy. At finite temperature, these nontrivial mechanics lead to non trivial statistical properties, which do not exist in compatible ribbons. We calculate explicitly and find that different statistical geometrical measures, such as the persistence length $\ell_p$, the torsional correlation length $\ell_\tau$, and the Kuhn length $\ell_k$ vary non-monotonically with ribbon width and with temperature. The  volume occupied by a ribbon varies in temperature and width in a non trivial way. Part of this variation is presented as three different phases (Fig. \ref{fig:Phases}), which we term Ideal Chain (at high temperatures), Plain Ergodic (narrow, cold ribbons), and Random Structured (wide, cold ribbons).
	While the first is common to all elastic ribbons, the second is purely geometrical in nature and we would expect all ribbons (including compatible) with some underling structure to have a similar phase. The third phase exists only in positively curved incompatible ribbons and arises purely due to the unique residual stresses in the problem we studied. In this phase, degeneracy in the bulk energy of the ribbon leads to continuous family of (mechanical-) equilibrium configurations. This, at low but finite temperatures, leads the ribbon to look as a random coil on large scales, yet with segments that have a definite and well defined shape, in contrast to compatible ribbons where there is a single equilibrium shape. Though the details of the transitions and the characteristic configurations are unique to positively curved ribbons, we expect a qualitatively similar phase space to appear in other cases of incompatible ribbons. The mechanics and geometry of all such systems are non-trivial and dominated by the competition between the residual bending and stretching energies. In this sense, our results point to a general phase space, that should be expected even in relatively simple self assembled systems.
	It should be emphasized that the work on such ribbons is far from closure as such ribbons (especially in Plain Ergodic phase) are densely coiled, and self avoidance effects should be important. Other important corrections may arise at high temperatures, where the Gaussian approximation should fail. It is worth to note that the resulting phase diagram is reminiscent of quantum phase transition, in the sense of the existence of a critical point even in zero temperature. While somewhat similar connections between elastic  and quantum systems were made in the past \cite{Moroz1997}, the extent of this similarity in the context of this paper is ill understood and requires further investigation.

	\section{Acknowledgment}
	This research was supported by the  US-Israel Binational Science Foundation \# 2014310. D. G. was also supported by The Harvey M. Kruger Family Center of Nanoscience and Nanotechnology.

	\clearpage
	\newpage
	\appendix
	\section{Formalism}\label{ch:Formalism}	
	A ribbon is a thin, narrow, and long, sheet with thickness, width and length satisfy $t\ll W \ll L$ respectively. We choose coordinates on the ribbon $(x,y)\in \left[0,L\right]\times\left[-\frac{W}{2},\frac{W}{2}\right]$, where $L$ is the ribbon's length and $W\ll L$ is its width. Thus, the ribbon's configuration is $\vec{r}(x,y)$, and it's mid-line is given by $\vec{r}(x,0)$.

	In order to calculate the statistics of a ribbon, we need an energetic model. Nevertheless, we may write formal expression describing the various statistical measures of the ribbon even without one. Following Panyukov and Rabin \cite{Panyukov2000a}, we begin  by defining the Frame Correlation Matrix (FCM)
	\begin{align}\label{eq:frameCorr}
	C_{ij}(x,x')& \equiv \langle \hat{v}_i(x)\hat{v}_j(x') \rangle = T_x \left[e^{-\int_{x'}^x \mathbf{\Lambda}(x'') \dt x''}\right]_{ij}
	\end{align}
	for $x>x'$ and $\langle X \rangle$ mark a thermal averaging of $X$. In some sense, $\mathbf{\Lambda}$ is the generator of $\mathbf{C}(x,x')$, and in general
	$\int_{x'}^x\mathbf{\Lambda}(x'') \dt x''=\int_{x'}^x\mathbf{\Lambda}(x'';x,x',T) \dt x''$. From $\mathbf{C}$ alone,  we may extract various statistical measure of the thermal ribbon (see section \ref{app:Stat_Meas}for details): 
	\begin{align} 
	\ell_p \colon & C_{33}(x,x')  \propto e^{-(x-x')/{\ell_p}} \label{eq:ell_p}\\
	\ell_k \equiv & \lim\limits_{L\rightarrow\infty} \frac{1}{2 L} \langle r^2(L) \rangle = \lim\limits_{L\rightarrow\infty} \frac{1}{L}  \left[\int_0^L\int_{0}^{x} C_{33}(x,x') \dt x' \dt x \right]  \label{eq:ell_k}\\
	R_g^2(L)  \equiv& \frac{1}{L} \int_{0}^L \langle r^2(x) \rangle \dt x - \langle r_0^2 \rangle \label{eq:R_g}\\ \nonumber
	\text{\scriptsize $\langle r_0^2 \rangle =$}&\text{\scriptsize $\left[\frac{2}{L^2} \int_{0}^{L}\int_{0}^{x}\left[\int_{x'}^x \int_{0}^{x'} C_{33}(y,y') \dt y' \dt y  + 2 \int_0^{x'}\int_0^y C_{33}(y,y') \dt y' \dt y\right]\dt x' \dt x\right]$}.
	\end{align}
	where $\vec{r}(L) =\vec{r}(L,0)$ is the end-to-end vector, $r_0= \frac{1}{L}\int_0^L \vec{r}(x) \dt x$ is the ribbon's center-of mass. $\ell_p$ is the persistence length which is the scale on which tangent-tangent correlations decay, hence equation (\ref{eq:ell_p}) above is not an equation defining $\ell_p$, rather it is a relation $\ell_p$ satisfies. $\ell_k$ is the Kuhn length of the ribbon, it is defined as the segment length a random freely-jointed chain with the same end-to-end length. Classically  (\cite{Rubinstein2003}) $\ell_k \propto \ell_p$, however it is readily seen that in general, $\ell_k, \ell_p$ do not satisfy any simple relation, only when $\Lambda_{13}=\Lambda_{23}=\Lambda_{31}=\Lambda_{32}=0$, and $\Lambda_{33} =const$, do we recover the classical results.
	Another statistical measure, which is somewhat less common is the torsional correlation length, $\ell_\tau$ {(\cite{Giomi2010})}, defined as the scale along we lose bi-normal-bi-normal correlations
	
	\begin{align}
	\ell_\tau \colon  & C_{11}(x,x')  \propto e^{-(x-x')/{\ell_\tau}}.
	\end{align}
	
	While the measures in Eqs. \ref{eq:ell_p} - \ref{eq:R_g}, are important and useful, they are scalars, and as such cannot capture the dimensions of a gyrating ribbon well. A better parameters  would therefore be a tensorial quantity, this is the gyration tensor describing the ellipsoid containing the ribbon. It is a symmetric tensor, defined as
	\begin{align} 
	R_{ij}(L) = \frac{1}{L} \int_{0}^L  \left[\langle r_i(x) r_j(x) \rangle - \langle (r_0)_i(r_0)_j \rangle \right]\dt x.
	\end{align}
	where $r_i(x)= \vec{r}(x) \cdot \hat{v}_i(0)$. Another, somewhat more useful, measure (that appears in the calculation of $T$) is the Frame-Origin Correlation Tensor (FOCT).
	
	\begin{subequations}\label{eq:frame_origin_tensor}
		\begin{align} 
		\rho_{AB}(x,x') =& \rho_{a a' b b'}(x,x') =\langle O_{ab}(x,x') O_{a' b'}(x,x') \rangle
		\end{align}
		Where we used a multi-index notation as $\rho$ is an average of a Kronecker (tensorial) product of two rotation matrices, and may be represented by a matrix. Using an index-less notation we may write
		\begin{align}\label{eq:rhomat}
		\mathbf{\rho}(x,x')=&\langle \mathbf{O}(x,x') \otimes \mathbf{O}(x,x') \rangle \equiv e^{\int_{x'}^x \mathbf{\Gamma} \dt x''} ,
		\end{align}
		where $\mathbf{\Gamma}= \mathbf{\Gamma}(x'';x,x',T)$, similarly to $\mathbf{\Lambda}$.
	\end{subequations}
	If we define the partial trace
	$ PTr[\rho]_{ab} = \sum_k \rho_{a k k b}$, it can be seen that $PTr[\rho](x,x')= \mathbf{C}(x,x')$.

	Without any specification regarding the elastic model it is hard to advance, since (at least in principle) the markings $\mathbf{\Gamma,~\Lambda}$ are only formal in the sense that they may depend on many different parameters (including the integration limits $x,~x'$). We will therefore assume that one may expand the energy as a bilinear form of the deviation from equilibrium values (as is often the case)
	i.e-
	\begin{align}\label{eq:energy_exp}
	E_{el}=& \int \left[ A_{ll} \Delta l^2 (x) +A_{mm} \Delta m^2 (x)+A_{nn} \Delta n^2 (x) \right.\\ \nonumber &\left. + A_{lm} \Delta l(x)\Delta m(x) +A_{ln} \Delta l(x)\Delta n(x) +A_{nm} \Delta n(x)\Delta m(x) \right] \dt x
	\end{align}
	where $\Delta X = X- \langle X \rangle$. If the model includes higher orders or even derivatives of the curvatures, then one must modify the following equations.
	
	Under Eq. \ref{eq:energy_exp}, $\langle  \Delta l (x) \Delta l (x') \rangle \propto \delta(x-x')$, and similarly for every other combination of $\left( \Delta l, \Delta m , \Delta n\right)$.
	\begin{subequations}\label{eq:general_corrs_mat}
		\begin{align}
		\mathbf{\Lambda}(x)& = -\lim\limits_{y\rightarrow x}\frac{\langle\mathbf{ O}(y,x) \rangle}{y-x} = \langle \mathbf{\Omega} \rangle -\frac{1}{2}\langle \mathbf{\Delta\Omega^2} \rangle \\ \label{eq:G_mat}
		\mathbf{\Gamma}(x) &= \lim\limits_{y \rightarrow x} \frac{\langle \mathbf{O}(y,x) \otimes \mathbf{O}(y,x) \rangle}{y-x} = - (\mathbf{\Lambda} \oplus \mathbf{\Lambda}) +\langle \mathbf{\Delta \Omega \otimes \Delta \Omega} \rangle,
		\end{align}
		
		where $A\oplus B= A \otimes I_B + I_A\otimes B$ is the Kronecker sum, defined for square matrices only and $I_A$ is a unit matrix with the same dimensions as the matrix $A$. Where we marked $\mathbf{\Delta\Omega= \Omega- \langle \Omega \rangle}$. From our assumption $\langle \mathbf{\Delta \Omega}(x) \mathbf{\Delta \Omega}(x') \rangle= \mathbf{\langle \Delta \Omega^2 \rangle} \delta(x-x')$ and  $\mathbf{\langle \Delta \Omega} (x) \otimes \mathbf{\Delta \Omega} (x')\rangle = \mathbf{\langle \Delta \Omega \otimes \Delta \Omega \rangle} \delta(x-x')$.
	\end{subequations}
	Direct calculation results with
	\begin{subequations}
		\begin{align}
		&\mathbf{\langle \Delta \Omega^2 \rangle} = \mathbf{M}^{-1}- I_{3\times 3} \Tr \mathbf{M}^{-1} \\ 
		&\langle  \Delta \Omega \otimes \Delta \Omega\rangle_{a a' b b'} = \epsilon_{abi}\epsilon_{a'b'j}(M^{-1})_{ij}
		\end{align}
		where $\epsilon_{ikj}$ is the Levi-Civita tensor, and
		\begin{align*}
		\mathbf{M}^{-1}&=\left(\begin{array}{ccc}
		\langle \Delta l^2\rangle & 0 &\langle \Delta l \Delta m\rangle \\
		0 & 0 & 0 \\
		\langle \Delta l \Delta m\rangle& 0& \langle  \Delta m^2 \rangle
		\end{array}\right).
		\end{align*}
	\end{subequations}
	
	We are now missing the last piece of the puzzle- an elastic model. We use a quasi-one dimensional Hamiltonian describing the elasticity of a residually stressed ribbon, derived in \cite{Grossman2016}. It is based on a dimensional reduction of previous work by Efrati et al. \cite{Efrati2009a} describing the elastic energy of residually stressed sheet. We assume a ribbon whose thickness ($t$) is much smaller that its width $W$, which in turn is much smaller than the ribbon's length ($t\ll W\ll L$). This model enables us to treat residually stressed (as well as non- residually stressed) ribbons analytically.

	\section{Statistical Measures}\label{app:Stat_Meas}
	Following \cite{Panyukov2000a}, and assuming an energetic model as in Eq. (\ref{eq:energy_exp}). We  start from the Frame Correlation Matrix definition:
	\begin{align}
	C_{ij}(x,x')&= \langle \hat{v}_i(x) \hat{v}_j(x')  \rangle
	\end{align}
	
	We start from the calculation of the gyration radius as it encompasses also that of $\ell_k$.
	
	\begin{align}
	R_g^2&= \langle \int_0^L \frac{\dt x}{L} \left[r(x)^2 - 2\vec{r}(x) \vec{r}_0+ r_0^2 \right] \rangle=- \langle r_0^2 \rangle + \frac{1}{L}\int_0^L \langle r(x)^2\rangle \dt x
	\end{align}
	Thus
	
	\begin{subequations}
		\begin{align}
		\langle r_0^2 \rangle &=  \frac{1}{L^2} \int_0^L \int_0^L\langle \vec{r}(x) \vec{r}(x')  \rangle \dt x \dt x'  = \frac{2}{L^2} \int_0^L \int_0^x\langle \vec{r}(x) \vec{r}(x')  \rangle \dt x \dt x' \\ \nonumber
		& = \frac{2}{L^2} \int_0^L \int_0^x \int_0^x \int_0^{x'} \langle \hat{v}_3(\tilde{x}) \hat{v}_3(\tilde{x}')  \rangle \dt x \dt x' \dt \tilde{x} \dt \tilde{x}' \\ \nonumber
		& = \frac{2}{L^2} \int_0^L \int_0^x \left[\int_{x'}^x \int_0^{x'} \langle \hat{v}_3(\tilde{x}) \hat{v}_3(\tilde{x}')  \rangle \dt \tilde{x} \dt \tilde{x}' + 
		2 \int_{0}^{x'} \int_0^{\tilde{x}} \langle \hat{v}_3(\tilde{x}) \hat{v}_3(\tilde{x}')  \rangle \dt \tilde{x} \dt \tilde{x}' \right] \dt x \dt x'  \\ \nonumber
		&=\frac{2}{L^2} \int_0^L \int_0^x \left[\int_{x'}^x \int_0^{x'}  C_{33}(\tilde{x},\tilde{x}')  \dt \tilde{x} \dt \tilde{x}' + 
		2 \int_{0}^{x'} \int_0^{\tilde{x}} C_{33}(\tilde{x},\tilde{x}')   \dt \tilde{x} \dt \tilde{x}' \right] \dt x \dt x'  \\ \nonumber
		& \stackrel{Eq. (\ref{eq:energy_exp})}{=}  \frac{2}{L^2} \int_0^L \int_0^x \left[\int_{x'}^x \int_0^{x'} (e^{-\Lambda(\tilde{x}-\tilde{x}')})_{33}\dt \tilde{x} \dt \tilde{x}' + 
		2 \int_{0}^{x'} \int_0^{\tilde{x}} (e^{-\Lambda(\tilde{x}-\tilde{x}')})_{33} \dt \tilde{x} \dt \tilde{x}' \right] \dt x \dt x' \\ \nonumber
		& = \frac{2}{L^2} \int_0^L \int_0^x \left[\int_{x'}^x \int_0^{x'} (e^{-\Lambda(\tilde{x}-\tilde{x}')})_{33}\dt \tilde{x} \dt \tilde{x}' + 
		2 \int_{0}^{x'} \int_0^{\tilde{x}} (e^{-\Lambda(\tilde{x}-\tilde{x}')})_{33} \dt \tilde{x} \dt \tilde{x}' \right] \dt x \dt x' \\ \nonumber
		&= \left[\frac{2}{3} \Lambda^{-1} L - \Lambda^{-2} - \frac{2}{L} \Lambda^{-3} e^{-\Lambda L} + \frac{2}{L^2} \Lambda^{-4} \left(1- e^{-\Lambda L}\right) \right]_{33}
		\end{align}
		
		\begin{align}
		\frac{1}{L}\int_0^L \langle r(x)^2\rangle \dt x &=  \frac{1}{L}\int_0^L  \int_0^x \int_0^x \langle \hat{v}_3(\tilde{x}) \hat{v}_3(\tilde{x}')\rangle \dt \tilde{x} \dt \tilde{x}'  \dt x  =  \frac{2}{L}\int_0^L  \int_0^x \int_0^{\tilde{x}} C(\tilde{x},\tilde{x}')_{33} \dt \tilde{x} \dt \tilde{x}'  \dt s \\ \nonumber
		&\stackrel{Eq. (\ref{eq:energy_exp})}{=}  \frac{2}{L}\int_0^L  \int_0^x \int_0^{\tilde{x}} \left( e^{-\Lambda (\tilde{x}-\tilde{x}') } \right)_{33} \dt \tilde{x} \dt \tilde{x}'  \dt s \\ \nonumber & = \left[ \Lambda^{-1} L - 2 \Lambda^{-2} +\frac{2}{L} \Lambda^{-3}\left(1- e^{-\Lambda L}\right) \right]_{33}
		\end{align}
	\end{subequations}
	
	Hence
	\begin{align}
	\ell_k \equiv & \lim\limits_{L\rightarrow\infty} \frac{1}{2 L} \langle r^2(L) \rangle = \lim\limits_{L\rightarrow\infty} \frac{1}{L}  \left[\int_0^L\int_{0}^{x} C_{33}(x,x') \dt x' \dt x \right] \stackrel{Eq. (\ref{eq:energy_exp})}{=}   \left(\Lambda^{-1}\right)_{33}
	\end{align}

	Finally we conclude this appendix with the calculation of the gyration tensor $R_{ij}(L)$ (and the Frame Origin Correlation Tensor $\rho_{AB}$.
	
	\subsection{Gyration Tensor}
	As before, we are interested in the thermal average
	\begin{equation}
	R_{ij}=  \int_{0}^{L} \frac{\dt s}{L} \left[r^i(s) r^J(s)- r^i(s)r^j_0-r^{j}(s)r^i_0 + r^i_0r^j_0\right]= -\langle r^i_0 r^j_0 \rangle + \frac{1}{L}\int_0^L \langle r^i(s)r^j(s) \rangle \dt s
	\end{equation}
	In all following results one must understand all equalities as being taken under the symmetry operator (i.e taking only the symmetric part, in relation to the indicies ,i and j)
	
	Similarly to the previous calculation
	\begin{subequations}
		\begin{align}
		\langle r^i_0r^j_0 \rangle &=  \frac{2}{L^2} \int_0^L \int_0^x\langle r^i(x) r^j(x')  \rangle \dt x \dt x' \\ \nonumber
		& = \frac{2}{L^2} \left[ \int_0^L \int_0^x \int_{x'}^x \int_0^{x'}\langle \hat{v}_3^i(\tilde{x}) \hat{v}_3^j(\tilde{x}')  \rangle \dt x \dt x' \dt \tilde{x} \dt \tilde{x}' +2 \int_0^L \int_0^x \int_{0}^{x'} \int_0^{\tilde{x}}\langle \hat{v}_3^i(\tilde{x}) \hat{v}_3^j(\tilde{x}')  \rangle \dt x \dt x' \dt \tilde{x} \dt \tilde{x}' \right]  \\ \nonumber
		&=\frac{2}{L^2}\left[I_1 +2 I_2\right]
		\end{align}
		
		\begin{align}
		\frac{1}{L}\int_0^L \langle r^i(x)r^j(x)\rangle \dt x &=    \frac{2}{L}\int_0^L  \int_0^x \int_0^{\tilde{x}} \langle {v}_3^i(\tilde{x}) {v}_3^j(\tilde{x}')\rangle \dt \tilde{x} \dt \tilde{x}'  \dt x \\ \nonumber
		\end{align}
	\end{subequations}
	While this may seem as a relatively simple calculation, much  like the previous one, this is not the case. In order to calculate the tensor we must have an external coordinates (in the embedding space). However, all our formalism is defined on the ribbon's manifold using a Darboux frame. We use the symmetry for solid body rotations to define the external coordinate system using the Darboux frame at the beginning of the ribbon.  Thus
	\begin{align}
	\langle {v}_3^i(x) {v}_3^j(x') \rangle &=\langle \left( \hat{v}_3(x) \hat{v}_i(0)\right) \left(\hat{v}_3(x') \hat{v}_j(0) \right) \rangle =  \langle O_{3 i}(x,0) O_{3j}(x',0)\rangle
	\end{align}
	Where $O(x,x')= e^{-\int_{x'}^x \Omega(x'')\dt x''}$, where the exponent is time (position) ordered. Notice that the trace of this quantity recalls the previous result $ \langle \hat{v}_3(x) \hat{v}_3(x) \rangle = \langle O_{33}(x,x')\rangle$.
	Looking at the more general tensor
	\begin{equation}
	\langle{v}^i_\alpha (x) {v}^j_\beta (x) \rangle = \langle O_{\alpha i}(x,0) O_{ \beta j}(x',0)\rangle = \langle O_{\alpha \mu}(x,x') \rangle \langle O_{\mu i}(x,0) O_{\beta j}(x',0)\rangle,
	\end{equation}
	and we identify/define $\rho(x,0)= \langle O(x,0)\otimes O(x,0)\rangle = \langle e^{\int_0^x -\Omega \dt x' \oplus \int_0^x -\Omega \dt x'} \rangle$ (or any other position rather than $0$), where we $\otimes$ is the outer product and $\oplus$ is the Kronecker sum. From the distribution of the curvatures it is clear that  $\rho (x,x') = \rho(x,x'') \rho(x'',x')$ where $x>x''>x'$, ans we need to understand the  multiplication between those tensor by the rules of multiplication of the outer product. 
	\begin{align}
	\rho(x,x')_{a a' b b'}&= \rho(x,x'')_{a a' c c'}\rho(x'',x)_{c c' b b'} \\ \nonumber
	\rho(x,x')_{a a' b b'}& = \langle O(x,x')_{a b} O(x,x')_{a' b'} \rangle
	\end{align}
	We may thus write $\rho$ in a multi-index manner $\rho_{a a' b b'}\equiv \rho_{A B}$, where $\rho_{A B}$ satisfies the regular matrix rules. In order to find the functional form of $\rho$ we follow \cite{Panyukov2000}-
	\begin{align*}
	\rho (x,x') &= \rho(x,x'') \rho(x'',x') \\ 
	&\Downarrow\\
	\dot{\rho}(x,x') &= \frac{\partial \rho(x,x')}{\partial x} =\lim\limits_{x'' \rightarrow x} \frac{\rho(x,x'')}{x-x''}  \rho(x'',x')
	\end{align*}
	Thus
	\begin{align*}
	\mathbf{\Gamma}(x)= &= \lim\limits_{x'' \rightarrow x} \frac{\mathbf{\rho}(x,x'')}{x-x''} = \lim\limits_{x'' \rightarrow x} \frac{\langle \mathbf{O}(x,x'') \otimes  \mathbf{O}(x,x'') \rangle }{x-x''} \\
	&=  \lim\limits_{x'' \rightarrow x} \frac{\langle\left(-\int_s''^x \mathbf{\Omega} (\tilde{x}) \dt \tilde{x} + \int_s''^x\int_s''^x\int_s''^x \mathbf{\Omega} (\tilde{x})\mathbf{\Omega} (\hat{x}) \dt \tilde{x}\dt \hat{x} + \dots \right) \otimes  \left(-\int_s''^x \mathbf{\Omega} (\tilde{x}) \dt \tilde{x} + \int_s''^x\int_s''^x\int_s''^x \mathbf{\Omega} (\tilde{x})\mathbf{\Omega} (\hat{x}) \dt \tilde{x}\dt \hat{x} + \dots \right) \rangle }{x-x''}\\
	&= \left[-\mathbf{\Lambda} (x)\right] \oplus \left[ - \mathbf{\Lambda} (x) \right] + \mathbf{g}(x)
	\end{align*}
	Where
	
	\begin{align*}
	\mathbf{\Lambda} &=  \langle \mathbf{\Omega} \rangle -\frac{1}{2} \left(\mathbf{H}^{-1} -I_{3 \times 3}\Tr \mathbf{H}^{-1}\right) \\
	\mathbf{g}&= \lim\limits_{x'' \rightarrow x} \frac{\int_s''^x\int_s''^x \langle \mathbf{\Omega} (\tilde{x}) \otimes \mathbf{\Omega}(\hat{x}) \rangle \dt \tilde{x}\dt \hat{x}}{x-x''}\\
	g_{a a' b b'}&= \epsilon_{a b i}\epsilon_{a' b' j} H^{-1}_{ij}
	\end{align*}
	
	where $\mathbf{H}^{-1}$ is the correlation matrix of the curvatures ($l,~m$), $\epsilon_{ijk}$ is the Levi-Civita tensor and $I_{3 \times 3}$ is the ${3 \times 3}$ unit matirx.    Therefore-
	$$\rho(x,x') = e^{\int_{x'}^x \mathbf{\Gamma}(\tilde{x}) \dt \tilde{x}}    $$

	Also notice that $\mathbf{\Gamma}$ is not invertible. Hence there exist some   transformation $\mathbf{U}$ such that 
	\begin{align}\mathbf{U} \mathbf{\Gamma} \mathbf{U}^T = \left( \begin{array}{c|c}
	\mathbf{G} &0 \\ \hline
	0  & 0
	\end{array}\right),
	\end{align}
	where $\mathbf{G}$ is some $n \times n,~~ n<9$  matrix. Thus, 
	$$ e^{\int \mathbf{\Gamma} \dt x} =  \mathbf{U}^T \left( \begin{array}{c|c}
	e^{\int \mathbf{G} \dt x} &0 \\ \hline
	0  & I_{9-n \times 9-n}  \end{array}\right) \mathbf{U}$$.
	
	Finally we may then write-
	\begin{equation}
	\langle {v}_a^i(x) v_b^j(x') \rangle  = \left[e^{-\int_{{x'}}^{x} \mathbf{\Lambda} \otimes I \dt {x''}}  e^{\int_{0}^{{x'}} \mathbf{\Gamma} \dt x''}\right]_{a b i j}
	\end{equation}
	In our case all the matrices are constant so it makes calculation easier-
	\begin{equation}
	\langle {v}_a^i(x) {v}_b^j(x') \rangle= \left[e^{- \mathbf{\Lambda} \otimes \mathbf{I} \left(x-x'\right)}  e^{\mathbf{\Gamma} x' }\right]_{a b i j} =
	\left[e^{-\mathbf{A} (x-x')}  e^{\mathbf{\Gamma} x' }\right]_{a b i j}=\left[e^{- \mathbf{A} x} e^{\mathbf{A} x'}  e^{\mathbf{\Gamma} x' }\right]_{a b i j}
	\end{equation}
	where we defined $\mathbf{A}= \mathbf{\Lambda} \otimes \mathbf{I}$
	In what follows we omit indicies, and  use the following markings:
	\begin{align*}
	\mathbf{F}(x) &= \int_0^x \dt x' e^{\mathbf{\Gamma} x'} = \mathbf{U}^T \left( \begin{array}{c|c}
	\mathbf{G}^{-1}\left(e^{\mathbf{G} x}  -1 \right)&0 \\ \hline
	0  & x \end{array}\right) \mathbf{U} \\
	\textbf{D}(x) &= \int_0^x \dt x' \mathbf{F}(x') = \mathbf{U}^T \left( \begin{array}{c|c}
	\mathbf{G}^{-2}\left(e^{\mathbf{G} x}  -1 \right) - \mathbf{G}^{-1} x &0 \\ \hline
	0  & \raisebox{-0.5ex}{$\frac{1}{2}x^2$} \end{array}\right) \mathbf{U}\\
	\mathbf{J}(x) &= \int_0^x \dt x' \mathbf{D}(x') = \mathbf{U}^T \left( \begin{array}{c|c}
	\mathbf{G}^{-3}\left(e^{\mathbf{G} x}  -1 \right)-\mathbf{G}^{-2} x - \frac{1}{2}\mathbf{G}^{-1} x^2 &0 \\ \hline
	0  & \raisebox{-0.5ex}{$\frac{1}{6}x^3$} \end{array}\right) \mathbf{U}\\
	\overbrace{X} &\equiv \mathbf{A}^{-1}\sum_{n=0}^{\infty} (-1)^{n} \mathbf{A}^{-n} X \mathbf{\Gamma}^n
	\end{align*}
	In the following we omit for simplicity the index $_{3i3j}$ that we need to take from all the tensorial expressions.
	\begin{subequations}
		\begin{align}
		I_1&=\int_0^L \dt x \int_0^x \dt x' \int_{x'}^x \dt \tilde{x} \int_0^{x'} \dt \tilde{x}'\langle \hat{v}_3^i(\tilde{x}) \hat{v}_3^j(\tilde{x}')  \rangle = \int_0^L \dt x \int_0^x \dt x' \int_{x'}^x \dt \tilde{x} \int_0^{x'} \dt \tilde{x}' e^{- \mathbf{A} \tilde{x}} e^{ \mathbf{A} \tilde{x}'} e^{\mathbf{\Gamma} \tilde{x}' }   \\ \nonumber
		&= \int_0^L \dt x \int_0^x \dt x' ~\mathbf{A}^{-1}\left(e^{-\mathbf{A} x'}- e^{- \mathbf{A} x}\right) \overbrace{e^{\mathbf{A} x'}e^{\mathbf{\Gamma} x'}-1}   \\ \nonumber
		&= \int_0^L \dt x \int_0^x \dt x' ~\mathbf{A} ^{-1} \left[ \overbrace{e^{\mathbf{\Gamma} x'}-e^{-\mathbf{A} x'}} -e^{- \mathbf{A} x}\overbrace{e^{\mathbf{A} x'}e^{\mathbf{\Gamma} x'}-1}\right] \\ \nonumber 
		&=  \int_0^L \dt ~ \mathbf{A}^{-1} \overbrace{\left[ \mathbf{F}(x)+\mathbf{A}^{-1}\left(e^{-\mathbf{A} x}-1 \right) -\overbrace{e^{\mathbf{\Gamma} x}}+\overbrace{e^{- \mathbf{A} x}}+e^{- \mathbf{A} x}x \right]}   \\ \nonumber  
		&= \mathbf{A}^{-1} \overbrace{\left[ \mathbf{D}(L)-\mathbf{A}^{-1}L+\mathbf{A}^{-2}\left(1-e^{-\mathbf{A} L}\right) -\overbrace{\mathbf{F}(L)}+\mathbf{A}^{-1}\overbrace{\left(1-e^{-\mathbf{A} L}\right)}- \mathbf{A}^{-1} e^{-\mathbf{A} L}L + \mathbf{A}^{-2}\left(1-e^{-\mathbf{A} L}\right) \right]}\\ \nonumber
		&= \mathbf{A}^{-1} \overbrace{\left[ \mathbf{D}(L) -\mathbf{A}^{-1}L - \overbrace{\mathbf{F}(L)} -\mathbf{A}^{-1} e^{-\mathbf{A} L}L  +\mathbf{A}^{-1}\left(1-e^{-\mathbf{A} L}\right) \left(\overbrace{1}+2A^{-1} \right)  \right]}.
		\end{align}
	\end{subequations}
	In the last row we merely reordered the expressions from highest to lowest orders in $L$.
	\begin{subequations}
		\begin{align}
		I_2&=\int_0^L \dt x \int_0^x \dt x' \int_{0}^{x'} \dt \tilde{x} \int_0^{\tilde{x}} \dt \tilde{x}'\langle \hat{v}_3^i(\tilde{x}) \hat{v}_3^j(\tilde{x}')  \rangle = \int_0^L \dt x \int_0^x \dt x' \int_{0}^{x'} \dt \tilde{x} \int_0^{\tilde{x}} \dt \tilde{x}'~ e^{- \mathbf{A} \tilde{x}} e^{ \mathbf{A} \tilde{x}'} e^{\mathbf{\Gamma} \tilde{x}' }   \\ \nonumber
		&= \int_0^L \dt x \int_0^x \dt x' \int_{0}^{x'} \dt \tilde{x}  \overbrace{\left[e^{\mathbf{\Gamma} \tilde{x}} - e^{-\mathbf{A} \tilde{x}} \right]} \\ \nonumber
		&=\int_0^L \dt x \int_0^x \dt x' \overbrace{\left[\mathbf{F}(x') + \mathbf{A}^{-1} \left(e^{-\mathbf{A} x'}-1\right) \right]} \\ \nonumber
		&=\int_0^L \dt x  \overbrace{\left[\mathbf{D}(x) - \mathbf{A}^{-1}x + \mathbf{A}^{-2} \left(1- e^{-\mathbf{A} x}\right) \right]} \\ \nonumber
		&= \overbrace{\left[\mathbf{J}(L) - \frac{1}{2}\mathbf{A}^{-1}L^2 + \mathbf{A}^{-2} L- \mathbf{A}^{-3} \left(1-  e^{-\mathbf{A} L}\right) \right]}
		\end{align}
	\end{subequations}
	Thus we find that 
	
	\begin{subequations}
		\begin{align}
		\langle r_0^i r_0^j \rangle =& \frac{2}{L^2} \overbrace{\left[2 \left( \mathbf{J}(L) - \frac{1}{2}\mathbf{A}^{-1}L^2 + \mathbf{A}^{-2} L- \mathbf{A}^{-3} \left(1-  e^{-\mathbf{A} L}\right) \right)\right]} +  \\ \nonumber
		& \frac{2 \mathbf{A}^{-1}}{L^2} \overbrace{\left[ \mathbf{D}(L) -\mathbf{A}^{-1}L - \overbrace{\mathbf{F}(L)} -\mathbf{A}^{-1} e^{-\mathbf{A} L}L  +\mathbf{A}^{-1}\left(1-e^{-\mathbf{A} L}\right) \left(\overbrace{1}+2A^{-1} \right)   \right]} \\ \nonumber
		=& \frac{2}{L^2} \overbrace{\left[2  \mathbf{J}(L) + \mathbf{A}^{-1}\left( \mathbf{D}(L) - L^2\right)  \right]} +  \\ \nonumber
		& \frac{2 \mathbf{A}^{-1}}{L^2} \overbrace{\left[ -\overbrace{\mathbf{F}(L)} +  \mathbf{A}^{-1} \left(1- e^{-\mathbf{A} L}\right) L 
			+\mathbf{A}^{-1}\left(1-e^{-\mathbf{A} L}\right) \overbrace{1} \right]}
		\end{align}
	\end{subequations}

	Finally
	\begin{subequations}
		\begin{align}
		\frac{1}{L}\int_0^L \langle r^i(x)r^j(x)\rangle \dt x &=    \frac{2}{L}\int_0^L \dt x  \int_0^x \dt \tilde{x} \int_0^{\tilde{x}} \dt \tilde{x}' ~\langle \hat{v}_3^i(\tilde{x}) \hat{v}_3^j(\tilde{x}')\rangle  = \frac{2}{L}\int_0^L \dt x  \int_0^x \dt \tilde{x} \int_0^{\tilde{x}} \dt \tilde{x}' e^{-\mathbf{A} \tilde{x}} e^{\mathbf{A} \tilde{x}'}e^{\mathbf{\Gamma}\tilde{x}'} \\ \nonumber
		&= \frac{2}{L}\int_0^L \dt x  \int_0^x \dt \tilde{x} \overbrace{\left[e^{\mathbf{\Gamma} \tilde{x}} - e^{-\mathbf{A} \tilde{x}}\right]} \\ \nonumber
		&=\frac{2}{L}\int_0^L \dt x   \overbrace{\left[\mathbf{F}(x) + \mathbf{A}^{-1}\left( e^{-\mathbf{A} x}-1 \right)\right]} \\ \nonumber
		&= \frac{2}{L} \overbrace{\left[\mathbf{D}(L) - \mathbf{A}^{-1}L +\mathbf{A}^{-2}\left(1- e^{-\mathbf{A} L} \right)\right]}
		\end{align}
	\end{subequations}
	We therefore conclude
	
	\begin{subequations}
		\begin{align}
		R_{ij} &=  \frac{2}{L}{ \overbrace{\left[ \mathbf{D}(L) L -2 \mathbf{J}(L)-\mathbf{A}^{-1} \mathbf{D}(L) + \mathbf{A}^{-1} \overbrace{\mathbf{F}(L)}-\mathbf{A}^{-2}\left(1-e^{\mathbf{A} L}\right)\overbrace{1}  \right]}}_{3i3j}
		\end{align}
	\end{subequations}
	
	\section{Simple Gaussian approximation}\label{app:Specific_Calcs}
	The euqations describing the thermal averages within the simple gaussian approximation around mechanical equilibrium are:
	\begin{subequations}  \label{eq:fluctuations}
		\begin{align}
		\tilde{h}_{eq}-\tilde{h}_0 &= \left\{ \begin{array}{cc}
		0 & \tilde{w} \leq 1\\				
		\frac{\tilde{w} \left(1-\nu\right)\left[1+\text{Erf}(\sqrt{2 \Psi g \, \tilde{z}_{eq}^2})\right]}{2\left(1+\nu\right)^3\left(\sqrt{\frac{2 \Psi}{\pi g}}e^{-2 \Psi g \, \tilde{z}_{eq}^2}\right)+2 \Psi \tilde{z}_{eq} \left[1+\text{Erf}(\sqrt{2 \Psi g \, \tilde{z}_{eq}^2})\right]} & \tilde{w} > 1\\
		\end{array} \right.\\
		\langle  \left(\tilde{h}- \tilde{h}_{0}\right)^2 \rangle&= \left\{ \begin{array}{cc}
		\frac{1}{2 \tilde{w} \Psi \left((1+\nu)^3 + 12 (1-\nu) \tilde{w}^4 \tilde{h}_{eq}^2\right)}& \tilde{w} \leq 1\\				
		\sqrt{\frac{(1-\nu ) \tilde{z}_{eq}}{\Psi \left(1+(1-\nu ) \tilde{w}^4\right)}} \frac{(1-\nu )+2 (1-\nu )^2 \tilde{w}^4 +\sqrt{2 \pi } (\nu +1)^4 \tilde{w} \tilde{z}_{eq} \sqrt{\frac{\Psi}{g}} e^{2 g \Psi \tilde{z}_{eq}^2} \left[1+\text{Erf}\left(\sqrt{2 \Psi g \, \tilde{z}_{eq}^2}\right)\right]}{4 (\nu +1)^4 \tilde{w}^{3/2} \left(\sqrt{\frac{\Psi \tilde{z}_{eq}}{g}}+ \sqrt{2 \pi } \Psi \tilde{z}_{eq}^{3/2} e^{2 g \Psi \tilde{z}_{eq}^2} \left[\text{Erf}\left(\sqrt{2 \Psi g \, \tilde{z}_{eq}^2}\right)+1\right]\right)} & \tilde{w} > 1\\
		\end{array} \right.  \\
		\tilde{z}_{eq}-\tilde{z}_0 &= \left\{ \begin{array}{cc}
		\frac{\sqrt{\pi }}{2 \sqrt{(1-\nu) \Psi \tilde{w}  \left((1+\nu)^2-4 \tilde{w}^4 \tilde{h}_{eq}^2\right)}}& \tilde{w} \leq 1\\				
		\frac{\sqrt{\pi } (1+\nu) \left[1+\text{Erf}\left(\sqrt{2 g \Psi \tilde{z}_{eq}^2}\right)\right]}{4 \tilde{w} \sqrt{g (1-\nu ) \Psi \tilde{z}_{eq}} \left(e^{-2 g \Psi \tilde{z}_{eq}^2}+ \sqrt{\pi } \sqrt{2 g \Psi \tilde{z}_{eq}^2} \left[1+\text{Erf}\left(\sqrt{2 g \Psi \tilde{z}_{eq}^2}\right)\right]\right) } & \tilde{w} > 1\\
		\end{array} \right.\\			
		\langle \left(\tilde{z}-\tilde{z}_{0}\right)^2 \rangle&= \left\{ \begin{array}{cc}
		\frac{1}{(1-\nu ) \tilde{w} \Psi \left((1+ \nu )^2-4 \tilde{w}^4 \tilde{h}_{eq}	^2\right)} & \tilde{w} \leq 1\\				
		\frac{(1-\nu )^2 \left(\sqrt{2} (1-\nu ) \Psi \sqrt{g \tilde{z}_{eq}} e^{-2 g \Psi \tilde{z}_{eq}^2}+\sqrt{\pi } \tilde{w} (\Psi \tilde{z}_{eq})^{3/2} \left((1+\nu)^4- (1-\nu )^2 g \tilde{w}^3\right) \left[1+\text{Erf}\left(\sqrt{2 g \Psi \tilde{z}_{eq}^2}\right)\right]\right)}{ \sqrt{\Psi} (1+\nu)^2 \tilde{w}^2 \left(2 \Psi g \tilde{z}_{eq}\right)^{3/2}  \left(e^{-2 g \Psi \tilde{z}_{eq}^2}+\sqrt{\pi } \sqrt{2 g \Psi \tilde{z}_{eq}^2} \left[1 +\text{Erf}\left(\sqrt{2 g \Psi \tilde{z}_{eq}^2}\right)\right]\right)} & \tilde{w} > 1\\
		\end{array} \right. 			
		\end{align}
	\end{subequations}
	Where $\Psi= \frac{E}{k_B T}=\frac{5^{1/4}}{ 3^{5/4}\sqrt{2}\left(1-\nu\right)^{3/4}(1+\nu)^{7/2}}	 \frac{Y t^{7/2} k_0^{1/2} }{k_B T}$, $g= \frac{(1+	\nu)^4 \tilde{w}}{(1-\nu ) \left[1+(1-\nu ) \tilde{w}^4\right]}$, $-1<\text{Erf}(x)<1$ is the error function. These results are depicted in Fig \ref{fig:moments}. Note that the equilibrium values diverge near the critical width, suggesting that this approximation fails close enough to the transition. Indeed one may approximate the equilibrium values using a saddle-point approximation. This is done in appendix \ref{app:Saddle_Point}. Nevertheless, the results here are valid at low temperatures, where these differences are not important. Furthermore, the more exact treatment, only changes the result quantitatively and not qualitatively.
	\section{Boundary Layer}\label{app:Bound_Layer}
	The boundary layer energy  for an arbitrary width $\tilde{w}>1$ and  arbitrary Poisson's  ratio is 
	\begin{align}
	H_{bound}&\propto -Y t^{7/2}k_0^{1/2} \int \frac{(1-\nu)^2 (1+\nu)^{3/2} \left(\left(\tilde{w}^4-1\right) \cos (2 \theta )+3 \tilde{w}^4+4 \sqrt{\tilde{w}^4-1} \tilde{w}^2 \cos (\theta )-1\right)}{4 \sqrt{2} \tilde{w}^3 \sqrt{\sqrt{\tilde{w}^4-1} \cos (\theta )+\tilde{w}^2}} \dt \tilde{x}
	\end{align}
	\section{Small Fluctuations Averaging of $\theta(q=0)$}\label{app:Theta_avg}
	The derivative term in the Hamiltonian  us given by
	\begin{align}
	H_{Der}&\propto 4 (1-\nu) \left(\Delta h'\right)^2+  \left(3 -2 \nu +  (1-2 \nu ) \cos (2 \theta_0 )\right) \left(\Delta z'\right)^2 +z_{eq}^2 \left(3-2 \nu-( 1-2 \nu) \cos (2 \theta_0 ) \right) \left(\Delta \theta '\right)^2 \\ \nonumber 
	& - 2 z_{eq} (1-2 \nu) \sin (2 \theta _0) \Delta \theta ' \Delta z' -8 (1 -\nu ) \Delta h' \left(\cos (\theta_0 ) \Delta z'-z_{eq}  \sin (\theta_0 ) \Delta \theta ' \right)
	\end{align}
	where $\theta_0$ is the uniform ($q\rightarrow0$) value. Using Fourier space we can now calculate the different averages in the main text. E.g-

	\begin{align*}
	\avg{l}&= \avg{h+\Delta h} + \avg{(z+\Delta z)\Re e^{(i\theta_0 +i \Delta \theta)}}= h +z\avg{\Re e^{(i \theta_0+ i \Delta\theta)}}+ \avg{\Delta z \Re e^{(i \theta_0+ i \Delta\theta)}}
	\end{align*}
	Since 
	\begin{align*}
	\avg{\Re e^{(i \theta_0+ i \Delta\theta)}} &= \Re \avg {e^{i \theta_0}  \avg {e^{i \Delta\theta}}_q}_0 =  \Re \avg {e^{i \theta_0}  e^{-\avg { \Delta\theta^2}_q}}_0
	\end{align*}
	where the average over $\Delta \theta$ obviously depends on $\theta_{0}$. However
	\begin{align*}
	\langle \Delta \theta ^2 \rangle_q=
	\int \frac{\dt q}{2\pi} \frac{1}{q^2(a+b\cos(2\theta_0))} 
	\end{align*}
	which formally diverges hence (taking the limit)  the above average vanishes.
	
	The second term is the given by
	\begin{align*}
	\avg{ \Delta z \Re e^{(i \theta_0+ i \Delta\theta)}} &= \Re \avg {e^{i \theta_0}  \avg {\Delta z e^{i \Delta\theta}}_q}_0 \simeq  \Re \avg {e^{i \theta_0}  \avg {\Delta z+ i \Delta z \Delta\theta - \Delta z\Delta\theta^2/2}_q}_0  = \Re \avg {e^{i \theta_0}  \avg {i \Delta z \Delta\theta}_q}_0 \\ &= \Re \avg {e^{i \theta_0} i \int \frac{\dt q}{2 \pi} \frac{1}{q^2 a \sin(2\theta_0)}}_0
	\end{align*}
	the last term is proportional to $\avg{\frac{\sin(\theta_0)}{\sin(2\theta_0)}}_0$ which is ill-defined but may be shown to be $0$ at the proper limit.
	
	The second moment is then given by
	\begin{align*}
	\avg {\Delta l_1\Delta l_2} &= \avg{\left(\Delta h_1 +(z+\Delta z_1)\cos(\theta_0+\Delta \theta_1)\right)\left(\Delta h_2 +(z+\Delta z_2)\cos(\theta_0+\Delta \theta_2)\right)}  \\ 
	&= \avg{\Delta h_1 \Delta h_2} +z^2\avg{\cos(\theta_0+\Delta \theta_1)
		\cos(\theta_0+\Delta \theta_2)} + \avg{\Delta z_1 \Delta z_2 \cos(\theta_0+\Delta \theta_1)\cos(\theta_0+\Delta \theta_2)} \\ & + z\avg{\Delta z_1  \cos(\theta_0+\Delta \theta_1)\cos(\theta_0+\Delta \theta_2)} + \avg{\Delta h_1 \Delta z_2 \cos(\theta_0+\Delta \theta_2)}+ z \avg{\Delta h_1 \cos(\theta_0+\Delta \theta_1)\cos(\theta_0+\Delta \theta_2)} 
	\end{align*}
	where we omitted terms containing the intercahnging of $1$ and $2$.
	using
	\begin{align*}
	\cos(\theta_0+\Delta\theta_1)\cos(\theta_0+\Delta\theta_2)&= \frac{1}{2} \left(\cos(2\theta_0 + \Delta \theta_1 + \Delta\theta_2) + \cos(\Delta \theta_1 - \Delta\theta_2)\right)
	\end{align*}
	We now note that 
	\begin{align*}
	\avg{\cos(\theta_0+\Delta \theta_1)	\cos(\theta_0+\Delta \theta_2)} &=  \frac{1}{2}\avg{\cos(2\theta_0 + \Delta \theta_1 + \Delta\theta_2) + \cos(\Delta \theta_1 - \Delta\theta_2)}= \frac{1}{2}\Re \avg{e^{2i\theta_0}e^{i(\Delta \theta_1 + \Delta\theta_2)} + e^{i(\Delta \theta_1 - \Delta\theta_2)}  } \\&= \frac{1}{2} e^{-\avg{(\Delta \theta_1-\Delta \theta_2)^2}/2}
	\end{align*}
	Also, marking $\alpha_+=2$ ,$\alpha_-=0$ we may calculate-
	\begin{align*}
	\avg{\Delta z_1 \Delta z_2 \cos(\alpha_{\pm}\theta_0+\Delta\theta_1 \pm \Delta \theta_2)} &=\Re\avg{e^{\alpha_{\pm} i\theta_{0}}\avg{\Delta z_1 \Delta z_2 e^{i(\Delta \theta_1 \pm\Delta \theta_2)}}_q}_0\\&= \Re\avg{e^{\alpha_{\pm}i\theta_{0}} \left(\avg{\Delta z_1 \Delta z_2}_q\avg{e^{i(\Delta \theta_1 \pm\Delta \theta_2)}}_q + \avg{\Delta z i\Delta \theta}_q\avg{\Delta z i \Delta \theta}_q +...\right)}_0 \\ &=
	\Re \langle \frac{1}{2} e^{-\langle (\Delta\theta_1 -\Delta \theta_2)^2 \rangle_q/2} \langle \Delta z_1\Delta z_2 \rangle_q \rangle_0
	\end{align*}
	This term, after averaging over $\theta_0$, differs only slightly from those calculated using the "naive" approximation in the main text.
	
	Similarly, terms like $\avg{\Delta z \cos(\alpha_{\pm}\theta_0 + \Delta\theta_1 \pm \Delta\theta_2)}= \avg{\Delta h \cos(\alpha_{\pm}\theta_0 + \Delta\theta_1 \pm \Delta\theta_2)} =0$. The last terms we need is then
	\begin{align}
	\avg{\Delta h_1 \Delta z_2 \cos(\theta_0+\Delta \theta_2)} &= \Re \avg{\Delta h_1 \Delta z_2 e^{i\theta_0 + i\Delta \theta_1}}= \Re \avg{e^{i\theta_0}\avg{\Delta h_1 \Delta z_2 e^{i \Delta \theta_1}}_q}_0 \\ &= Re \avg{e^{i\theta_0}\avg{\Delta h_1 \Delta z_2}_q \avg{e^{i \Delta \theta_1}}_q}_0 =0 
	\end{align}
	other terms are obviously zero. For  a specific realization of $\theta_0$ we do not need to average of course.  To the required order and accuracy, such results may as well be evaluated with a Hamiltonian of the form \ref{eq:second_order_fourier}.
	\section{Saddle Point Approximation}\label{app:Saddle_Point}
	
	One way to describe the statistical nature of elastic ribbons is to calculate the fluctuations in their curvatures. These, being our order parameters, are a natural choice. Calculation is done in a similar manner to the one described in \cite{Grossman2016}, we limit ourselves to  the Gaussian approximation. We start from the partition function $\mathcal{Z}= \int e^{-\beta H\left[z(x),h(x)\right]}\prod_x  z(x)\dt z(x) \dt h(x) \dt \theta$, where we note the non-trivial measure.  We define the free energy $F$ and free enrgy density $\mathcal{F}$
	\begin{align}
	-\Psi F=-\Psi \int  \mathcal{F}(\tilde{z},\tilde{h},\theta) \dt \tilde{x}&=  -\Psi \int \mathcal{H}(\tilde{z},\tilde{h},\theta) \dt \tilde{x} + \int \ln( {\tilde{z}}) \dt \tilde{x}. \\ \nonumber
	&\Downarrow \\ \nonumber
	\mathcal{F} &= \mathcal{H} - \frac{1}{\Psi} \ln(\tilde{z}).
	\end{align}.
	Where $\Psi= \frac{5^{1/4}}{ 3^{5/4}\sqrt{2}\left(1-\nu\right)^{3/4}(1+\nu)^{7/2}}	 \frac{Y t^{7/2} k_0^{1/2} }{k_B T}$. And we may write  $\mathcal{Z}= \int e^{-\Psi F \left[z(x),h(x)\right] \dt \tilde{x}}\prod_x  \dt z(x) \dt h(x) \dt \theta$
	
	It is clear that the  minimum of $\mathcal{F}$, unless $T=0$ differ from those in Eq. \ref{eq:equi}.  We use saddle point approximation (SPA) to solve for the thermal equilibrium  values (marked $\tilde{h}_{eq}$ and $\tilde{z}_{eq}$). These are given

	We then expand $F$ to second order about the equilibrium values
	\begin{align}
	F \simeq F_{eq} + F_{(2)} = F_{eq} +\int \mathcal{F}_{(2)} \dt x
	\end{align}
	
	In principle, $F_{(2)} =F_{(2)}\left[z(x),h(x),\partial_x z,\partial_x h,\partial_x \theta\right]$. However, as mentioned earlier, we will assume for simplicity that we may omit the derivatives (as we've shown in the past that these do not change the results significantly, and their main effect in this case it they give rise to finite correlation lengths). Thus 
	\begin{align}\label{eq:second_order_no_der}
	\mathcal{F}_{(2)} &= \mathcal{H}_{(2)} + \frac{1}{2\Psi }\left(\frac{\Delta \tilde{z}}{\tilde{z} _{eq}}\right)^2 = \frac{1}{2} \left[\left(8 \tilde{w}^5\left(1-\nu\right)\left(3 \tilde{h}^2_{eq} -\tilde{z}^2_{eq}\right)+2\tilde{w} \left(1+\nu\right)^3 \right)\Delta \tilde{h}^2  \right. \\ \nonumber 
	&\left. +\left(8 \tilde{w}^5\left(1-\nu\right)\left(3 \tilde{z}^2_{eq} -\tilde{h}^2_{eq}\right)+2\tilde{w} \left(1-\nu\right)\left(1+\nu\right)^2 + \frac{1}{\Psi \tilde{z}_{eq}^2}\right)\Delta \tilde{z}^2 - 32 \tilde{w}^5 \left(1-\nu\right)\tilde{h}_{eq} \tilde{z}_{eq}\Delta \tilde{z} \Delta \tilde{h}\right]. \\ \nonumber
	\end{align}

	The average of a quantity Q is given by the functional integral
	\begin{align}
	\langle Q \rangle = \frac{1}{\mathcal{Z}}\int Q e^{-\beta F_{(2)}\left[z(x),h(x)\right]}  \prod_x  \dt z(x) \dt h(x) \dt \theta(x). 
	\end{align}
	
	Where we redefined the partition function  $\mathcal{Z}= \int e^{-\beta F_{2}\left[z(x),h(x)\right]}\prod_x \dt z(x) \dt h(x) \dt \theta$.  At the given approximation, integration over the angle is trivial. Omitting derivatives means that there are no correlation at different positions. i.e-
	$ \langle Q(\tilde{x})Q(\tilde{x}')\rangle = \delta(\tilde{x}-\tilde{x}')\langle Q^2 \rangle$. Hence, the averages and fluctuations of the curvatures are given by.
	
	Hence, the averages and fluctuations of the curvatures are given by.
	\begin{subequations} \label{eq:avgs_freeEner}
		\begin{align}
		\langle \tilde{l} \rangle& = \langle \tilde{h} \rangle+ \langle \tilde{z}\rangle \langle \cos \theta \rangle = \langle \tilde{h} \rangle =\tilde{h}_{eq}=\langle \tilde{n} \rangle \\
		\langle \tilde{m} \rangle &=0 \\
		\langle  \Delta\tilde{l}(\tilde{x})  \Delta \tilde{l}(\tilde{x}')\rangle &=\langle \tilde{l}(\tilde{x}) \tilde{l}(\tilde{x}') \rangle - \langle \tilde{l} \rangle^2 =  \delta (\tilde{x}-\tilde{x}') \langle \Delta \tilde{l}^2 \rangle\\ \nonumber
		\langle \Delta \tilde{l}^2 \rangle &= \langle\Delta \tilde{h}^2 \rangle + \frac{1}{2} \left(  \langle\Delta \tilde{z}^2 \rangle \right) = \langle \Delta \tilde{n}^2 \rangle\\
		\langle \Delta \tilde{m}^2 \rangle &=  \frac{1}{2}   \langle\Delta \tilde{z}^2 \rangle \\	\nonumber
		\langle \Delta \tilde{h}^2\rangle&= \langle \left(\tilde{h}- \tilde{h}_{eq} \right)^2 \rangle \\ \nonumber
		\langle \Delta \tilde{z}^2\rangle&= \langle \left(\tilde{z}- \tilde{z}_{eq} \right)^2 \rangle 
		\end{align}	
	\end{subequations}
	
	\begin{subequations} \label{eq:fluctuations_freeEner}
		\begin{align}
		\langle\Delta \tilde{h}^2\rangle &=\frac{2 \left(8 (1-\nu) \tilde{w}^5 \left(3 \tilde{z}_{eq}^2-\tilde{h}_{eq}^2\right)+2 (1-\nu) (1+\nu)^2 \tilde{w}+\frac{1}{\Psi  \tilde{z}_{eq}^2}\right)}{\Psi \tilde{w} \Gamma } \\
		\langle \Delta \tilde{z}^2 \rangle &= \frac{4 \left(4 (1-\nu) \tilde{w}^4 \left(3 \tilde{h}_{eq}^2 - \tilde{z}_{eq}^2 \right)+(1+\nu)^3\right)}{\Psi \Gamma }
		\end{align}
		where
		{\scriptsize
			\begin{align*}
			\Gamma &= 2 \left(12 \tilde{h}_{eq}^2 (1-\nu) \tilde{w}^4+(1+\nu)^3-4 (1-\nu) \tilde{w}^4 \tilde{z}_{eq}^2\right) \left(8 (1-\nu) \tilde{w}^5 \left(3 \tilde{z}_{eq}^2-\tilde{h}_{eq}^2\right)+2 (1-\nu) (1+\nu)^2 \tilde{w}+\frac{1}{\Psi  \tilde{z}_{eq}^2}\right)-256(1-\nu)^2 \tilde{w}^9 \tilde{h}_{eq}^2 \tilde{z}_{eq}^2
			\end{align*}}
	\end{subequations}
	Where $\Psi= \frac{5^{1/4}}{ 3^{5/4}\sqrt{2}\left(1-\nu\right)^{3/4}(1+\nu)^{7/2}}	 \frac{Y t^{7/2} k_0^{1/2} }{k_B T}$, $g= \frac{(1+	\nu)^4 \tilde{w}}{(1-\nu ) \left[1+(1-\nu ) \tilde{w}^4\right]}$, $-1<\text{Erf}(x)<1$ is the error function. These results are depcited in Fig \ref{fig:moments}

	\begin{subequations}
		\begin{align}\label{eq:saddle_point}
		\tilde{h}_{eq}&=\frac{(1+\nu)}{(1-\nu ) \tilde{w}^4}
		\left(1-2 \tilde{z}_{eq} ^2 \tilde{w} \Psi  \left(5 \nu +(1-\nu ) \left(\nu ^2-4 \tilde{z}_{eq} ^2 (\nu +1)^2 \tilde{w}^5 \Psi  \left(4 \tilde{z}_{eq} ^2-(\nu +1)^2\right) \right.\right.\right. \\\nonumber&
		\left.\left.\left.+\tilde{w}^4 \left(4 \tilde{z}_{eq} ^2+(\nu +1)^2\right)+4 \tilde{z}_{eq} ^2 (\nu +1)^4 \tilde{w} \Psi \right)+3\right)\right)
		\end{align}
		\begin{align}
		\tilde{z}_{eq}&=\sqrt{\frac{2  \tilde{w}^3+ \Psi (1+ \nu)^2 \left(\tilde{w}^4 \left(8 \sqrt{\zeta}+(1 +\nu)^2+8 ~\text{sgn}(\tilde{w}-1)\; \sqrt{\phi} \right)-(1+ \nu)^2\right)}{16 \Psi  (1+\nu)^2 \tilde{w}^4}}
		\end{align}
		where
		{\small
			\begin{align*}
			\zeta&=\frac{1}{256} \left| \frac{\left(\tilde{w}^4-1\right)^3 (\nu -1) (\nu +1)^{12}+\frac{2 \tilde{w}^3 \left(\tilde{w}^4-1\right) \left(3 (\nu -1) \tilde{w}^4+\nu -5\right) (\nu +1)^8}{\Psi }+\frac{8 \tilde{w}^6 \left((\nu -1) \tilde{w}^4+2\right) (\nu +1)^4}{\Psi ^2}}{\tilde{w}^{12} (\nu -1) (\nu +1)^6 \sqrt{\phi }}\right| -\mu +2 \upsilon\\
			\phi&= \mu +\upsilon\\
			\upsilon&=\frac{\frac{4 (\nu -1) \tilde{w}^6}{\Psi ^2}+3 (\nu -1) (\nu +1)^8 \left(\tilde{w}^4-1\right)^2+\frac{4 (\nu +1)^4 \tilde{w}^3 \left(\nu +3 (\nu -1) \tilde{w}^4-5\right)}{\Psi }}{192 (\nu -1) (\nu +1)^4 \tilde{w}^8}\\
			\mu&=\frac{\sqrt[3]{2} \chi ^2+\frac{2 \tilde{w}^4 \left(\frac{(\nu -1)^2 \tilde{w}^6}{\Psi ^2}+(\nu +1)^8 \left((\nu +1)^2+3 \left(\nu ^2-6 \nu +5\right) \tilde{w}^4\right)+\frac{2 (\nu -1) (\nu +1)^5 \tilde{w}^3}{\Psi }\right)}{\Psi ^2}}{48\ 2^{2/3} (\nu -1) (\nu +1)^4 \tilde{w}^7 \chi }\\
			\chi&=\frac{1}{\Psi}\left\{\Psi ^3\eta +\tau\right\}^{1/3} \\
			\tau&=\tilde{w}^6 \left(\frac{2 (\nu -1)^3 \tilde{w}^9}{\Psi ^3}+\frac{6 (\nu -1)^2 (\nu +1)^5 \tilde{w}^6}{\Psi ^2}+2 (\nu +1)^{12} \left((\nu +1)^3+27 (\nu -1)^2 \tilde{w}^8-9 (\nu -1) ((\nu -1) \nu +4) \tilde{w}^4\right) \right. \\ &\left.+\frac{3 (\nu -1) (\nu +1)^8 \tilde{w}^3 \left(2 (\nu +1)^2+3 (\nu -1) (\nu +19) \tilde{w}^4\right)}{\Psi }\right)\\
			\eta&=\frac{1}{\Psi^3}\left\{\tau^2-4 \tilde{w}^{12}\left(\frac{(\nu -1)^2 \tilde{w}^6}{\Psi ^2}+(\nu +1)^8 \left((\nu +1)^2+3 (\nu -5) (\nu -1) \tilde{w}^4\right)+\frac{2 (\nu -1) (\nu +1)^5 w^3}{\Psi }\right)^3\right\}^{1/2}
			\end{align*}}
	\end{subequations}

	\begin{align}\label{eq:second_order_free}
	\mathcal{F}_{(2)} &= \mathcal{H}_{(2)} + \frac{1}{2\Psi }\left(\frac{\Delta \tilde{z}}{\tilde{z} _{eq}}\right)^2 = \frac{1}{2} \left[\left(8 \tilde{w}^5\left(1-\nu\right)\left(3 \tilde{h}^2_{eq} -\tilde{z}^2_{eq}\right)+2\tilde{w} \left(1+\nu\right)^3 \right)\Delta \tilde{h}^2  \right. \\ \nonumber 
	&\left. +\left(8 \tilde{w}^5\left(1-\nu\right)\left(3 \tilde{z}^2_{eq} -\tilde{h}^2_{eq}\right)+2\tilde{w} \left(1-\nu\right)\left(1+\nu\right)^2 + \frac{1}{\Psi \tilde{z}_{eq}^2}\right)\Delta \tilde{z}^2 - 32 \tilde{w}^5 \left(1-\nu\right)\tilde{h}_{eq} \tilde{z}_{eq}\Delta \tilde{z} \Delta \tilde{h}\right] \\ \nonumber
	& + 8 \sqrt{ \frac{5}{3} \left(1-\nu\right) } \frac{k_0 t}{1+\nu} \left[\frac{(1+\nu)^2\tilde{w}^3}{48} \left(4\left(1-\nu\right) \Delta\dot{\tilde{h}}^2+ \left(3-2\nu\right)\left(\tilde{z}_{eq}^2 \Delta \dot{\theta}^2 + \Delta\dot{\tilde{z}}^2\right)\right)\right]
	\end{align}

	\begin{figure}
		\begin{tabular}{cc}
			\begin{subfigure}{0.40\textwidth}
				\centering
				\includegraphics[width=\textwidth]{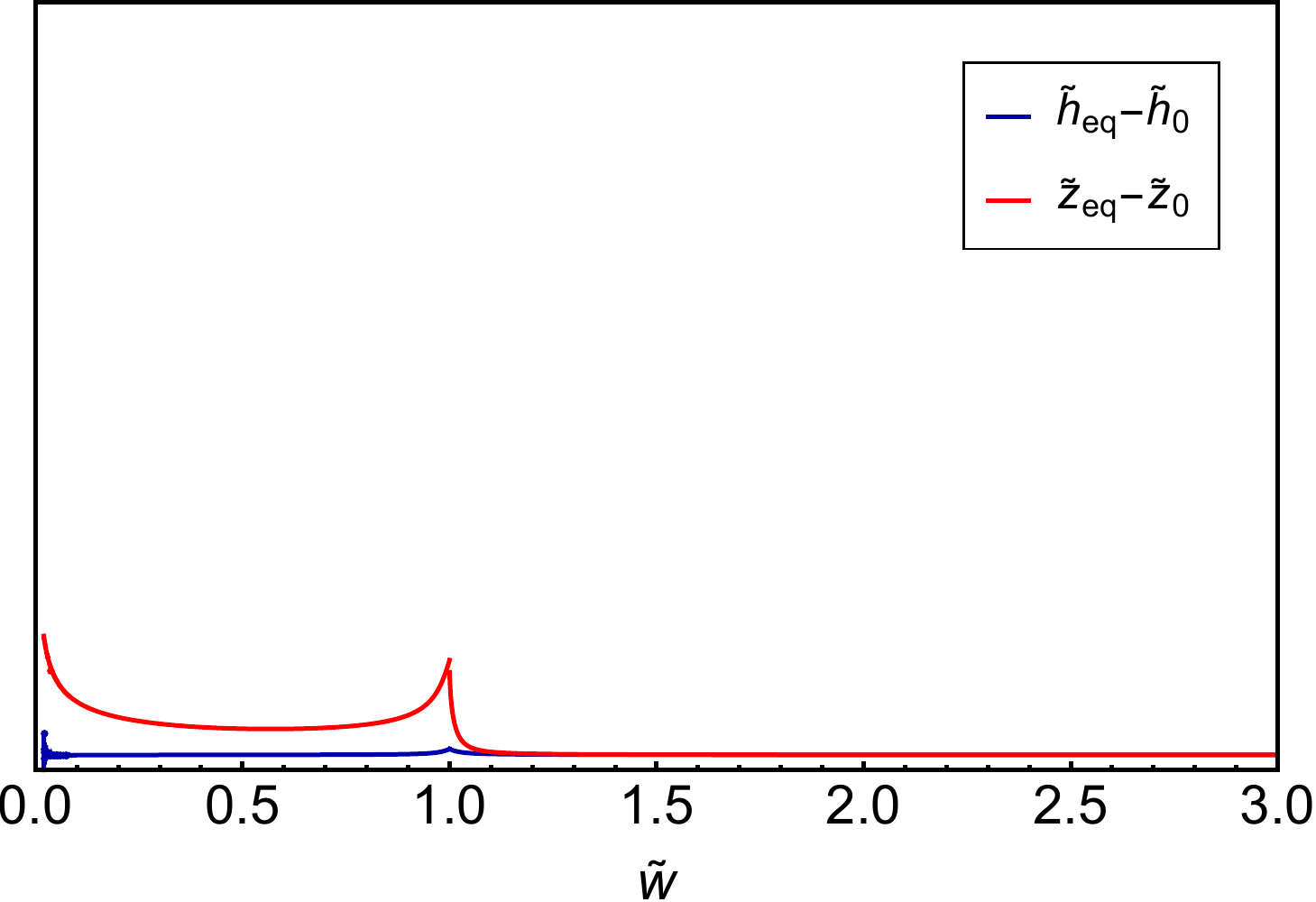}
				\caption{$\Psi=10^{-3}$}
			\end{subfigure} &
			\begin{subfigure}{0.40\textwidth}
				\centering
				\includegraphics[width=\textwidth]{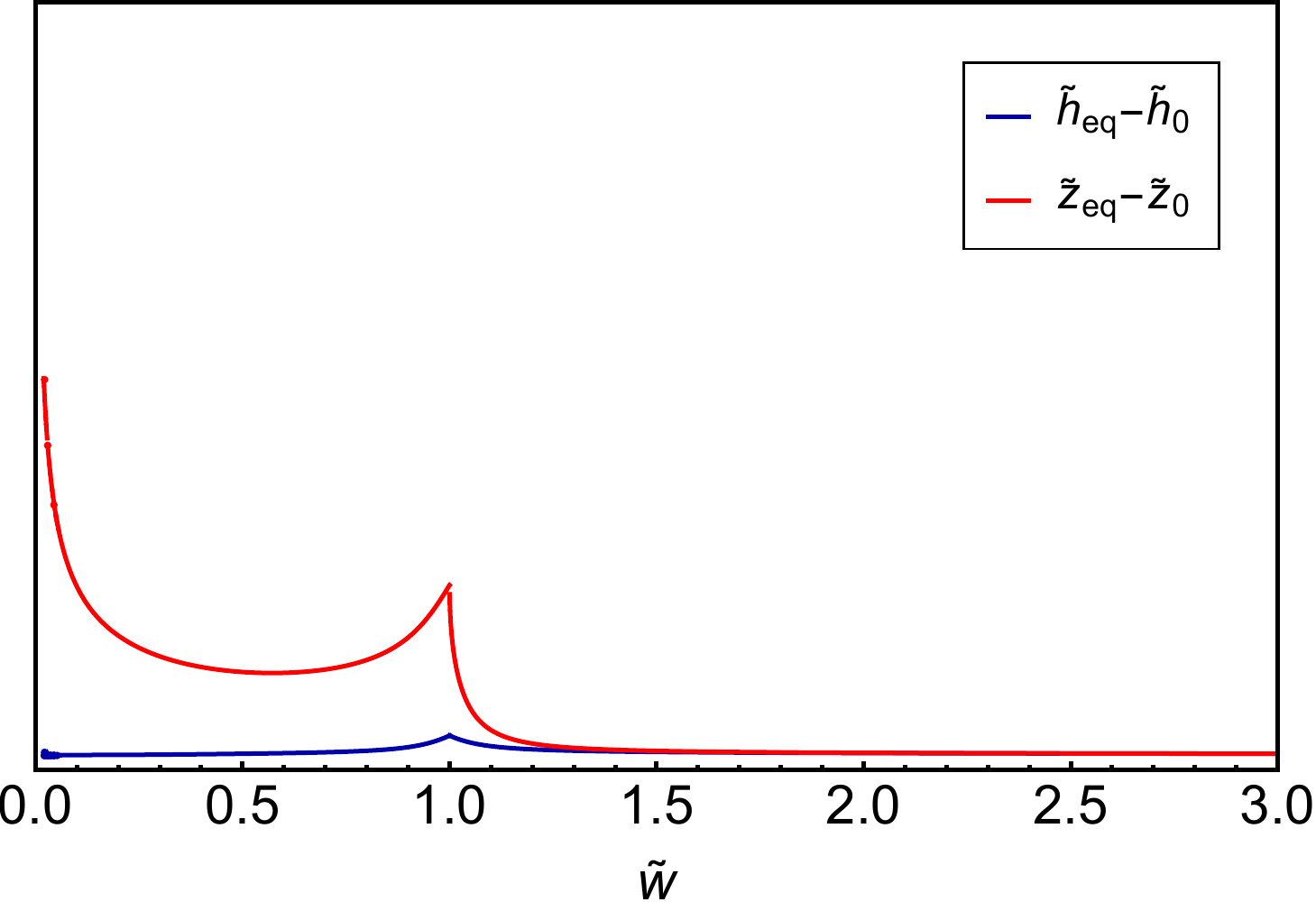}
				\caption{$\Psi=10^{-2}$}
			\end{subfigure} \\
			\begin{subfigure}{0.40\textwidth}
				\centering
				\includegraphics[width=\textwidth]{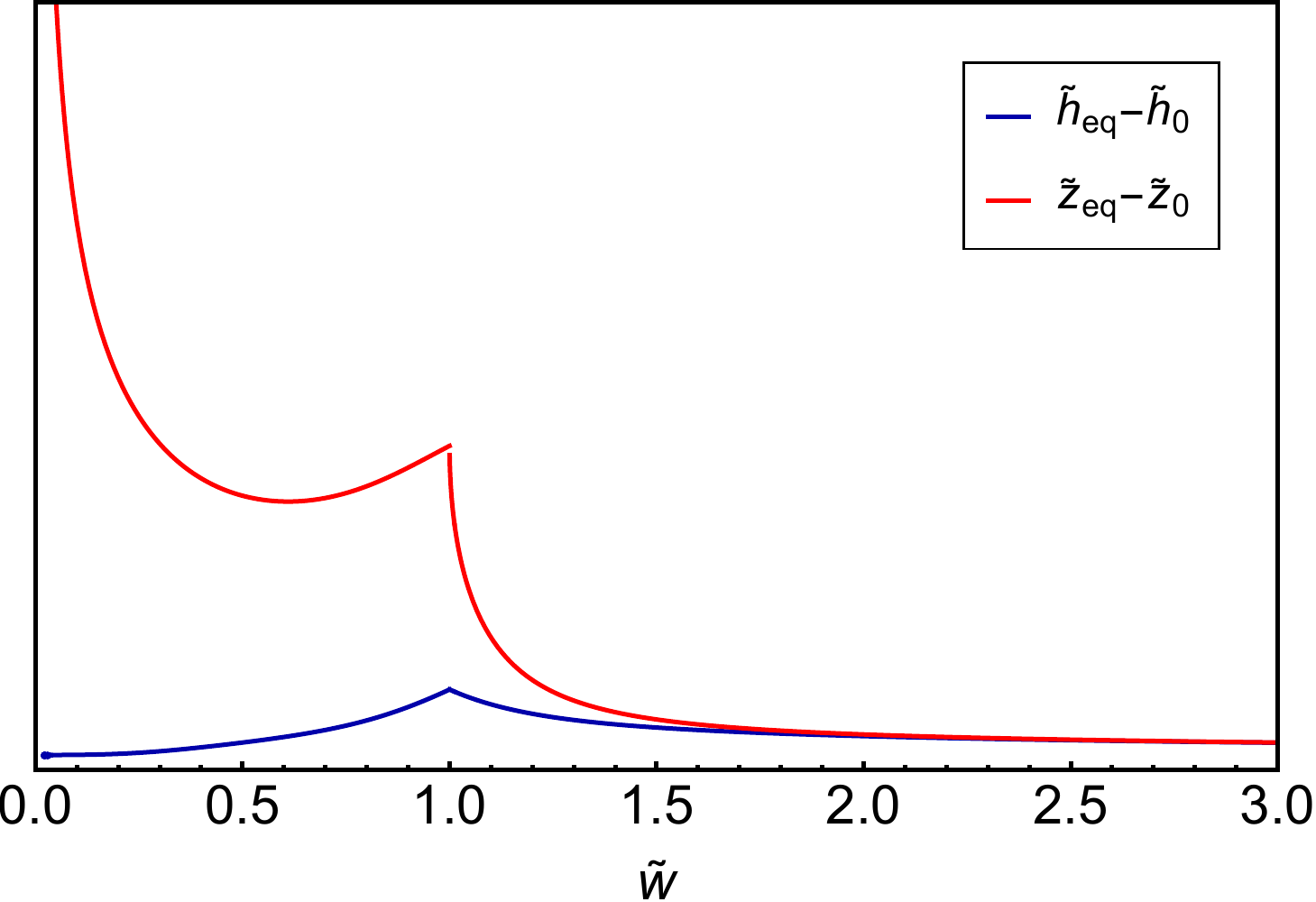}
				\caption{$\Psi=10^{-1}$}
			\end{subfigure}
			&
			\begin{subfigure}{0.40\textwidth}
				\centering
				\includegraphics[width=\textwidth]{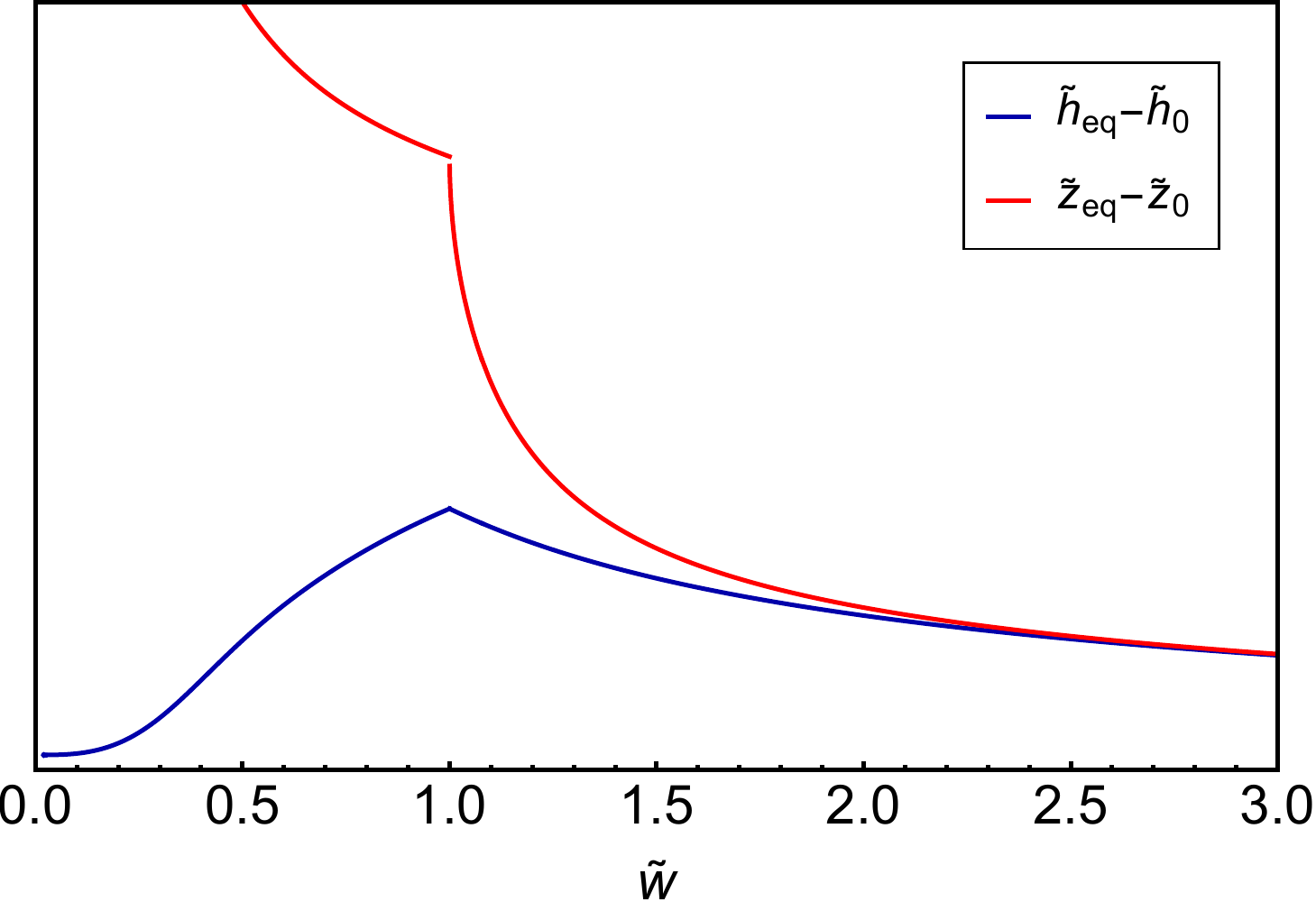}
				\caption{$\Psi=10^{0}$}
			\end{subfigure}\\
			\begin{subfigure}{0.40\textwidth}
				\centering
				\includegraphics[width=\textwidth]{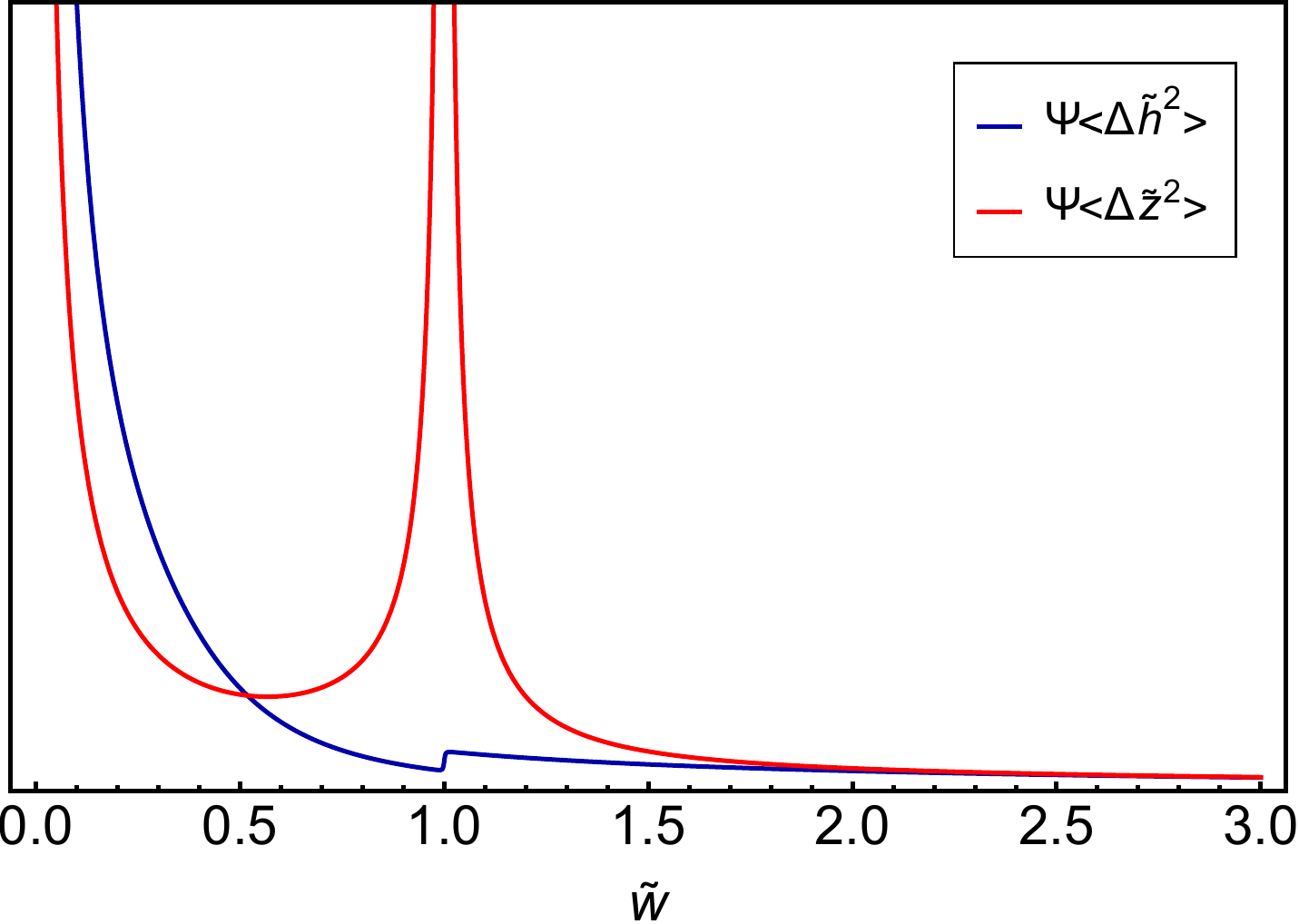}
				\caption{$\Psi=10^{-5}$}
			\end{subfigure} &
			\begin{subfigure}{0.40\textwidth}
				\centering
				\includegraphics[width=\textwidth]{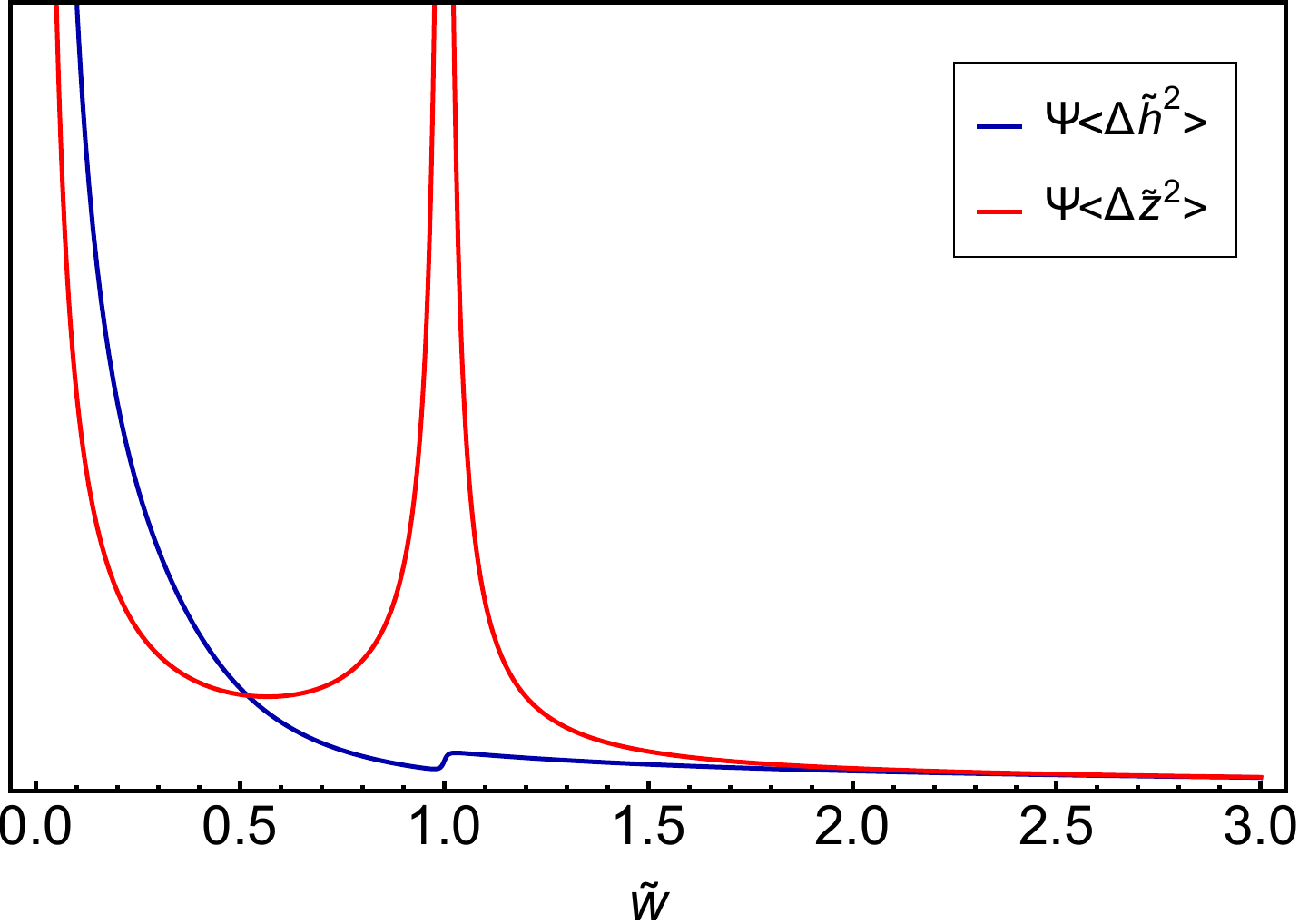}
				\caption{$\Psi=10^{-4}$}
			\end{subfigure} \\
			\begin{subfigure}{0.40\textwidth}
				\centering
				\includegraphics[width=\textwidth]{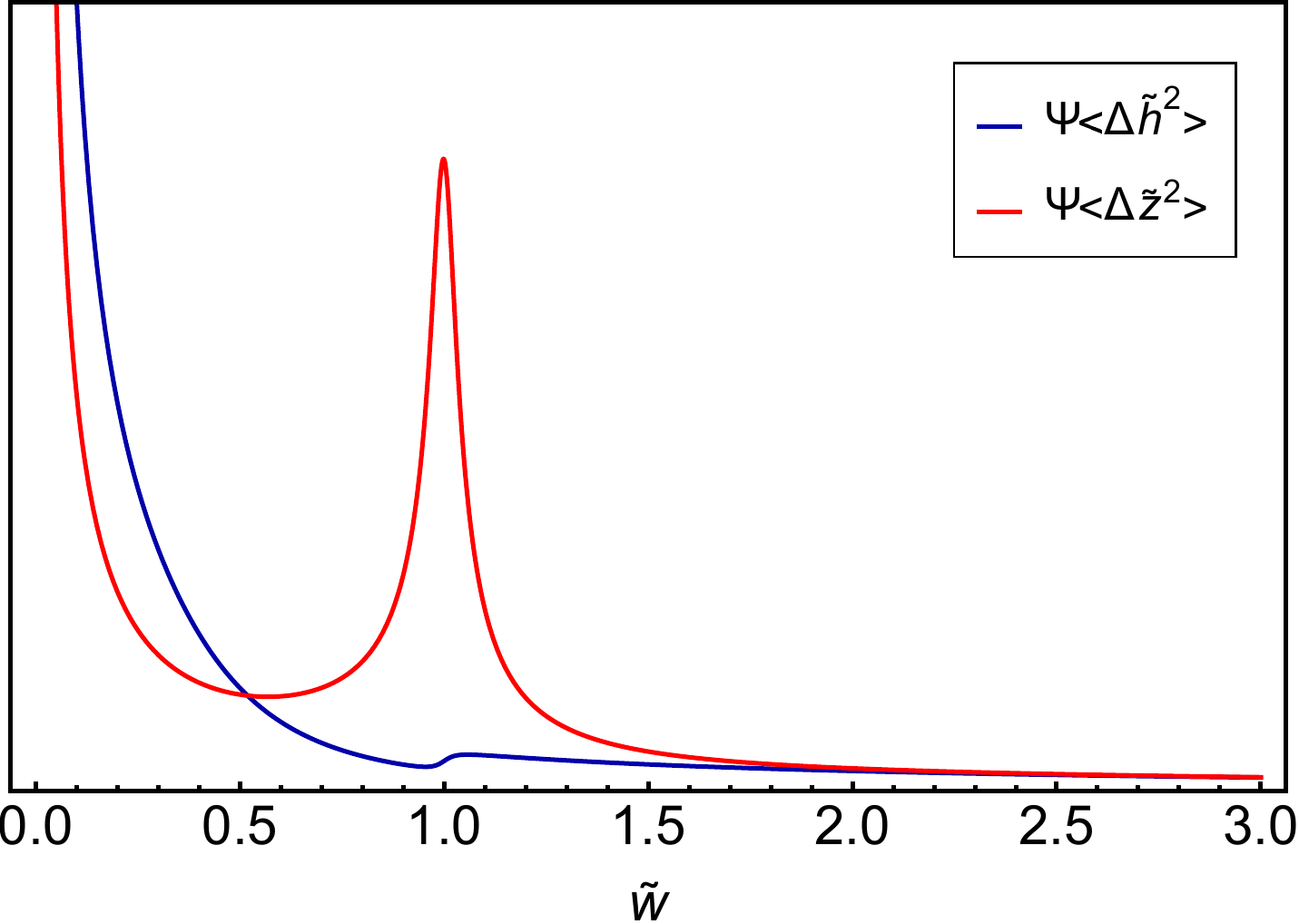}
				\caption{$\Psi=10^{-3}$}
			\end{subfigure}
			&
			\begin{subfigure}{0.40\textwidth}
				\centering
				\includegraphics[width=\textwidth]{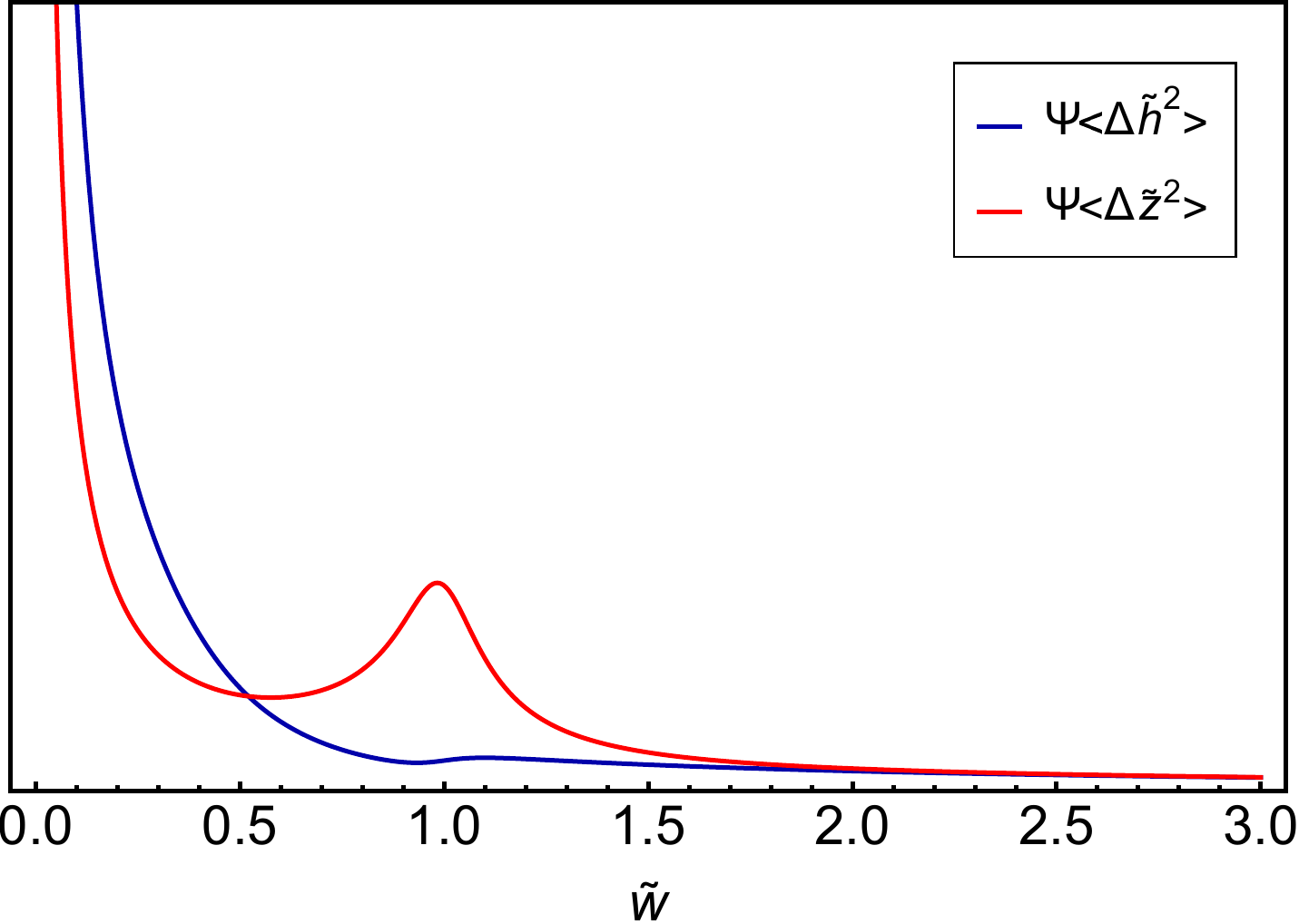}
				\caption{$\Psi=10^{-2}$}
			\end{subfigure}\\
		\end{tabular}
		\caption{}
	\end{figure}

	\clearpage

	\bibliography{C:/Users/Doron/Documents/PhD_work/Papers_And_Notes/Bibtex/short,C:/Users/Doron/Documents/PhD_work/Papers_And_Notes/Bibtex/elastic_stat,footnotes}

\begin{thebibliography}{29}%
\makeatletter
\providecommand \@ifxundefined [1]{%
 \@ifx{#1\undefined}
}%
\providecommand \@ifnum [1]{%
 \ifnum #1\expandafter \@firstoftwo
 \else \expandafter \@secondoftwo
 \fi
}%
\providecommand \@ifx [1]{%
 \ifx #1\expandafter \@firstoftwo
 \else \expandafter \@secondoftwo
 \fi
}%
\providecommand \natexlab [1]{#1}%
\providecommand \enquote  [1]{``#1''}%
\providecommand \bibnamefont  [1]{#1}%
\providecommand \bibfnamefont [1]{#1}%
\providecommand \citenamefont [1]{#1}%
\providecommand \href@noop [0]{\@secondoftwo}%
\providecommand \href [0]{\begingroup \@sanitize@url \@href}%
\providecommand \@href[1]{\@@startlink{#1}\@@href}%
\providecommand \@@href[1]{\endgroup#1\@@endlink}%
\providecommand \@sanitize@url [0]{\catcode `\\12\catcode `\$12\catcode
  `\&12\catcode `\#12\catcode `\^12\catcode `\_12\catcode `\%12\relax}%
\providecommand \@@startlink[1]{}%
\providecommand \@@endlink[0]{}%
\providecommand \url  [0]{\begingroup\@sanitize@url \@url }%
\providecommand \@url [1]{\endgroup\@href {#1}{\urlprefix }}%
\providecommand \urlprefix  [0]{URL }%
\providecommand \Eprint [0]{\href }%
\providecommand \doibase [0]{http://dx.doi.org/}%
\providecommand \selectlanguage [0]{\@gobble}%
\providecommand \bibinfo  [0]{\@secondoftwo}%
\providecommand \bibfield  [0]{\@secondoftwo}%
\providecommand \translation [1]{[#1]}%
\providecommand \BibitemOpen [0]{}%
\providecommand \bibitemStop [0]{}%
\providecommand \bibitemNoStop [0]{.\EOS\space}%
\providecommand \EOS [0]{\spacefactor3000\relax}%
\providecommand \BibitemShut  [1]{\csname bibitem#1\endcsname}%
\let\auto@bib@innerbib\@empty
\bibitem [{\citenamefont {Bouchiat}\ and\ \citenamefont
  {M\'ezard}(1998)}]{Bouchiat1998}%
  \BibitemOpen
  \bibfield  {author} {\bibinfo {author} {\bibfnamefont {C.}~\bibnamefont
  {Bouchiat}}\ and\ \bibinfo {author} {\bibfnamefont {M.}~\bibnamefont
  {M\'ezard}},\ }\href {\doibase 10.1103/PhysRevLett.80.1556} {\bibfield
  {journal} {\bibinfo  {journal} {Phys. Rev. Lett.}\ }\textbf {\bibinfo
  {volume} {80}},\ \bibinfo {pages} {1556} (\bibinfo {year}
  {1998})}\BibitemShut {NoStop}%
\bibitem [{\citenamefont {Adamcik}\ \emph {et~al.}(2011)\citenamefont
  {Adamcik}, \citenamefont {Castelletto}, \citenamefont {Bolisetty},
  \citenamefont {Hamley},\ and\ \citenamefont {Mezzenga}}]{Adamcik2011}%
  \BibitemOpen
  \bibfield  {author} {\bibinfo {author} {\bibfnamefont {J.}~\bibnamefont
  {Adamcik}}, \bibinfo {author} {\bibfnamefont {V.}~\bibnamefont
  {Castelletto}}, \bibinfo {author} {\bibfnamefont {S.}~\bibnamefont
  {Bolisetty}}, \bibinfo {author} {\bibfnamefont {I.~W.}\ \bibnamefont
  {Hamley}}, \ and\ \bibinfo {author} {\bibfnamefont {R.}~\bibnamefont
  {Mezzenga}},\ }\href {\doibase 10.1002/anie.201100807} {\bibfield  {journal}
  {\bibinfo  {journal} {Angewandte Chemie (International ed. in English)}\
  }\textbf {\bibinfo {volume} {50}},\ \bibinfo {pages} {5495} (\bibinfo {year}
  {2011})}\BibitemShut {NoStop}%
\bibitem [{\citenamefont {Smith}\ \emph {et~al.}(2001)\citenamefont {Smith},
  \citenamefont {Zastavker},\ and\ \citenamefont {Benedek}}]{Smith2001}%
  \BibitemOpen
  \bibfield  {author} {\bibinfo {author} {\bibfnamefont {B.}~\bibnamefont
  {Smith}}, \bibinfo {author} {\bibfnamefont {Y.}~\bibnamefont {Zastavker}}, \
  and\ \bibinfo {author} {\bibfnamefont {G.}~\bibnamefont {Benedek}},\ }\href
  {\doibase 10.1103/PhysRevLett.87.278101} {\bibfield  {journal} {\bibinfo
  {journal} {Phys. Rev. Lett.}\ }\textbf {\bibinfo {volume} {87}},\ \bibinfo
  {pages} {278101} (\bibinfo {year} {2001})}\BibitemShut {NoStop}%
\bibitem [{\citenamefont {Zhan}\ \emph {et~al.}(2005)\citenamefont {Zhan},
  \citenamefont {Gao},\ and\ \citenamefont {Liu}}]{Zhan2005}%
  \BibitemOpen
  \bibfield  {author} {\bibinfo {author} {\bibfnamefont {C.}~\bibnamefont
  {Zhan}}, \bibinfo {author} {\bibfnamefont {P.}~\bibnamefont {Gao}}, \ and\
  \bibinfo {author} {\bibfnamefont {M.}~\bibnamefont {Liu}},\ }\href@noop {}
  {\bibfield  {journal} {\bibinfo  {journal} {Chemical communications}\ ,\
  \bibinfo {pages} {462}} (\bibinfo {year} {2005})}\BibitemShut {NoStop}%
\bibitem [{\citenamefont {Oda}\ \emph {et~al.}(2008)\citenamefont {Oda},
  \citenamefont {Artzner}, \citenamefont {Laguerre}, \citenamefont {Huc},
  \citenamefont {Escarpit}, \citenamefont {Cedex},\ and\ \citenamefont
  {Rennes}}]{Oda2008}%
  \BibitemOpen
  \bibfield  {author} {\bibinfo {author} {\bibfnamefont {R.}~\bibnamefont
  {Oda}}, \bibinfo {author} {\bibfnamefont {F.}~\bibnamefont {Artzner}},
  \bibinfo {author} {\bibfnamefont {M.}~\bibnamefont {Laguerre}}, \bibinfo
  {author} {\bibfnamefont {I.}~\bibnamefont {Huc}}, \bibinfo {author}
  {\bibfnamefont {R.}~\bibnamefont {Escarpit}}, \bibinfo {author}
  {\bibfnamefont {F.-P.}\ \bibnamefont {Cedex}}, \ and\ \bibinfo {author}
  {\bibfnamefont {I.~D. P.~D.}\ \bibnamefont {Rennes}},\ }\href@noop {}
  {\bibfield  {journal} {\bibinfo  {journal} {JACS}\ ,\ \bibinfo {pages}
  {14705}} (\bibinfo {year} {2008})}\BibitemShut {NoStop}%
\bibitem [{\citenamefont {Ziserman}\ \emph {et~al.}(2011)\citenamefont
  {Ziserman}, \citenamefont {Mor}, \citenamefont {Harries},\ and\ \citenamefont
  {Danino}}]{Ziserman2011}%
  \BibitemOpen
  \bibfield  {author} {\bibinfo {author} {\bibfnamefont {L.}~\bibnamefont
  {Ziserman}}, \bibinfo {author} {\bibfnamefont {A.}~\bibnamefont {Mor}},
  \bibinfo {author} {\bibfnamefont {D.}~\bibnamefont {Harries}}, \ and\
  \bibinfo {author} {\bibfnamefont {D.}~\bibnamefont {Danino}},\ }\href
  {\doibase 10.1103/PhysRevLett.106.238105} {\bibfield  {journal} {\bibinfo
  {journal} {Phys. Rev. Lett.}\ }\textbf {\bibinfo {volume} {106}},\ \bibinfo
  {pages} {238105} (\bibinfo {year} {2011})}\BibitemShut {NoStop}%
\bibitem [{\citenamefont {Sadowsky}(1930)}]{Sadowsky1930}%
  \BibitemOpen
  \bibfield  {author} {\bibinfo {author} {\bibfnamefont {M.}~\bibnamefont
  {Sadowsky}},\ }\href@noop {} {\bibfield  {journal} {\bibinfo  {journal}
  {Sitzungsber. Preuss. Akad. Wiss.}\ }\textbf {\bibinfo {volume} {22}},\
  \bibinfo {pages} {412} (\bibinfo {year} {1930})}\BibitemShut {NoStop}%
\bibitem [{\citenamefont {Panyukov}\ and\ \citenamefont
  {Rabin}(2000{\natexlab{a}})}]{Panyukov2000a}%
  \BibitemOpen
  \bibfield  {author} {\bibinfo {author} {\bibfnamefont {S.}~\bibnamefont
  {Panyukov}}\ and\ \bibinfo {author} {\bibfnamefont {Y.}~\bibnamefont
  {Rabin}},\ }\href {\doibase 10.1103/PhysRevE.62.7135} {\bibfield  {journal}
  {\bibinfo  {journal} {Phys. Rev. E}\ }\textbf {\bibinfo {volume} {62}},\
  \bibinfo {pages} {7135} (\bibinfo {year} {2000}{\natexlab{a}})}\BibitemShut
  {NoStop}%
\bibitem [{\citenamefont {Giomi}\ and\ \citenamefont
  {Mahadevan}(2010)}]{Giomi2010}%
  \BibitemOpen
  \bibfield  {author} {\bibinfo {author} {\bibfnamefont {L.}~\bibnamefont
  {Giomi}}\ and\ \bibinfo {author} {\bibfnamefont {L.}~\bibnamefont
  {Mahadevan}},\ }\href {\doibase 10.1103/PhysRevLett.104.238104} {\bibfield
  {journal} {\bibinfo  {journal} {Phys. Rev. Lett.}\ }\textbf {\bibinfo
  {volume} {104}},\ \bibinfo {pages} {238104} (\bibinfo {year}
  {2010})}\BibitemShut {NoStop}%
\bibitem [{\citenamefont {Rubinstein}\ and\ \citenamefont
  {Colby}(2003)}]{Rubinstein2003}%
  \BibitemOpen
  \bibfield  {author} {\bibinfo {author} {\bibfnamefont {M.}~\bibnamefont
  {Rubinstein}}\ and\ \bibinfo {author} {\bibfnamefont {R.~H.}\ \bibnamefont
  {Colby}},\ }\href@noop {} {\emph {\bibinfo {title} {Polymer Physics}}}\
  (\bibinfo  {publisher} {Oxford University Press, Oxford, England},\ \bibinfo
  {year} {2003})\BibitemShut {NoStop}%
\bibitem [{\citenamefont {Armon}\ \emph {et~al.}(2011)\citenamefont {Armon},
  \citenamefont {Efrati}, \citenamefont {Kupferman},\ and\ \citenamefont
  {Sharon}}]{Armon2011}%
  \BibitemOpen
  \bibfield  {author} {\bibinfo {author} {\bibfnamefont {S.}~\bibnamefont
  {Armon}}, \bibinfo {author} {\bibfnamefont {E.}~\bibnamefont {Efrati}},
  \bibinfo {author} {\bibfnamefont {R.}~\bibnamefont {Kupferman}}, \ and\
  \bibinfo {author} {\bibfnamefont {E.}~\bibnamefont {Sharon}},\ }\href
  {\doibase 10.1126/science.1203874} {\bibfield  {journal} {\bibinfo  {journal}
  {Science}\ }\textbf {\bibinfo {volume} {333}},\ \bibinfo {pages} {1726}
  (\bibinfo {year} {2011})}\BibitemShut {NoStop}%
\bibitem [{\citenamefont {Armon}\ \emph {et~al.}(2014)\citenamefont {Armon},
  \citenamefont {Aharoni}, \citenamefont {Moshe},\ and\ \citenamefont
  {Sharon}}]{Armon2014}%
  \BibitemOpen
  \bibfield  {author} {\bibinfo {author} {\bibfnamefont {S.}~\bibnamefont
  {Armon}}, \bibinfo {author} {\bibfnamefont {H.}~\bibnamefont {Aharoni}},
  \bibinfo {author} {\bibfnamefont {M.}~\bibnamefont {Moshe}}, \ and\ \bibinfo
  {author} {\bibfnamefont {E.}~\bibnamefont {Sharon}},\ }\href {\doibase
  10.1039/c3sm52313f} {\bibfield  {journal} {\bibinfo  {journal} {Soft Matter}\
  }\textbf {\bibinfo {volume} {10}},\ \bibinfo {pages} {2733} (\bibinfo {year}
  {2014})}\BibitemShut {NoStop}%
\bibitem [{\citenamefont {Guest}\ \emph {et~al.}(2011)\citenamefont {Guest},
  \citenamefont {Kebadze},\ and\ \citenamefont {Pellegrino}}]{Guest2011}%
  \BibitemOpen
  \bibfield  {author} {\bibinfo {author} {\bibfnamefont {S.}~\bibnamefont
  {Guest}}, \bibinfo {author} {\bibfnamefont {E.}~\bibnamefont {Kebadze}}, \
  and\ \bibinfo {author} {\bibfnamefont {S.}~\bibnamefont {Pellegrino}},\
  }\href@noop {} {\bibfield  {journal} {\bibinfo  {journal} {J. Mech. Mat.
  Struct.}\ }\textbf {\bibinfo {volume} {6}},\ \bibinfo {pages} {203} (\bibinfo
  {year} {2011})}\BibitemShut {NoStop}%
\bibitem [{\citenamefont {Levin}\ and\ \citenamefont
  {Sharon}(2016)}]{Levin2016}%
  \BibitemOpen
  \bibfield  {author} {\bibinfo {author} {\bibfnamefont {I.}~\bibnamefont
  {Levin}}\ and\ \bibinfo {author} {\bibfnamefont {E.}~\bibnamefont {Sharon}},\
  }\href@noop {} {\bibfield  {journal} {\bibinfo  {journal} {Phys. Rev. Lett.}\
  }\textbf {\bibinfo {volume} {116}},\ \bibinfo {pages} {035502} (\bibinfo
  {year} {2016})}\BibitemShut {NoStop}%
\bibitem [{\citenamefont {Aggeli}\ \emph {et~al.}(1997)\citenamefont {Aggeli},
  \citenamefont {Bell}, \citenamefont {Boden}, \citenamefont {Keen},
  \citenamefont {Knowles}, \citenamefont {McLeish}, \citenamefont
  {Pitkeathly},\ and\ \citenamefont {Radford}}]{Aggeli1997}%
  \BibitemOpen
  \bibfield  {author} {\bibinfo {author} {\bibfnamefont {A.}~\bibnamefont
  {Aggeli}}, \bibinfo {author} {\bibfnamefont {M.}~\bibnamefont {Bell}},
  \bibinfo {author} {\bibfnamefont {N.}~\bibnamefont {Boden}}, \bibinfo
  {author} {\bibfnamefont {J.}~\bibnamefont {Keen}}, \bibinfo {author}
  {\bibfnamefont {P.}~\bibnamefont {Knowles}}, \bibinfo {author} {\bibfnamefont
  {T.}~\bibnamefont {McLeish}}, \bibinfo {author} {\bibfnamefont
  {M.}~\bibnamefont {Pitkeathly}}, \ and\ \bibinfo {author} {\bibfnamefont
  {S.}~\bibnamefont {Radford}},\ }\href {\doibase 10.1038/386259a0} {\bibfield
  {journal} {\bibinfo  {journal} {Nature (London)}\ }\textbf {\bibinfo {volume}
  {386}},\ \bibinfo {pages} {259} (\bibinfo {year} {1997})}\BibitemShut
  {NoStop}%
\bibitem [{\citenamefont {Oda}\ \emph {et~al.}(1999)\citenamefont {Oda},
  \citenamefont {Huc}, \citenamefont {Schmutz}, \citenamefont {Candau},\ and\
  \citenamefont {MacKintosh}}]{Oda1999}%
  \BibitemOpen
  \bibfield  {author} {\bibinfo {author} {\bibfnamefont {R.}~\bibnamefont
  {Oda}}, \bibinfo {author} {\bibfnamefont {I.}~\bibnamefont {Huc}}, \bibinfo
  {author} {\bibfnamefont {M.}~\bibnamefont {Schmutz}}, \bibinfo {author}
  {\bibfnamefont {S.~J.}\ \bibnamefont {Candau}}, \ and\ \bibinfo {author}
  {\bibfnamefont {F.~C.}\ \bibnamefont {MacKintosh}},\ }\href@noop {}
  {\bibfield  {journal} {\bibinfo  {journal} {Nature (London)}\ }\textbf
  {\bibinfo {volume} {399}},\ \bibinfo {pages} {566} (\bibinfo {year}
  {1999})}\BibitemShut {NoStop}%
\bibitem [{\citenamefont {Grossman}\ \emph {et~al.}(2016)\citenamefont
  {Grossman}, \citenamefont {Sharon},\ and\ \citenamefont
  {Diamant}}]{Grossman2016}%
  \BibitemOpen
  \bibfield  {author} {\bibinfo {author} {\bibfnamefont {D.}~\bibnamefont
  {Grossman}}, \bibinfo {author} {\bibfnamefont {E.}~\bibnamefont {Sharon}}, \
  and\ \bibinfo {author} {\bibfnamefont {H.}~\bibnamefont {Diamant}},\ }\href
  {\doibase 10.1103/PhysRevLett.116.258105} {\bibfield  {journal} {\bibinfo
  {journal} {Phys. Rev. Lett.}\ }\textbf {\bibinfo {volume} {116}},\ \bibinfo
  {pages} {258105} (\bibinfo {year} {2016})}\BibitemShut {NoStop}%
\bibitem [{\citenamefont {Hu}\ \emph {et~al.}(2016)\citenamefont {Hu},
  \citenamefont {Kahn}, \citenamefont {Guo}, \citenamefont {Huang},
  \citenamefont {Fadeev}, \citenamefont {Harries},\ and\ \citenamefont
  {Willner}}]{Hu2016}%
  \BibitemOpen
  \bibfield  {author} {\bibinfo {author} {\bibfnamefont {Y.}~\bibnamefont
  {Hu}}, \bibinfo {author} {\bibfnamefont {J.~S.}\ \bibnamefont {Kahn}},
  \bibinfo {author} {\bibfnamefont {W.}~\bibnamefont {Guo}}, \bibinfo {author}
  {\bibfnamefont {F.}~\bibnamefont {Huang}}, \bibinfo {author} {\bibfnamefont
  {M.}~\bibnamefont {Fadeev}}, \bibinfo {author} {\bibfnamefont
  {D.}~\bibnamefont {Harries}}, \ and\ \bibinfo {author} {\bibfnamefont
  {I.}~\bibnamefont {Willner}},\ }\href {\doibase 10.1021/jacs.6b10458}
  {\bibfield  {journal} {\bibinfo  {journal} {Journal of the American Chemical
  Society}\ }\textbf {\bibinfo {volume} {138}},\ \bibinfo {pages} {16112}
  (\bibinfo {year} {2016})},\ \bibinfo {note} {pMID: 27960351},\ \Eprint
  {http://arxiv.org/abs/http://dx.doi.org/10.1021/jacs.6b10458}
  {http://dx.doi.org/10.1021/jacs.6b10458} \BibitemShut {NoStop}%
\bibitem [{\citenamefont {Zhang}\ \emph {et~al.}(2017)\citenamefont {Zhang},
  \citenamefont {Mourran},\ and\ \citenamefont {Moller}}]{Zhang2017}%
  \BibitemOpen
  \bibfield  {author} {\bibinfo {author} {\bibfnamefont {H.}~\bibnamefont
  {Zhang}}, \bibinfo {author} {\bibfnamefont {A.}~\bibnamefont {Mourran}}, \
  and\ \bibinfo {author} {\bibfnamefont {M.}~\bibnamefont {Moller}},\ }\href
  {\doibase 10.1021/acs.nanolett.7b00015} {\bibfield  {journal} {\bibinfo
  {journal} {Nano Letters}\ }\textbf {\bibinfo {volume} {17}},\ \bibinfo
  {pages} {2010} (\bibinfo {year} {2017})},\ \bibinfo {note} {pMID: 28181437},\
  \Eprint {http://arxiv.org/abs/http://dx.doi.org/10.1021/acs.nanolett.7b00015}
  {http://dx.doi.org/10.1021/acs.nanolett.7b00015} \BibitemShut {NoStop}%
\bibitem [{\citenamefont {Yesylevskyy}\ and\ \citenamefont
  {Ramseyer}(2014)}]{Yesylevskyy2014}%
  \BibitemOpen
  \bibfield  {author} {\bibinfo {author} {\bibfnamefont {S.}~\bibnamefont
  {Yesylevskyy}}\ and\ \bibinfo {author} {\bibfnamefont {C.}~\bibnamefont
  {Ramseyer}},\ }\href@noop {} {\bibfield  {journal} {\bibinfo  {journal}
  {Physical Chemistry Chemical Physics}\ }\textbf {\bibinfo {volume} {16}},\
  \bibinfo {pages} {17052} (\bibinfo {year} {2014})}\BibitemShut {NoStop}%
\bibitem [{\citenamefont {Masuda}\ and\ \citenamefont
  {Shimizu}(2004)}]{Masuda2004}%
  \BibitemOpen
  \bibfield  {author} {\bibinfo {author} {\bibfnamefont {M.}~\bibnamefont
  {Masuda}}\ and\ \bibinfo {author} {\bibfnamefont {T.}~\bibnamefont
  {Shimizu}},\ }\href {\doibase 10.1021/la049085y} {\bibfield  {journal}
  {\bibinfo  {journal} {Langmuir}\ }\textbf {\bibinfo {volume} {20}},\ \bibinfo
  {pages} {5969} (\bibinfo {year} {2004})},\ \bibinfo {note} {pMID: 16459618},\
  \Eprint {http://arxiv.org/abs/http://dx.doi.org/10.1021/la049085y}
  {http://dx.doi.org/10.1021/la049085y} \BibitemShut {NoStop}%
\bibitem [{\citenamefont {Marson}\ \emph {et~al.}(2014)\citenamefont {Marson},
  \citenamefont {Phillips}, \citenamefont {Anderson},\ and\ \citenamefont
  {Glotzer}}]{Marson2014}%
  \BibitemOpen
  \bibfield  {author} {\bibinfo {author} {\bibfnamefont {R.~L.}\ \bibnamefont
  {Marson}}, \bibinfo {author} {\bibfnamefont {C.~L.}\ \bibnamefont
  {Phillips}}, \bibinfo {author} {\bibfnamefont {J.~A.}\ \bibnamefont
  {Anderson}}, \ and\ \bibinfo {author} {\bibfnamefont {S.~C.}\ \bibnamefont
  {Glotzer}},\ }\href@noop {} {\bibfield  {journal} {\bibinfo  {journal} {Nano
  letters}\ }\textbf {\bibinfo {volume} {14}},\ \bibinfo {pages} {2071}
  (\bibinfo {year} {2014})}\BibitemShut {NoStop}%
\bibitem [{\citenamefont {Panyukov}\ and\ \citenamefont
  {Rabin}(2000{\natexlab{b}})}]{Panyukov2000}%
  \BibitemOpen
  \bibfield  {author} {\bibinfo {author} {\bibfnamefont {S.}~\bibnamefont
  {Panyukov}}\ and\ \bibinfo {author} {\bibfnamefont {Y.}~\bibnamefont
  {Rabin}},\ }\href {\doibase 10.1103/PhysRevLett.85.2404} {\bibfield
  {journal} {\bibinfo  {journal} {Phys. Rev. Lett.}\ }\textbf {\bibinfo
  {volume} {85}},\ \bibinfo {pages} {2404} (\bibinfo {year}
  {2000}{\natexlab{b}})}\BibitemShut {NoStop}%
\bibitem [{\citenamefont {Efrati}\ \emph
  {et~al.}(2009{\natexlab{a}})\citenamefont {Efrati}, \citenamefont {Sharon},\
  and\ \citenamefont {Kupferman}}]{Efrati2009a}%
  \BibitemOpen
  \bibfield  {author} {\bibinfo {author} {\bibfnamefont {E.}~\bibnamefont
  {Efrati}}, \bibinfo {author} {\bibfnamefont {E.}~\bibnamefont {Sharon}}, \
  and\ \bibinfo {author} {\bibfnamefont {R.}~\bibnamefont {Kupferman}},\ }\href
  {\doibase 10.1016/j.jmps.2008.12.004} {\bibfield  {journal} {\bibinfo
  {journal} {Journal of the Mechanics and Physics of Solids}\ }\textbf
  {\bibinfo {volume} {57}},\ \bibinfo {pages} {762} (\bibinfo {year}
  {2009}{\natexlab{a}})}\BibitemShut {NoStop}%
\bibitem [{\citenamefont {Pezzulla}\ \emph {et~al.}(2016)\citenamefont
  {Pezzulla}, \citenamefont {Smith}, \citenamefont {Nardinocchi},\ and\
  \citenamefont {Holmes}}]{Pezzulla2016}%
  \BibitemOpen
  \bibfield  {author} {\bibinfo {author} {\bibfnamefont {M.}~\bibnamefont
  {Pezzulla}}, \bibinfo {author} {\bibfnamefont {G.~P.}\ \bibnamefont {Smith}},
  \bibinfo {author} {\bibfnamefont {P.}~\bibnamefont {Nardinocchi}}, \ and\
  \bibinfo {author} {\bibfnamefont {D.~P.}\ \bibnamefont {Holmes}},\
  }\href@noop {} {\bibfield  {journal} {\bibinfo  {journal} {Soft matter}\
  }\textbf {\bibinfo {volume} {12}},\ \bibinfo {pages} {4435} (\bibinfo {year}
  {2016})}\BibitemShut {NoStop}%
\bibitem [{\citenamefont {Efrati}\ \emph
  {et~al.}(2009{\natexlab{b}})\citenamefont {Efrati}, \citenamefont {Sharon},\
  and\ \citenamefont {Kupferman}}]{Efrati2009}%
  \BibitemOpen
  \bibfield  {author} {\bibinfo {author} {\bibfnamefont {E.}~\bibnamefont
  {Efrati}}, \bibinfo {author} {\bibfnamefont {E.}~\bibnamefont {Sharon}}, \
  and\ \bibinfo {author} {\bibfnamefont {R.}~\bibnamefont {Kupferman}},\ }\href
  {\doibase 10.1103/PhysRevE.80.016602} {\bibfield  {journal} {\bibinfo
  {journal} {Phys. Rev. E}\ }\textbf {\bibinfo {volume} {80}},\ \bibinfo
  {pages} {016602} (\bibinfo {year} {2009}{\natexlab{b}})}\BibitemShut
  {NoStop}%
\bibitem [{\citenamefont {Ghafouri}\ and\ \citenamefont
  {Bruinsma}(2005)}]{Ghafouri2005}%
  \BibitemOpen
  \bibfield  {author} {\bibinfo {author} {\bibfnamefont {R.}~\bibnamefont
  {Ghafouri}}\ and\ \bibinfo {author} {\bibfnamefont {R.}~\bibnamefont
  {Bruinsma}},\ }\href {\doibase 10.1103/PhysRevLett.94.138101} {\bibfield
  {journal} {\bibinfo  {journal} {Phys. Rev. Lett.}\ }\textbf {\bibinfo
  {volume} {94}},\ \bibinfo {pages} {138101} (\bibinfo {year}
  {2005})}\BibitemShut {NoStop}%
\bibitem [{Big()}]{Bigtheta}%
  \BibitemOpen
  \href@noop {} {}\bibinfo {note} {In the $\Theta(T)$ sense- these function are
  bounded above and below by $T$ up to some factor, for every $T$.}\BibitemShut
  {Stop}%
\bibitem [{\citenamefont {Moroz}\ and\ \citenamefont
  {Nelson}(1997)}]{Moroz1997}%
  \BibitemOpen
  \bibfield  {author} {\bibinfo {author} {\bibfnamefont {J.~D.}\ \bibnamefont
  {Moroz}}\ and\ \bibinfo {author} {\bibfnamefont {P.}~\bibnamefont {Nelson}},\
  }\href@noop {} {\bibfield  {journal} {\bibinfo  {journal} {Proceedings of the
  National Academy of Sciences}\ }\textbf {\bibinfo {volume} {94}},\ \bibinfo
  {pages} {14418} (\bibinfo {year} {1997})}\BibitemShut {NoStop}%
\end{thebibliography}%
\end{document}